\documentclass[aps,prd,preprint,groupedaddress]{revtex4}

\usepackage{epsfig}

\begin{document}

\def\d{{\rm d}}
\def\p{I\!\!P}

\def\lp{\left. }
\def\rp{\right. }
\def\lr{\left( }
\def\rr{\right) }
\def\le{\left[ }
\def\re{\right] }
\def\lg{\left\{ }
\def\rg{\right\} }
\def\lb{\left| }
\def\rb{\right| }

\def\beq{\begin{equation}}
\def\eeq{\end{equation}}
\def\bea{\begin{eqnarray}}
\def\eea{\end{eqnarray}}

\preprint{DESY 10-088}
\preprint{LPSC 10-078}
\title{Suppression factors in diffractive photoproduction of dijets}
\author{Michael Klasen}
\email[]{klasen@lpsc.in2p3.fr}
\affiliation{Laboratoire de Physique Subatomique et de Cosmologie,
 Universit\'e Joseph Fourier/CNRS-IN2P3/INPG, 53 Avenue des Martyrs,
 F-38026 Grenoble, France}
\author{Gustav Kramer}
\affiliation{{II.} Institut f\"ur Theoretische Physik, Universit\"at
 Hamburg, Luruper Chaussee 149, D-22761 Hamburg, Germany}
\date{\today}
\begin{abstract}
After new publications of H1 data for the diffractive photoproduction of 
dijets, which overlap with the earlier published H1 data and the recently
published data of the ZEUS collaboration, have appeared, we have recalculated
the cross sections for this process in next-to-leading order (NLO) of
perturbative QCD to see whether they can be interpreted consistently.
The results of these calculations are compared to the data of both 
collaborations. We find that the NLO cross sections disagree with the data,
showing that factorization breaking occurs at that order. If direct and
resolved contributions are both suppressed by the same amount, the global
suppression factor depends on the transverse-energy cut. However, by 
suppressing only the resolved contribution, also reasonably good agreement
with all the data is found with a suppression factor independent of the 
transverse-energy cut.
\end{abstract}
\maketitle

\section{Introduction}

\begin{figure}
 \centering
 \includegraphics[width=\columnwidth]{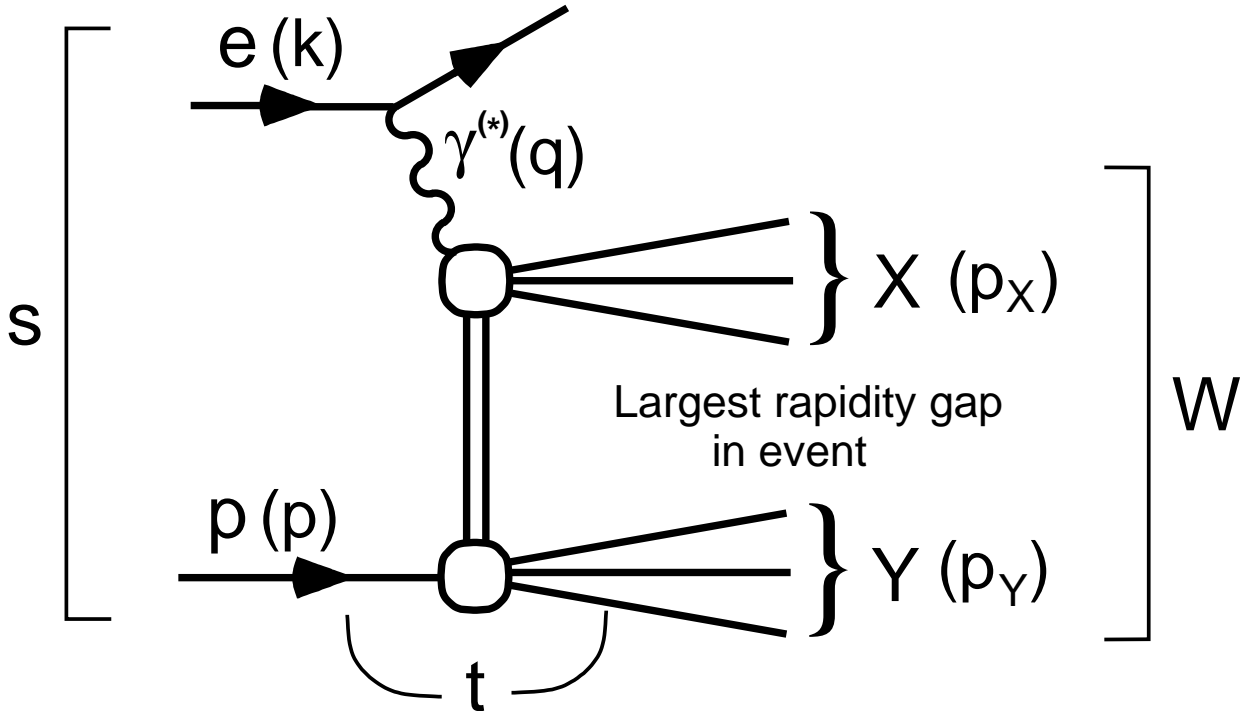}
 \caption{\label{fig:1}Diffractive scattering process $ep\to eXY$, where
 the hadronic systems $X$ and $Y$ are separated by the largest rapidity
 gap in the final state.}
\end{figure}

At high-energy colliders such as the $ep$ collider HERA at DESY and the 
$p\bar{p}$ collider Tevatron at Fermilab, diffractive processes are known to
constitute  an important fraction of all scattering events. These events are
defined experimentally by the presence of a forward-going hadronic system $Y$
with four-momentum $p_Y$, low mass $M_Y$ (typically a proton that remained
intact or a low-lying nucleon resonance), small four-momentum transfer
$t=(p-p_Y)^2$, and small longitudinal momentum transfer $x_{\p} = q(p-p_Y)/
(qp)$ from the incoming proton with four-momentum $p$ to the central hadronic
system $X$ (see Fig.\ \ref{fig:1} for the case of $ep \to eXY$).
%
%
Experimentally a large rapidity gap separates the hadronic system $X$ with
invariant mass $M_X$ from the final-state system $Y$.

Theoretically diffractive interactions are described in the framework of Regge
theory \cite{1} as the exchange of a trajectory with vacuum quantum numbers,
the Pomeron ($\p $) trajectory. Then the object exchanged between the systems
$X$ and $Y$, as indicated in Fig.\ \ref{fig:1}, is the Pomeron (or additional
lower-lying Regge poles), and the upper vertex of $ep \rightarrow eXY$, i.e.
$e \p \rightarrow eX$, can be interpreted as deep-inelastic scattering (DIS)
on the Pomeron target for the case that the virtuality of the exchanged photon
$Q^2 =-q^2$ is sufficiently large. In analogy to DIS on a proton target,
$ep \rightarrow eX$, the cross section for the process $e\p \rightarrow eX$ in
the DIS region can be expressed as the convolution of partonic cross sections 
and universal parton distribution functions (PDFs) of the Pomeron. The partonic
cross  sections are the same as for $ep$ DIS. The Pomeron PDFs are usually
multiplied with vertex functions for the lower vertex in Fig.\ \ref{fig:1},
yielding the diffractive parton distribution functions (DPDFs). The $Q^2$
evolution of the DPDFs is calculated with the usual DGLAP \cite{2} evolution
equations known from $ep \rightarrow eX$ DIS. Except for the $Q^2$ evolution,
the DPDFs can not be calculated in the framework of perturbative QCD and must be
determined from experiment. Such DPDFs \cite{3,4,5,6} have been 
obtained from the HERA inclusive measurements of the diffractive structure 
function $F^D_2$ \cite{3,4}, defined in analogy with the proton structure 
function $F_2$.

The presence of a hard scale such as the squared photon virtuality $Q^2=-q^2$
in deep-inelastic scattering or a large transverse jet energy
$E_T^{jet}$ in the photon-proton center-of-momentum frame should then allow
for calculations of partonic cross sections for the central system $X$ using
perturbative QCD. Such diffractive processes with the presence of a hard
scale are usually called hard diffractive processes.
The central problem in hard diffraction is the problem of QCD
factorization, i.e.\ the question whether diffractive cross sections are
factorisable into universal diffractive parton density functions and
partonic cross sections.

For DIS processes, factorization has indeed been proven to hold \cite{7},
and DPDFs have been extracted at low and intermediate $Q^2$ \cite{3,4} from
high-precision inclusive measurements of the process $ep \rightarrow eXY$  
using the usual DGLAP evolution equations. The proof of the factorization 
formula, usually referred to as the validity of QCD factorization in hard 
diffraction, also appears to be valid for the direct part of photoproduction  
$Q^2 \simeq 0$ or low-$Q^2$ electroproduction of jets \cite{7}. However, 
factorization does not hold for hard processes in diffractive hadron-hadron 
scattering. The problem is that soft interactions between the ingoing hadrons 
and their remnants occur in both the initial and final states. This agrees with 
experimental measurements at the Tevatron \cite{8}. Predictions of diffractive 
dijet cross sections for  collisions as measured by CDF using DPDFs determined
a few years ago \cite{9} and more recently \cite{4} by the H1 collaboration at
HERA overestimate the measured cross section by up to an order of magnitude
\cite{8,Klasen:2009bi}. This suppression of
the CDF cross section can be explained by the rescattering of the two
incoming hadron beams which, by creating additional hadrons, destroy the
rapidity gap \cite{10}.
%
\begin{figure}
 \centering
 \includegraphics[width=0.46\columnwidth]{fig1a}
 \includegraphics[width=0.49\columnwidth]{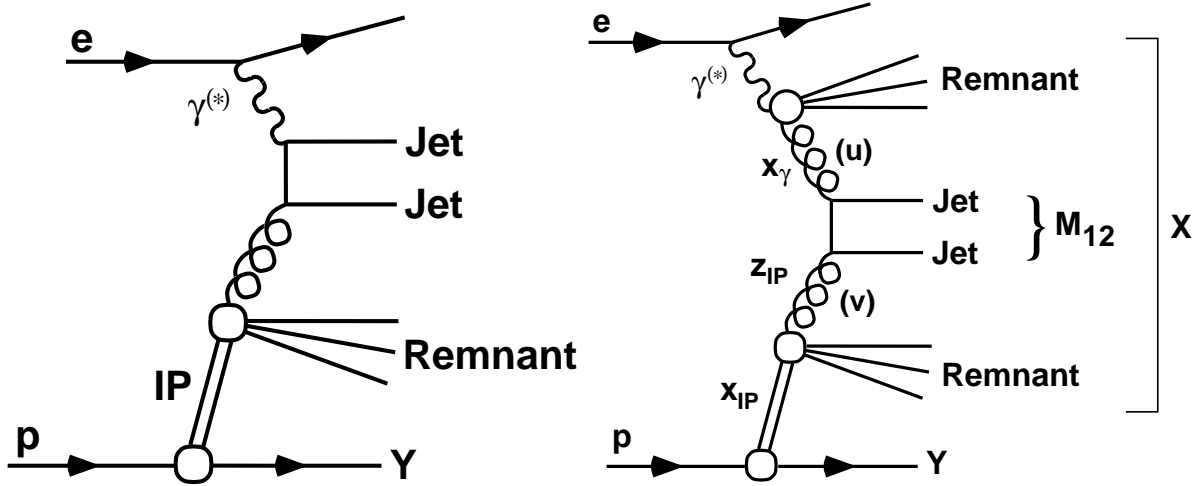}
 \caption{\label{fig:2}Diffractive production of dijets with invariant mass
 $M_{12}$ in direct (left) and resolved (right) photon-pomeron collisions,
 leading to the production of one or two additional remnant jets.}
\end{figure}
%
Jet production with real photons involves direct interactions of the
photon with quarks or gluons from the proton (or in our case from the pomeron)
as well as resolved photon contributions, leading to parton-parton
interactions and an additional remnant jet coming from the photon as reviewed 
in \cite{11} (see Fig.\ \ref{fig:2}). For the direct interactions, 
factorization is expected to be valid as in the case of DIS, whereas we expect 
it to fail for the resolved process as in hadron-hadron scattering. For this 
part of photoproduction one would therefore expect a similar suppression factor
due to rescattering effects of the ingoing partons. Introducing
vector-meson dominance photon fluctuations, such a suppression by about a
factor of three for resolved photoproduction at HERA was predicted \cite{12}.

On the experimental side, the first measurements of dijet cross sections in
diffractive photoproduction have been presented by the H1 collaboration as
contributions to two conferences \cite{13}. The kinematic range for these
data were $Q^2 < 0.01$ GeV$^2$, $x_{\p} < 0.03$, $E_T^{jet1} > 5$ GeV, 
$E_T^{jet2} > 4$ GeV and $165<W<240$ GeV, where jets were identified using the 
inclusive $k_T$-cluster algorithm. The measured cross sections as a function 
of $x_{\gamma}^{obs}$ and $z_{\p}^{obs}$ were compared to leading-order (LO) 
QCD predictions, using the RAPGAP Monte Carlo model \cite{14}. For the DPDFs 
the LO `H1 2002 fit' was used \cite{9}. It was found that these two cross 
sections were well described by the predictions in normalization and shape over
the whole range of $x_{\gamma}^{obs}$ and $z_{\p}^{obs}$, showing no breakdown 
of factorization in either the resolved or in the direct photoproduction. In 
addition, normalized cross sections as a function of various other variables 
were compared to the predictions with the result that all measured 
distributions were in good agreement.

Subsequently we calculated the next-to-leading order (NLO) corrections for
the cross section of diffractive dijet production using the same
kinematic cuts and with the same DPDFs as in the first H1 analysis
\cite{13} on the basis of our previous work on NLO corrections for
inclusive direct \cite{15} and resolved \cite{16} dijet photoproduction.
While at LO good agreement with the H1 data \cite{13} was found, consistent
with the finding in the H1 analysis \cite{13}, it was found that the NLO
corrections increase the cross section significantly \cite{17,18} and
require a suppression factor of the order of $R=0.5$. Since on theoretical
grounds only a suppression of the resolved cross section would be
acceptable, we demonstrated in \cite{17,18} that by multiplying the
resolved cross section with a suppression factor of $R = 0.34$, reasonably
good agreement with the preliminary H1 data \cite{13} could be achieved.
This value for the suppression factor turned out to be in good agreement
with the prediction of \cite{12}.

The first experimental data from the ZEUS collaboration were presented at
the DIS workshop in 2004 \cite{19}. The dijet cross sections were obtained
in the kinematic range $Q^2 < 1$ GeV$^2$, $x_{\p} < 0.025$ and $E_T^{jet1(2)}
> 7.5$ $(6.5)$ GeV. For these kinematic constraints NLO calculations were not
available in 2004. So, the measurements were compared to LO calculations, 
unfortunately with previous H1 DPDFs \cite{20} with the result that good 
agreement in the shape was achieved. However,
the normalization was off by a factor of
$0.6$, which was attributed later to the older DPDF input \cite{21}, so that 
the H1 and ZEUS results were consistent with each other. The situation 
concerning the agreement of H1 and ZEUS data and the influence of NLO 
corrections improved already considerably in the fall of 2004. At the ICHEP 
2004 both collaborations presented their new data and compared them with NLO 
cross section calculations. H1 compared their data with the predictions from 
the program of Frixione \cite{22} and ZEUS with our calculations along the 
lines of \cite{17,18}, where now only the different kinematic cuts of the ZEUS 
analysis had to be incorporated. Both collaborations also used the same DPDFs, 
namely the `H1 2002 fit' \cite{9}. The conclusion from the comparison of the 
respective NLO calculations with the H1 \cite{23} and ZEUS \cite{24} 
measurements were very similar. Both collaborations observed from their data 
that good agreement was achieved with the global suppression (of direct and 
resolved contribution) by a factor $0.5$. Concerning the model with suppression
of the resolved contribution only, the H1 collaboration concluded that this 
model was disfavored compared to the global suppression model. This result 
was obtained by comparing the differential cross sections 
$d\sigma/dx_{\gamma}^{obs}$ and $d\sigma/dy$ to a calculation, in which
the partonic cross sections were suppressed for $x_{\gamma}^{obs} < 0.9$ by a 
factor of $R=0.34$. In the ZEUS contribution to the ICHEP 2004,
differential cross 
sections of several observables had been shown and compared with our 
calculations. All five comparisons, namely for $y$, $x_{\p}$, 
$z_{\p}^{obs}$, $E_T^{jet1}$ and $\eta^{jet1}$, showed very good agreement 
both in shape and normalization of the cross section with the resolved 
photon suppression ($R = 0.34$). The only exception was the comparison of 
$d\sigma/dx_{\gamma}^{obs}$, which did not show agreement with the resolved 
photon suppression only. Here the suppression factor varied as a function of  
$x_{\gamma}^{obs}$ between 0.5 and 0.75 and the description of the shape was 
better without any suppression, which would signal a global suppression. We 
emphasize that at this stage of the analysis at the end of 2004, both 
the H1 \cite{23} and ZEUS \cite{24} collaborations showed that their data
were consistent with a global suppression of about a factor of two against the
sum of direct and resolved NLO QCD predictions. In addition, in the H1
contribution \cite{23} it was claimed explicitly that there was evidence 
for factorization breaking also in direct photoproduction. In 2005 the ZEUS
collaboration presented a more detailed comparison by dividing their data
sample into two subsets, $x_{\gamma}^{obs} > 0.75$ and 
$x_{\gamma}^{obs} < 0.75$, in 
order to be more sensitive to the suppression of the direct ($x_{\gamma}^{obs}
> 0.75$) and the resolved ($x_{\gamma}^{obs} < 0.75$) components. Together
with the result of H1 the overall conclusion at the DIS 2005 workshop was the
same as at the end of 2004, namely, that good agreement was achieved with the
global suppression of $0.5$, while a suppression of only the resolved
contribution at NLO was disfavored by the data \cite{25}.

The analysis of the ZEUS data with respect to the samples enriched in
direct and resolved processes was continued in a contribution to the
Uppsala Lepton-Photon conference in 2005 \cite{26}. For $x_{\gamma}^{obs} >
0.75$ the NLO 
predictions gave a good description of the shape of the measured cross section,
although the absolute normalization was a factor of two above the data. For 
$x_{\gamma}^{obs} < 0.75$ the NLO calculations were again above the data when 
no suppression ($R = 1$) was applied and below the data by a factor of two when
a suppression with $R =0.34$ was applied to the resolved photon processes. 
The ratio of the resolved-enriched to the direct-enriched samples was
reasonably well 
reproduced by the NLO predictions with $R = 1$, indicating that a suppression 
of the resolved sample with respect to the direct sample was not seen in any
particular kinematic region. This agreed with the earlier findings that a
uniform suppression for both resolved and direct process gives a better
description of the data. Of course, all these conclusions relied on the
fact, that the DPDFs as evaluated by H1 \cite{9} are really the correct
ones. The analysis in \cite{26} was based on the largest selection of
variables so far, namely $y, x_{\p},M_X,z_{\p}^{obs},E_T^{jet1}$ and 
$\eta^{jet1}$.

The conclusions above concerning the global overall suppression versus a
suppression in the resolved contribution only based solely on preliminary
data from H1 and ZEUS and the preliminary `H1 2002 fit' \cite{9} for the
DPDFs remained also after 2005 until the final publications of the H1 and
ZEUS analysis appeared in 2007. The comparison between the final
experimental results and the NLO theory used the new and final DPDFs
constructed by the H1 collaboration \cite{4}. This analysis was based
on the larger sample of the years 1997-2000 as compared to the previous
published PDF sets. In \cite{4} two NLO fits, `H1 2006 fit A' and `H1 2006
fit B' were presented, which both give a good description of inclusive
diffraction. These two sets of PDFs differ mainly in the gluon density at
large fractional parton momentum, which is poorly constrained by the
inclusive diffractive scattering data, since there is no direct coupling
of the photon to gluons, so that the gluon density is constrained only
through the evolution. The gluon density of fit A is peaked at the starting
scale at large fractional momentum, whereas the fit B is flat in that region.

The differential cross sections as measured in diffractive photoproduction
by H1 \cite{27} were compared with the NLO predictions obtained with the
Frixione program \cite{22}, interfaced to the `H1 2006 fit B' DPDFs. In
this publication \cite{27}, the conclusions deduced earlier from the
comparison with the preliminary data and the preliminary `H1 2002 fit'
\cite{9} are fully confirmed, now also with the new DPDF fits \cite{4}.
In particular, the global suppression is obtained, independent of the
DPDF fits used, i.e.\ fit A or fit B, by considering the ratio of measured
dijet cross section to NLO predictions in photoproduction in relation to
the same ratio in DIS. In this comparison the value of the suppression is 
$0.5 \pm 0.1$. In addition, by using the overall suppression factor 0.5, H1 
obtained a good description of all the measured distributions in the variables
$z_{\p}^{obs}$, $x_{\gamma}^{obs}$, $x_{\p}$, $W$, $E_T^{jet1}$,
$\bar{\eta}_{jet}^{lab}$, $|\Delta \eta_{jet}|$ and $M_{12}$ interfaced with the
`H1 2006 fit B' DPDFs and 
taking into account hadronization corrections \cite{27}. Finally, the 
H1 collaboration investigated how well the data are describable under the 
assumption that in the NLO calculation the cross section for 
$x_{\gamma}^{obs} > 0.9$ is not suppressed. The best agreement
in a fit was obtained for a suppression factor 0.44 for the NLO calculation
with $x_{\gamma}^{obs} < 0.9$, based on fitting the distributions for 
$x_{\gamma}^{obs},W,\bar{\eta}^{lab}_{jet}$ and $E_T^{jet1}$. In this comparison
they found disagreement for the largest $x_{\gamma}^{obs}$-bin and the lowest 
$\bar{\eta}^{lab}_{jet}$-bin (which are related), but better agreement in the 
$E_T^{jet1}$-distribution.
In \cite{27} this leads to the statement, that the assumption that the
direct cross section obeys factorization is strongly disfavored by their
analysis. In total, it is obvious that in the final H1 analysis \cite{27}
a global suppression in diffractive dijet photoproduction is clearly
established.

Just recently also the ZEUS collaboration presented their final result on
diffractive dijet photoproduction \cite{28}. As in their preliminary
analysis, the two jets with the highest transverse energies $E_T^{jet}$ were
required to satisfy $E_T^{jet1(2)} > 7.5$
$(6.5)$ GeV, which is higher than in the
H1 analysis with $E_T^{jet1(2)} > 5$ $(4)$ GeV \cite{27}. ZEUS compared their 
measurements with the NLO predictions for diffractive photoproduction of dijets
based on our program \cite{18}. Three sets of DPDFs were used, the ZEUS LPS
fit, determined from a NLO analysis of inclusive diffraction and diffractive 
charm-production data \cite{3}, and the two H1 fits, `H1 2006 fits A,B' 
\cite{4}. The NLO results obtained with the two H1 fits were scaled down by a 
factor of 0.87 \cite{4} since the H1 measurements used to derive the DPDFs 
include low-mass proton dissociative processes with $M_Y < 1.6$ GeV, 
which increases the photon-diffractive cross section by $1.15^{+0.15}_{-0.08}$
as compared to the pure proton final state as corrected to in the ZEUS
analysis. The comparison of the measured cross sections and the
theoretical predictions was based on the differential cross sections in
the variables $y$, $M_X$, $x_{\p}$, $z_{\p}^{obs}$, $E_T^{jet1}$, $\eta^{jet1}$
and  $x_{\gamma}^{obs}$. The data were reasonably well described in their shape
as a function of these variables and lay systematically below the
predictions. The predictions for the three DPDFs differed appreciably. The
cross sections for the `H1 2006 fit A' (`H1 2006 fit B') were the highest
(lowest), and the
one for the ZEUS LPS fit lay between the two, but nearer to the fit A than the
fit B predictions. For $d\sigma/dx_{\gamma}^{obs}$ ZEUS also showed the ratio
of the data and the NLO predictions using the ZEUS LPS fit. The ratio was 
consistent with a suppression factor of 0.7 independent of $x_{\gamma}^{obs}$.
This suppression factor depended
on the DPDFs and ranged between 0.6 (`H1 2006 fit A') and 0.9 (`H1 2006 fit B').
Taking into
account the scale dependence of the theoretical predictions the ratio was
outside the theoretical uncertainty for the ZEUS LPS fit and the `H1 2006 fit A',
but not for the `H1 2006 fit B'. In their conclusions the authors
of the ZEUS analysis \cite{28} made the statement that the NLO
calculations tend to overestimate the measured cross section, which would
mean that a suppression is present. Unfortunately, however, they continued,
that, within the large
uncertainties of the NLO calculations, the data were compatible with the QCD
calculations, i.e.\ with no suppression.

Such a statement clearly contradicts the result of the H1 collaboration
\cite{27} and casts doubts on the correctness of the H1 analysis. The
authors of \cite{28} attribute this discrepancy to the fact that the H1
measurements \cite{27} were carried out in a lower $E_T^{jet}$ and a higher
$x_{\p}$ range than those in the ZEUS study \cite{28}.
Besides the different $E_T^{jet}$ and $x_{\p}$  regions in \cite{27} and 
\cite{28} the two measurements suffer also from different experimental cuts of 
some other variables which makes it difficult to compare the two data sets 
directly (note also the lower center-of-mass energy for the H1 data). With
this in mind the H1 collaboration has done a second analysis \cite{Aaron:2010su,%
29}, in
which most of the experimental cuts are taken as in the ZEUS \cite{28}
analysis, i.e.\ the cuts on $x_{\p}$, $\eta^{jet1(2)}$ and on $E_T^{jet1(2)}$.
In addition they analyzed data sets also with the lower $E_T^{jet}$ cut,
namely $E_T^{jet1(2)} >
5$ $(4)$ GeV and with $x_{\p} < 0.03$ as in the previous H1 dijet analysis 
\cite{27}. Starting from these recent data \cite{Aaron:2010su,29} we have performed a new 
calculation of the NLO cross sections on the basis of \cite{18} for the new 
H1 \cite{Aaron:2010su,29} and the latest ZEUS \cite{28} analyses with the same DPDFs as 
input, in order to see whether we can confirm the different conclusions 
obtained from the older H1 \cite{27} and the ZEUS \cite{28} measurements. In 
this new comparison between the experimental and the theoretical results we 
shall concentrate on using the `H1 2006 fit B' as DPDF input, since it 
leads to smaller NLO cross sections than the DPDFs based on the 
`H1 2006 fit A' or the ZEUS LPS fit.

In section 2 we shall present, after defining the complete list of cuts on the
experimental variables and giving all the input used in the cross section
calculations, the comparison with the new H1 experimental data \cite{Aaron:2010su,29}.
In this comparison we shall concentrate on the main question, whether
there is a suppression in the photoproduction data at all. In addition we
shall investigate also whether a reasonable description of the data is
possible with suppression of the resolved cross section only, as we
studied it already in our previous work in 2004 \cite{17,18}. In section 3
the same comparison with the ZEUS data \cite{28} will be performed. In
section 4 we shall finish with a summary and our conclusions.

\section{Comparison with recent H1 data}

The recent H1 data for diffractive photoproduction of dijets \cite{Aaron:2010su,%
29} have
several advantages as compared to the earlier H1 \cite{27} and ZEUS
\cite{28} analyses. First, the integrated luminosity is three times higher
than in the previous H1 analysis \cite{27} comparable to the luminosity in
the ZEUS analysis \cite{28}. Second, H1 took data with low-$E_T^{jet}$
\cite{Aaron:2010su,29} and 
high-$E_T^{jet}$ \cite{29} cuts, which allows the comparison with 
\cite{27} and \cite{28}. The exact two kinematic ranges are given in Tab.\ 1.
%
\begin{table}
 \caption{Kinematic cuts applied in the most recent H1 analyses of diffractive
 dijet photoproduction \cite{Aaron:2010su,29}.}
 \begin{tabular}{c|c}
 H1 low-$E_T^{jet}$ cuts & H1 high-$E_T^{jet}$ cuts\\
 \hline
 $Q^2<$ 0.01 GeV$^2$   &  $Q^2<$ 0.01 GeV$^2$ \\
 0.3 $< y <$ 0.65     &  0.3 $< y <$ 0.65\\
 $E_T^{jet1}>$ 5 GeV & $E_T^{jet1} >$ 7.5 GeV \\
 $E_T^{jet2}>$ 4 GeV & $E_T^{jet2} >$ 6.5 GeV \\
 $-1 < \eta^{jet1(2)} < 2$ &  $-1.5 < \eta^{jet1(2)} < 1.5$ \\
 $z_{\p} <$ 0.8    &  $z_{\p} <$ 1  \\
 $x_{\p} <$ 0.03      & $x_{\p} <$ 0.025 \\
 $|t| <$ 1 GeV$^2$    &  $|t| <$ 1 GeV$^2$ \\
 $M_Y <$ 1.6 GeV      &  $M_Y <$ 1.6 GeV\\
 \end{tabular}
\end{table}
%
These ranges for the low-$E_T^{jet}$ cuts are as in the previous H1 analysis 
\cite{27} and for the high-$E_T^{jet}$ cuts are chosen as 
in the ZEUS analysis with two exceptions. In the ZEUS analysis the maximal cut 
on $Q^2$ is larger and the data are taken in an extended $y$ range. The 
definition of the various variables can be found in the H1 and ZEUS
publications
\cite{27,28} and in our previous work \cite{17,18}. Very important is
the cut on $x_{\p}$. It is kept small in both analysis in order for the 
pomeron exchange to be dominant. We base our analysis on the low-$E_T$ data
published in \cite{Aaron:2010su}, which differ from the data in \cite{29} in the
cut on $z_{\p}$ influencing not only the experimental data, but also the NLO
results
in all variables except the distribution in $z_{\p}$. The preliminary
low-$E_T^{jet}$ data \cite{29} have been compared previously to our theoretical
results at NLO in \cite{Kramer:2009zzc}.
In the experimental analysis as well as in the
NLO calculations, jets are defined with the inclusive $k_T$-cluster algorithm
with a distance parameter $d = 1$ \cite{30} in the laboratory frame. At
least two jets are required with the respective cuts on $E_T^{jet1}$ and 
$E_T^{jet2}$, where $E_T^{jet1(2)}$ refers to the jet with the largest 
(second largest) $E_T^{jet}$. As is well known, the lower
limits on the jet $E_T$ are chosen asymmetric in order to avoid an infrared
sensitivity in those NLO cross section computations, which are integrated
over $E_T^{jet}$ \cite{31}.

Before we confront the calculated cross sections with the experimental
data, we correct them for hadronization effects. The hadronization
corrections are calculated by means of the LO RAPGAP Monte Carlo generator
\cite{14}. The factors for the transformation of jets made up of stable
hadrons 
to parton jets were supplied by the H1 collaboration \cite{Aaron:2010su,29}. Most of our 
calculations are done with the `H1 2006 fit B' \cite{4} DPDFs, since they give the
smaller diffractive dijet cross sections as compared to the `H1 2006 fit A'.
The H1 and ZEUS collaborations constructed two more sets of DPDFs, which are
called `H1 2007 fit jets' and `ZEUS DPDF SJ'. These fits are obtained through
simultaneous fits to the diffractive inclusive and DIS dijet cross sections
\cite{32}. In these fits it is
assumed that there is no factorization breaking in the diffractive DIS dijet 
cross sections. Including these cross sections in the fits leads to additional 
constraints, mostly for the diffractive gluon distribution. On average the
`H1 2007 fit jets' is similar to the `H1 2006 fit B' except for the gluon
distribution at large momentum fraction and small factorization scale. In the
following analysis we shall disregard these new DPDF sets, since they would be
compatible with the factorization test of the photoproduction data only, if we
restricted these tests to the case that the resolved part has the breaking and 
not the direct part, which has the same theoretical structure as the DIS 
dijet cross section. Results with the `H1 2007 fit jets' can be found in
\cite{Aaron:2010su,29}. The `H1 2006 fits A,B' are based on $n_f=3$ massless 
flavors.
%
%
The production of
charm quarks was treated in the Fixed-Flavor Number Scheme (FFNS) in NLO
with non-zero charm-quark mass yielding a diffractive $F_2^c$. This
$F_2^c$ is contained in the 'H1 2006 fits A,B' parameterizations and is
then converted by us into a charm PDF
using the LO expression $F_2^c(x,Q^2)=2xe_c^2f_c(x,Q^2)$, where $e_c=2/3$ is
the electric charge of the charm quark.
%
%
%
 The bottom contribution was neglected. This 
assumption simplifies the calculations considerably. Since the charm 
contribution from the Pomeron is small, this should be a good approximation.
We then take $n_f=4$ with $\Lambda^{(4)}_{\overline{\rm MS}} = 0.347$ GeV, 
which corresponds to the value used in the DPDFs `H1 2006 fits A,B' \cite{4}. 
For the photon PDF we have chosen the NLO GRV parametrization transformed 
to the $\overline{\rm MS}$ scheme \cite{33}.

As it is clear from the discussion of the various preliminary analyses of
the H1 and ZEUS collaborations, there are two questions which we would
like to answer from the comparison with the recent H1 and the ZEUS data.
The first question is whether a suppression factor (sometimes also called 
rapidity-gap survival probability), which differs substantially from one, is 
needed to describe the data. The second question is whether the data are also
consistent with a suppression factor applied to the resolved cross section 
only. To give an answer to these two questions we calculated first the cross 
sections with no suppression factor ($R = 1$ in the following figures) with a
theoretical error obtained from varying the common scale of renormalization and
factorization by factors of 0.5  and 2 around the default value (highest
$E_T^{jet}$). In a second step we show the results  
for the same differential cross sections with a global suppression factor,
adjusted to $d\sigma/dE_T^{jet1}$ at the smallest $E_T^{jet1}$-bin. As in the
experimental analyses \cite{Aaron:2010su,29}, we consider the differential cross sections
in  the variables $x_{\gamma}^{obs}$, $z_{\p}^{obs}$, $\log_{10}(x_{\p})$,
$E_T^{jet1}$,  $M_X$, $M_{12}$, $\overline{\eta}^{jets}$, $|\Delta
\eta^{jets}|$ and $W$. The definition of the variables is given in the
experimental papers \cite{27,28,Aaron:2010su,29} or in our earlier work \cite{17,18,34}. In 
the latter references also the relevant formulas for the calculation of the 
dijet cross sections can be found.

For the low-$E_T^{jet}$ cuts, the resulting suppression factor is 
$R = 0.50\pm 0.09$, which gives in the lowest $E_T^{jet1}$-bin a cross section 
equal to the experimental data point. The error comes from the combined 
experimental statistical and systematic error. The theoretical error due to the
scale variation is taken into account when comparing to the various 
distributions. The result of this comparison is shown in Figs.\ 3a-i.
%
\begin{figure}
 \centering
 \includegraphics[width=0.325\columnwidth]{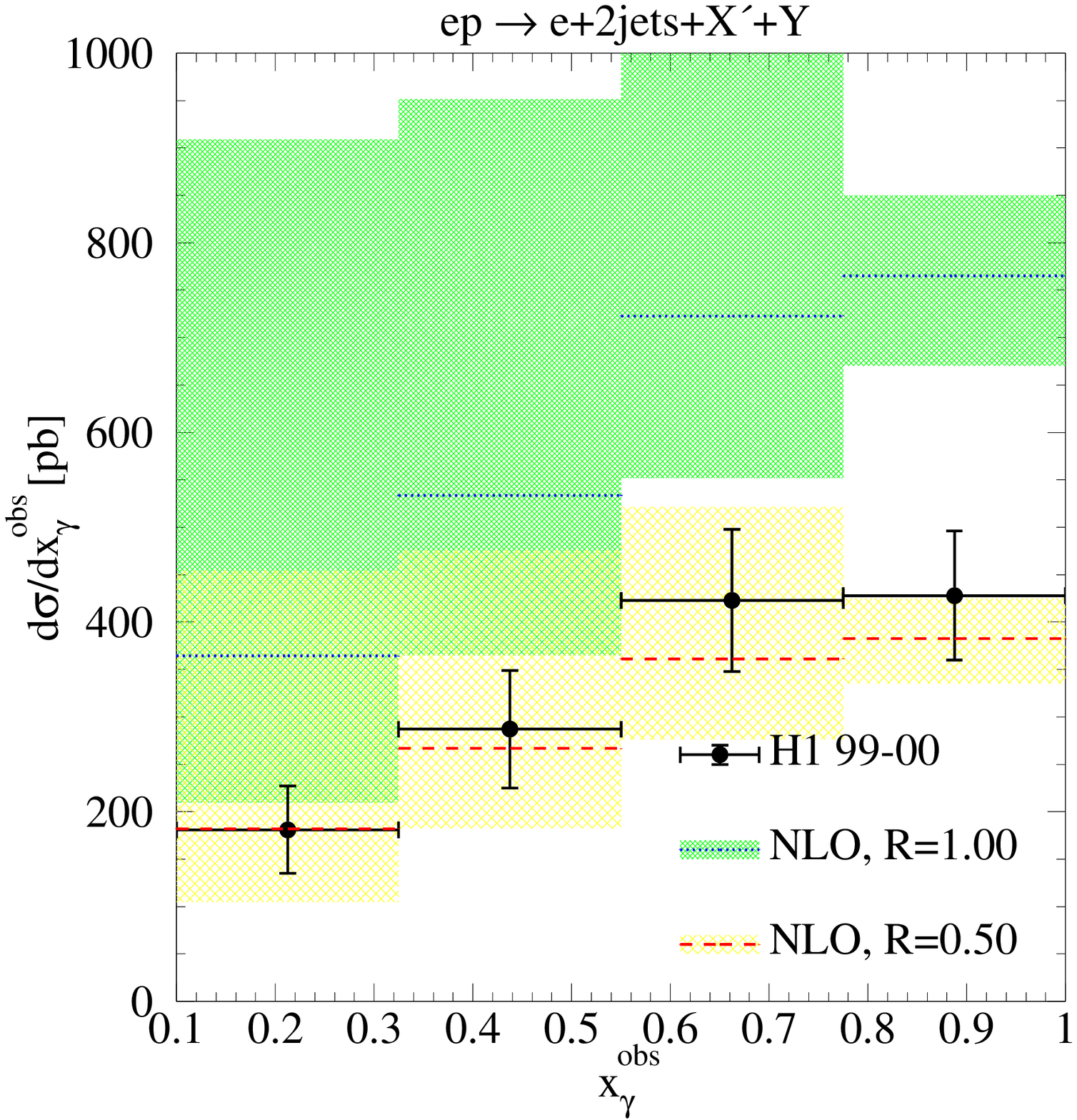}
 \includegraphics[width=0.325\columnwidth]{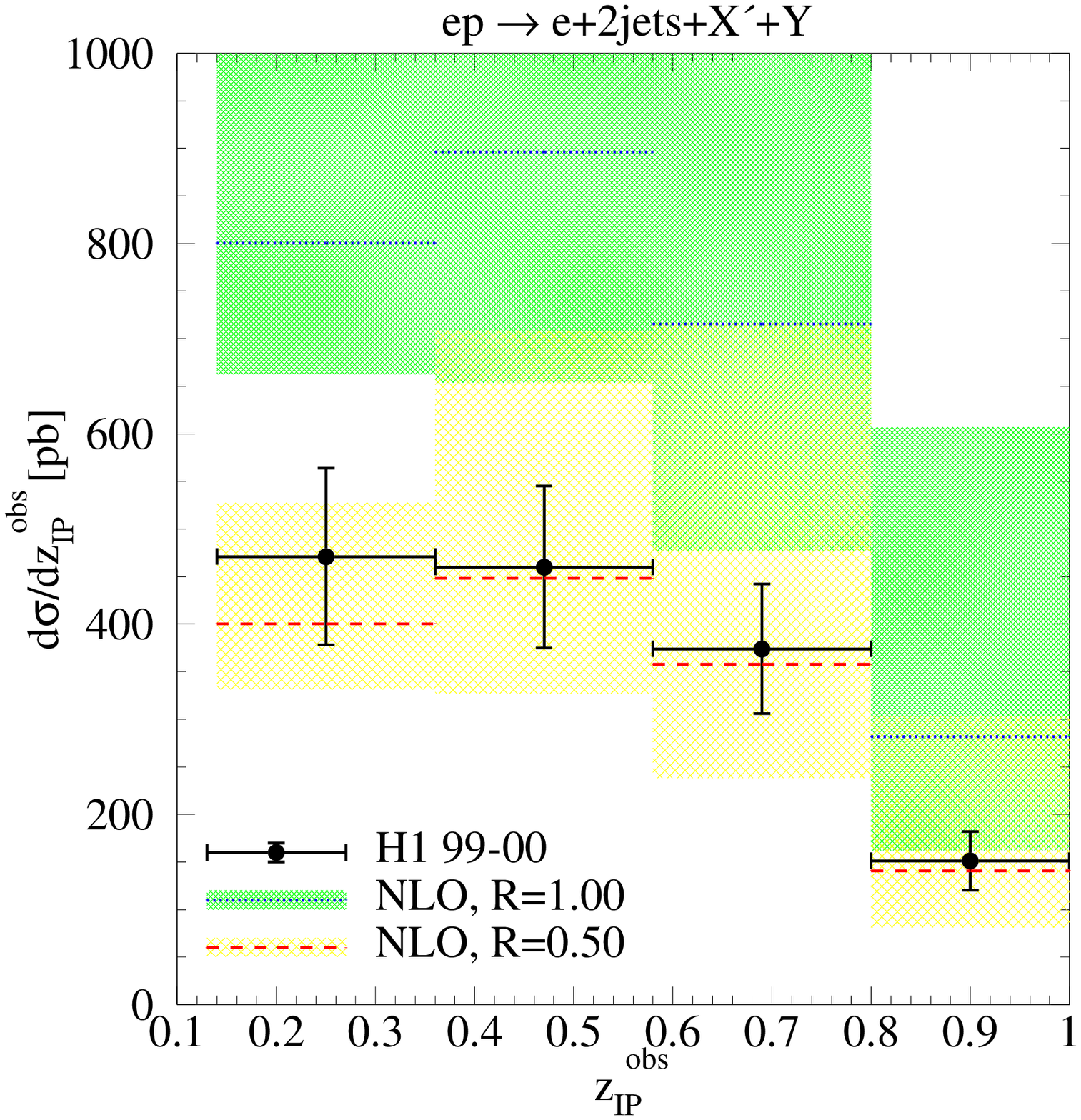}
 \includegraphics[width=0.325\columnwidth]{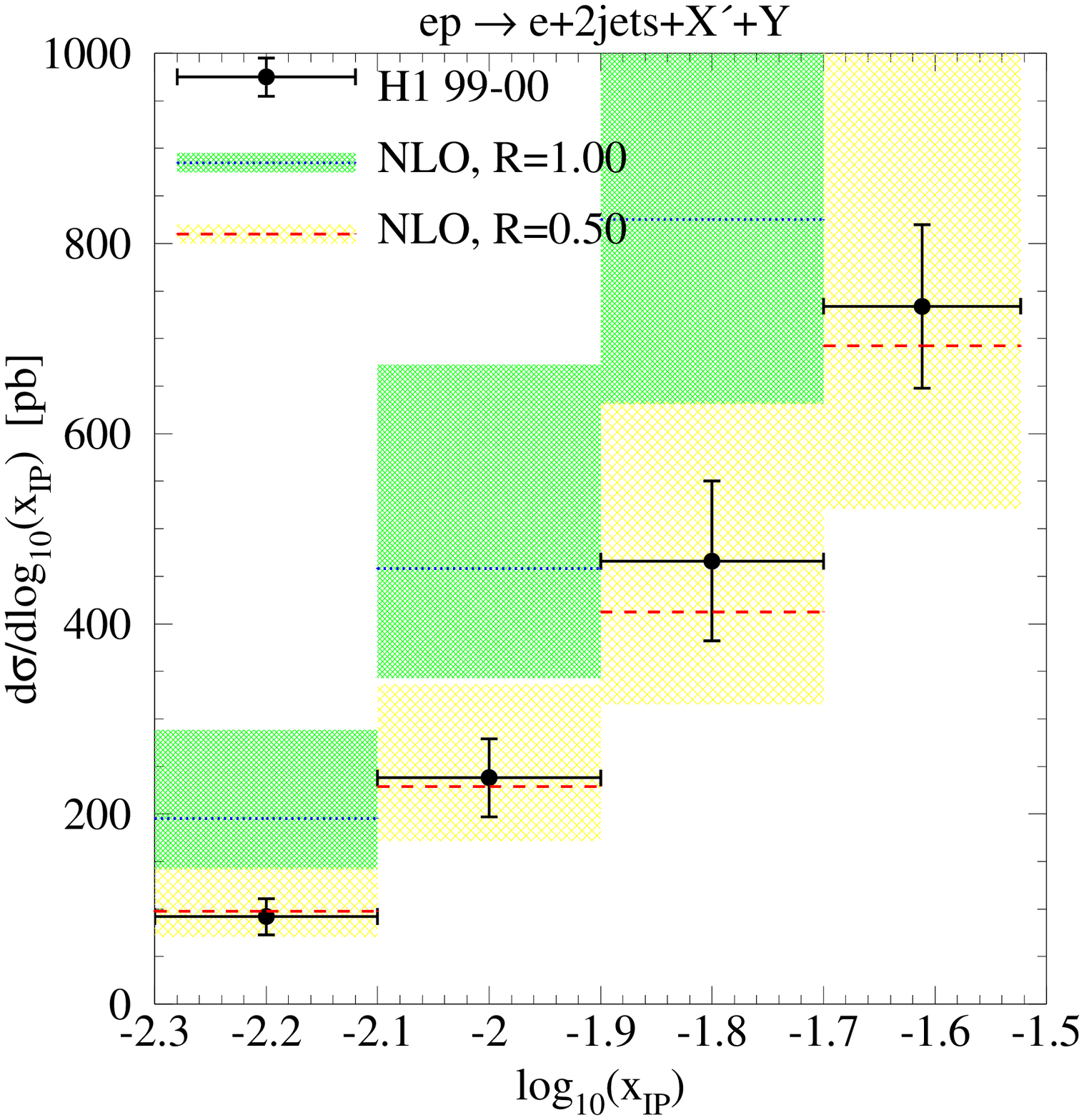}
 \includegraphics[width=0.325\columnwidth]{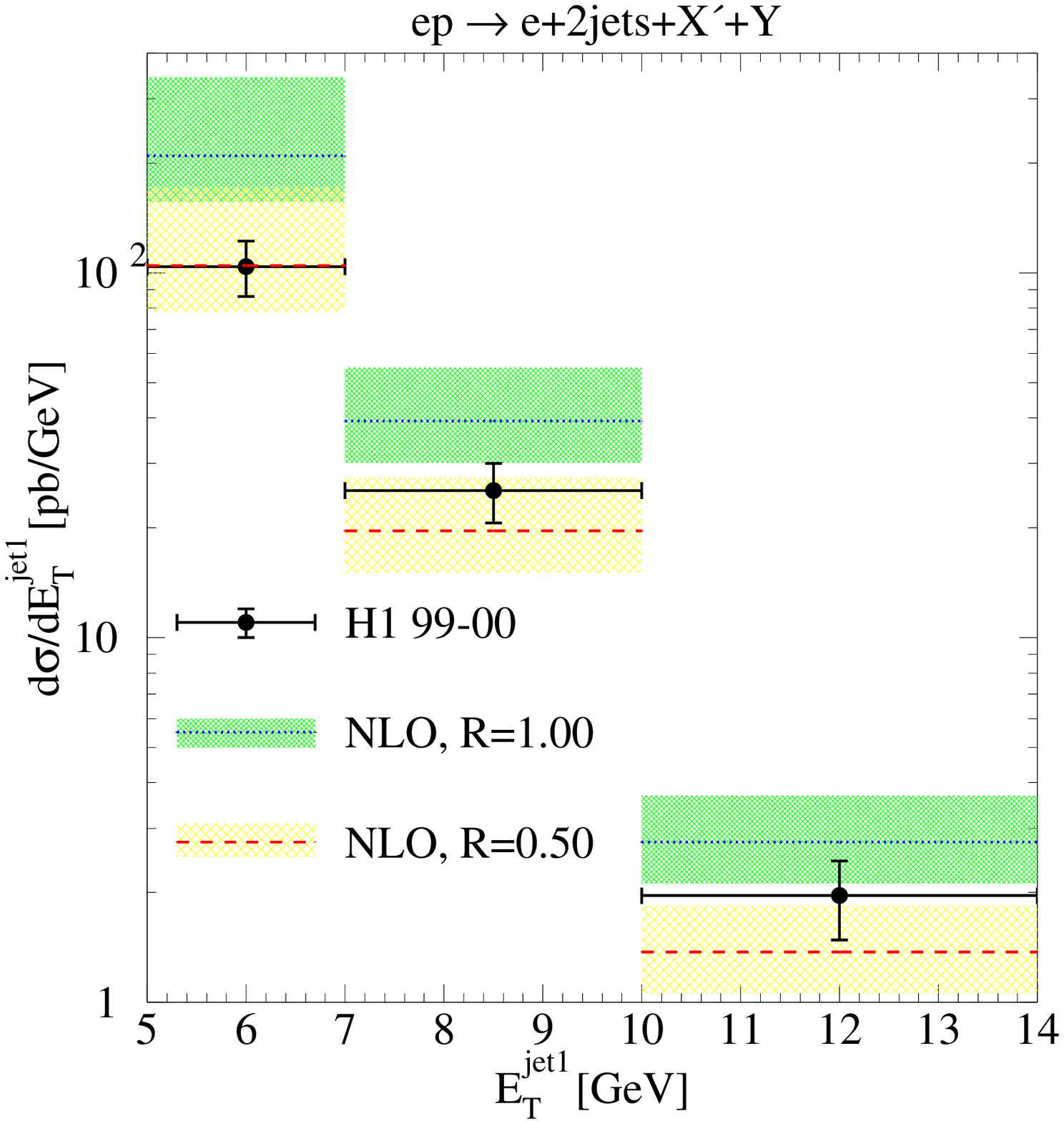}
 \includegraphics[width=0.325\columnwidth]{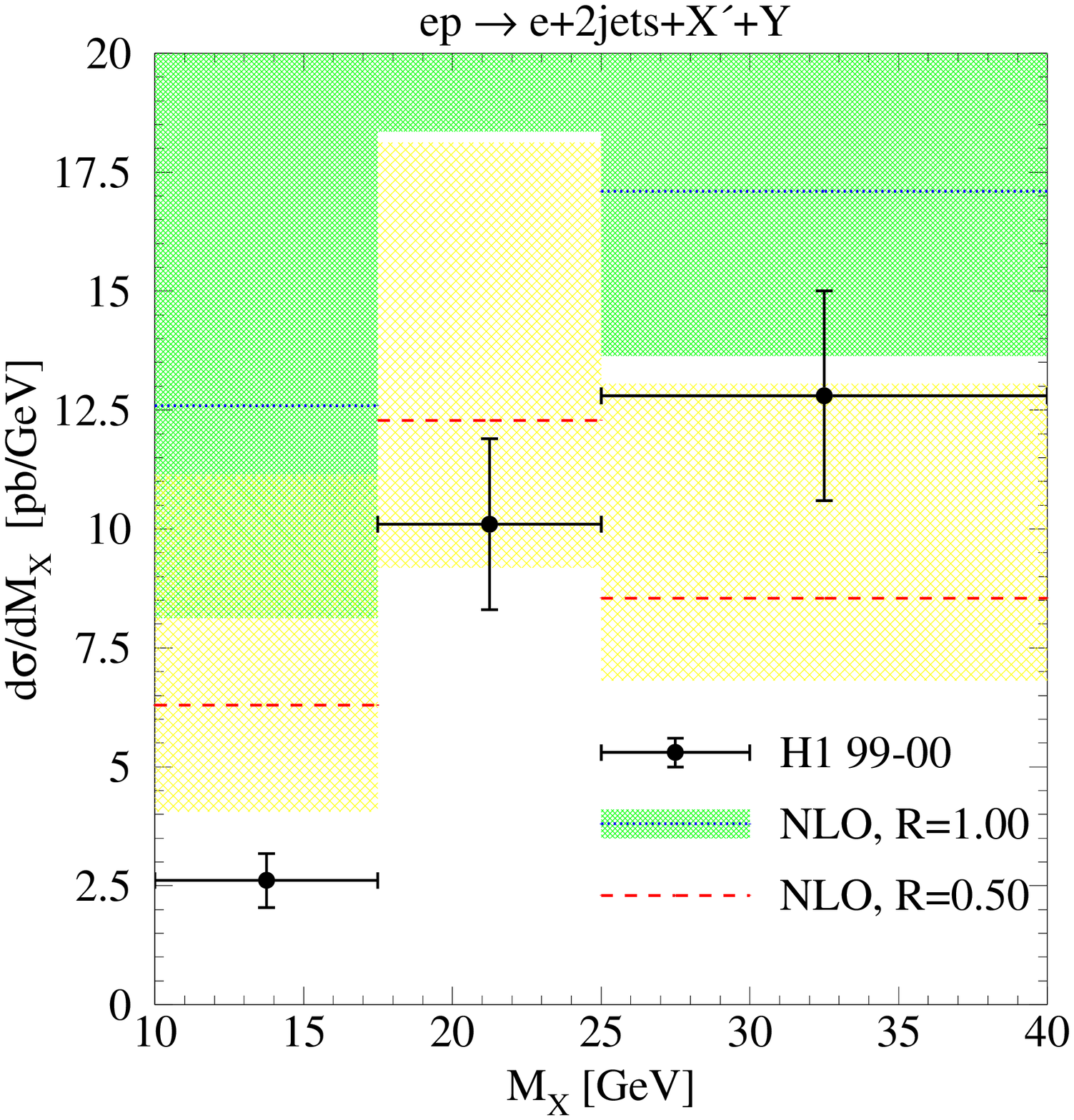}
 \includegraphics[width=0.325\columnwidth]{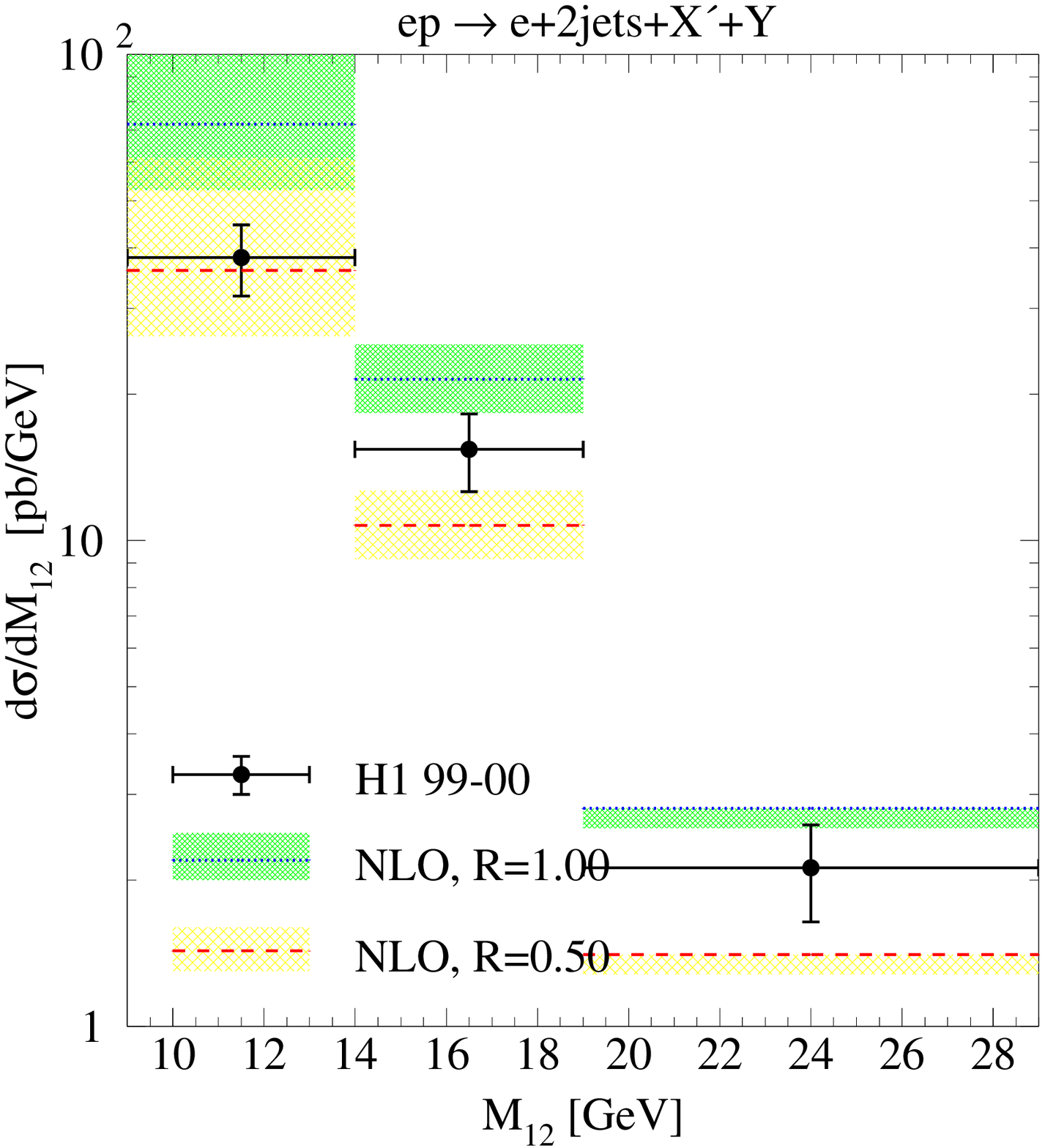}
 \includegraphics[width=0.325\columnwidth]{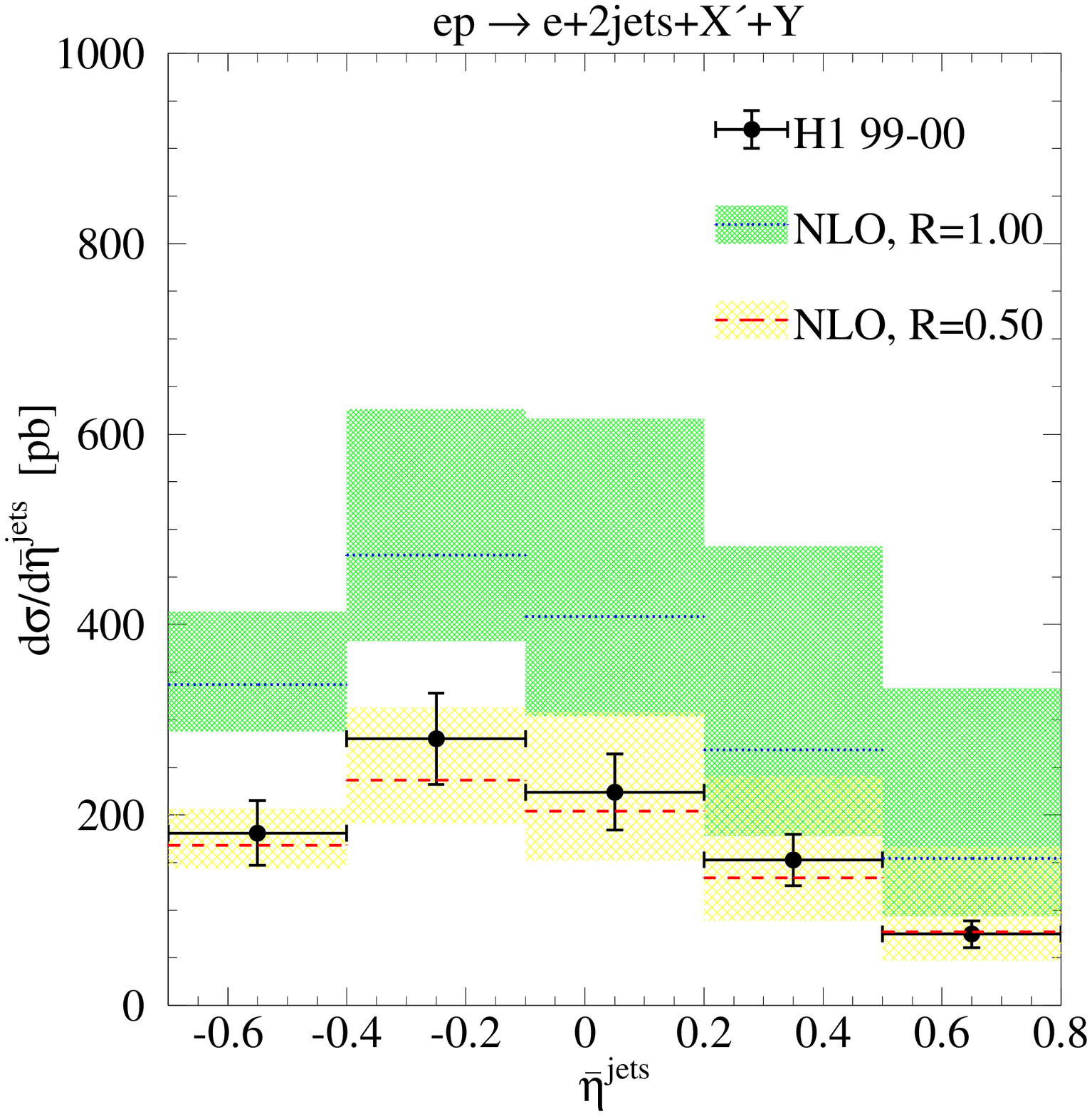}
 \includegraphics[width=0.325\columnwidth]{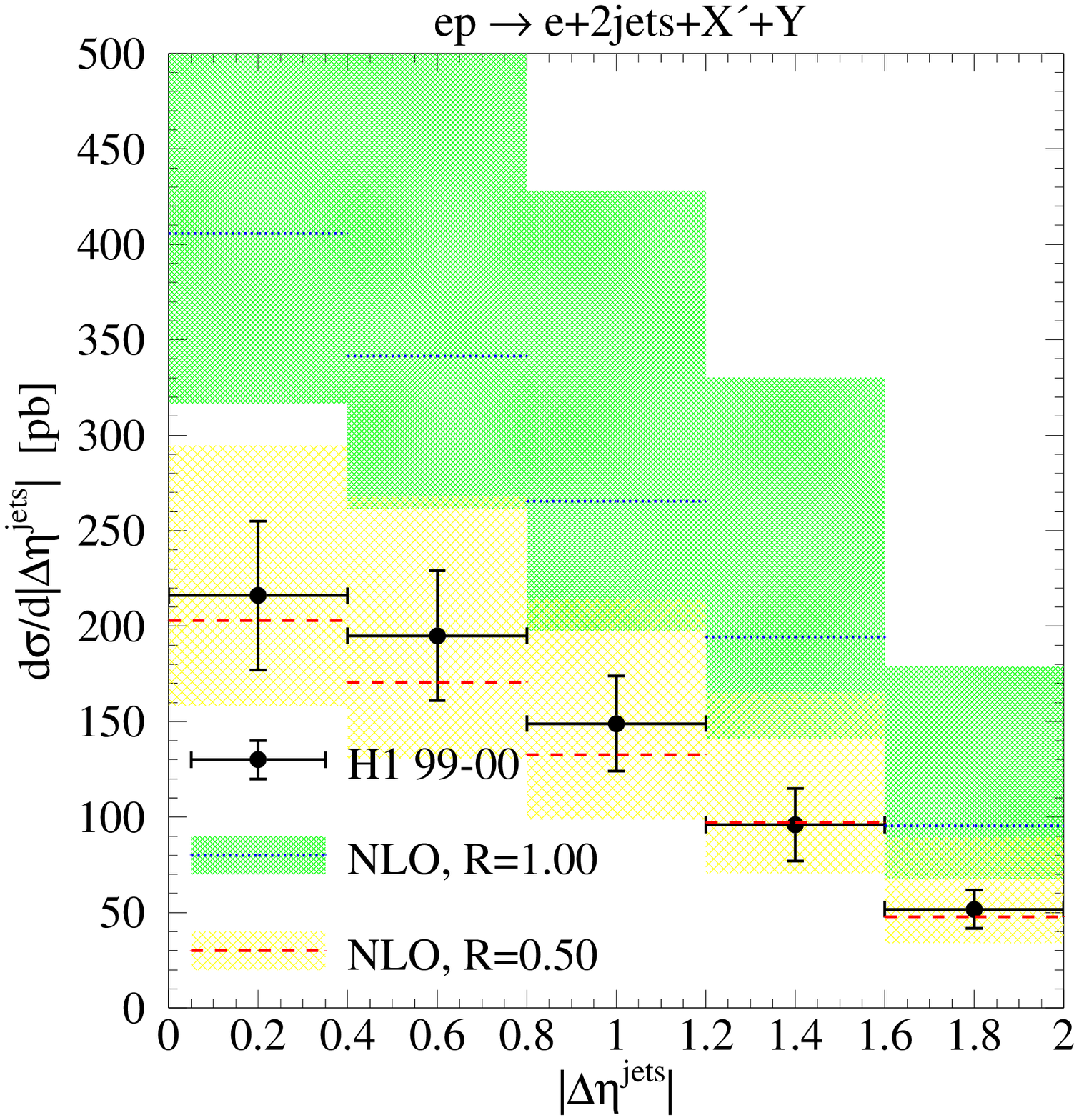}
 \includegraphics[width=0.325\columnwidth]{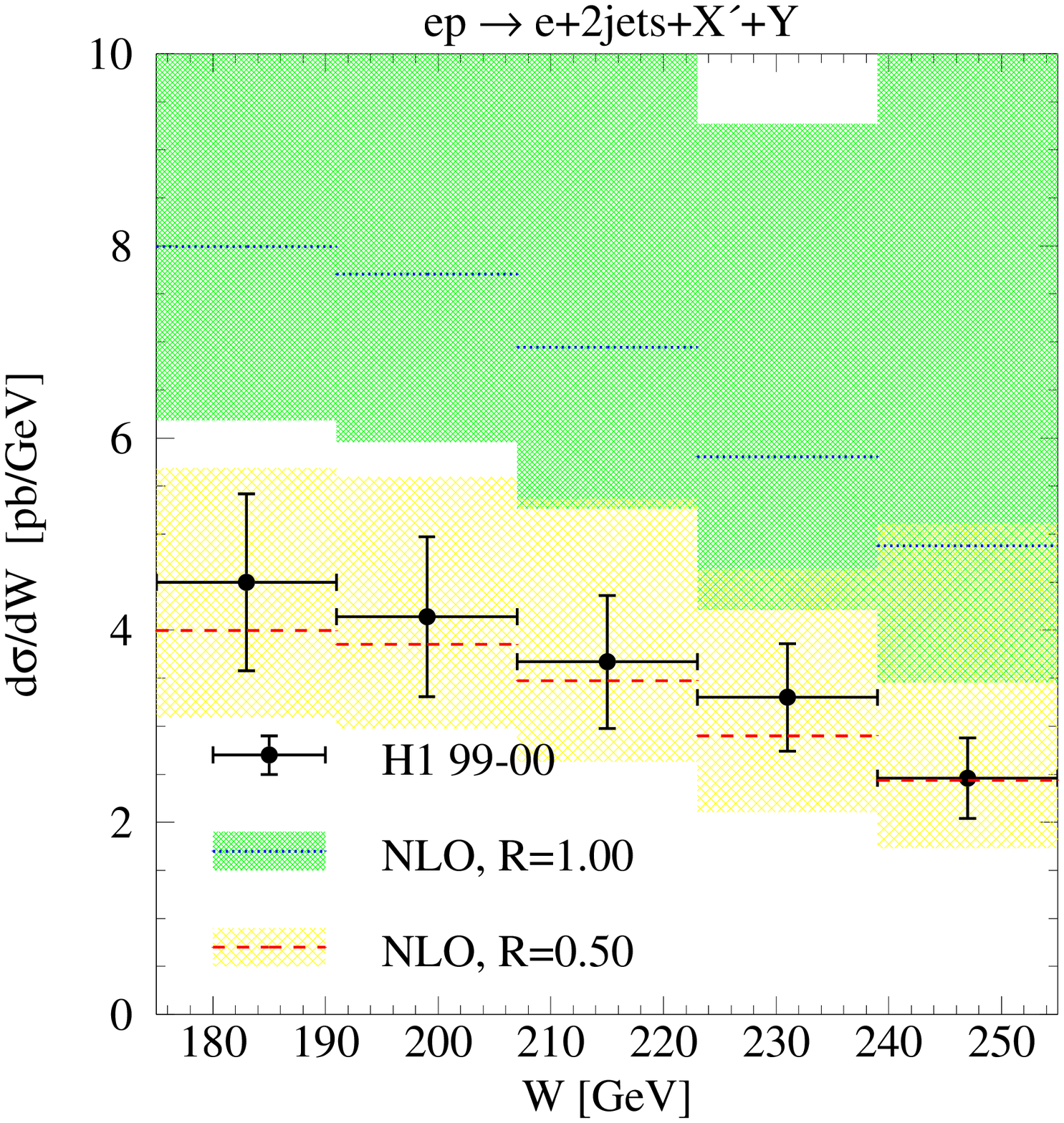}
 \caption{\label{fig:3}Differential cross sections for diffractive dijet
 photoproduction as measured by H1 with low-$E_T^{jet}$ cuts and compared to
 NLO QCD without ($R=1$) and with ($R=0.50$) global suppression 
 (color online).}
\end{figure}
%
With the exception of Figs.\ 3d and 3f, where the comparisons of 
$d\sigma/dE_T^{jet1}$ and $d\sigma/dM_{12}$ are shown, all other plots are 
such that the data points lie outside the error
band based on the scale variation for the unsuppressed case. However,
the predictions with suppression $R = 0.50$ agree nicely with the
data inside the error bands from the scale variation. Most of the data
points even agree with the $R = 0.50$ predictions inside the much smaller
experimental errors. The $M_X$-distribution agrees only for the second bin.
The reason for the disagreement of the two other $M_X$-bins might be that in
the theoretical results the variable $M_X$ is defined without the remnant
contributions, which, however, are taken into account in the experimental
definition of $M_X$. Further exceptions are the cross sections 
$d\sigma/dE_T^{jet1}$ and $\d\sigma/dM_{12}$, which are related. In 
$d\sigma/dE_T^{jet1}$ (see Fig.\ 3d) the predictions for the second and third 
bins lie practically outside the data points with their errors (note the
logarithmic scale). For $R = 1$ and
$R=0.50$ 
the cross sections fall off stronger with increasing $E_T^{jet1}$  than the 
data, the normalization being of course about two times larger for $R=1$.
In particular, the third data point agrees almost with the $R = 1$ prediction.
This means that the suppression decreases with increasing $E_T^{jet1}$. This
behavior was already apparent when we analyzed the first preliminary H1 data
\cite{17,18}. Such a behavior points in the direction that a suppression of 
the resolved cross section only would give better agreement with the data, as 
we shall see below. The same  observations can be made by looking at 
$d\sigma/dM_{12}$ in Fig.\ 3f. 
The survival probability $R = 0.50 \pm 0.09$ agrees with the result in 
\cite{Aaron:2010su}, which quotes $R=0.58$ $\pm$ 0.01 (stat.) $\pm$ 0.12 (syst.),
determined by fitting the integrated cross section. From our comparison we 
conclude that the low-$E_T^{jet}$  data show a global suppression of the order
of  two in complete agreement with the results in \cite{17, 18} and \cite{27} 
based on earlier preliminary \cite{13} and final H1 data \cite{27}. 

Next we want to answer the second question, whether the data could be
consistent with a suppression of the resolved component only, whose definition
is not unique, but rather factorization scale and scheme dependent. For this
purpose we have calculated the cross sections in two additional versions: (i)
suppression of the resolved cross section in the $\overline{\rm MS}$ scheme and
(ii) suppression of this resolved cross section plus that part of the NLO direct
part which depends on the factorization scale  at the photon vertex
\cite{35}. Of course, the needed suppression factors for the two versions will 
be different. We determine the suppression factors again by fitting the
measured $d\sigma/dE_T^{jet1}$ for the lowest $E_T^{jet1}$-bin (see Fig. 4d). 
Then, the suppression factor for version (i) is $R = 0.40$ (denoted res in the 
figures), and for version (ii) it is $R = 0.37$ (denoted res+dir-IS). The 
comparison with the H1 data of $d\sigma/dx_{\gamma}^{obs}$, 
$d\sigma/dz_{\p}^{obs}$, $d\sigma/d\log_{10}(x_{\p})$, $d\sigma/dE_T^{jet1}$, 
$d\sigma/dM_X$, $d\sigma/dM_{12}$, $d\sigma/d\bar{\eta}^{jets}$, 
$d\sigma/d|\Delta \eta^{jets}|$ and $d\sigma/dW$ is 
shown in Figs.\ 4a-i, where
%
\begin{figure}
 \centering
 \includegraphics[width=0.325\columnwidth]{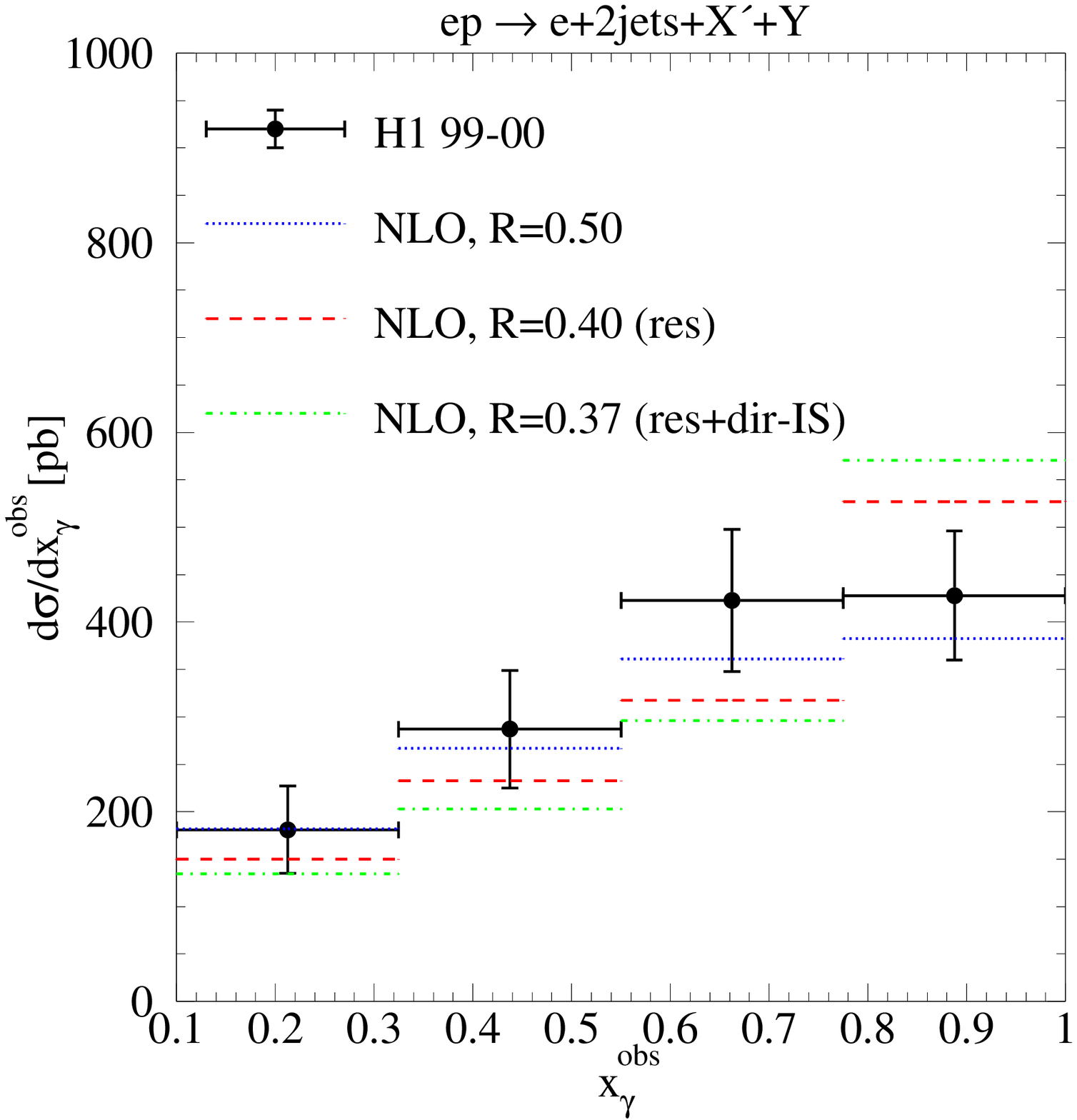}
 \includegraphics[width=0.325\columnwidth]{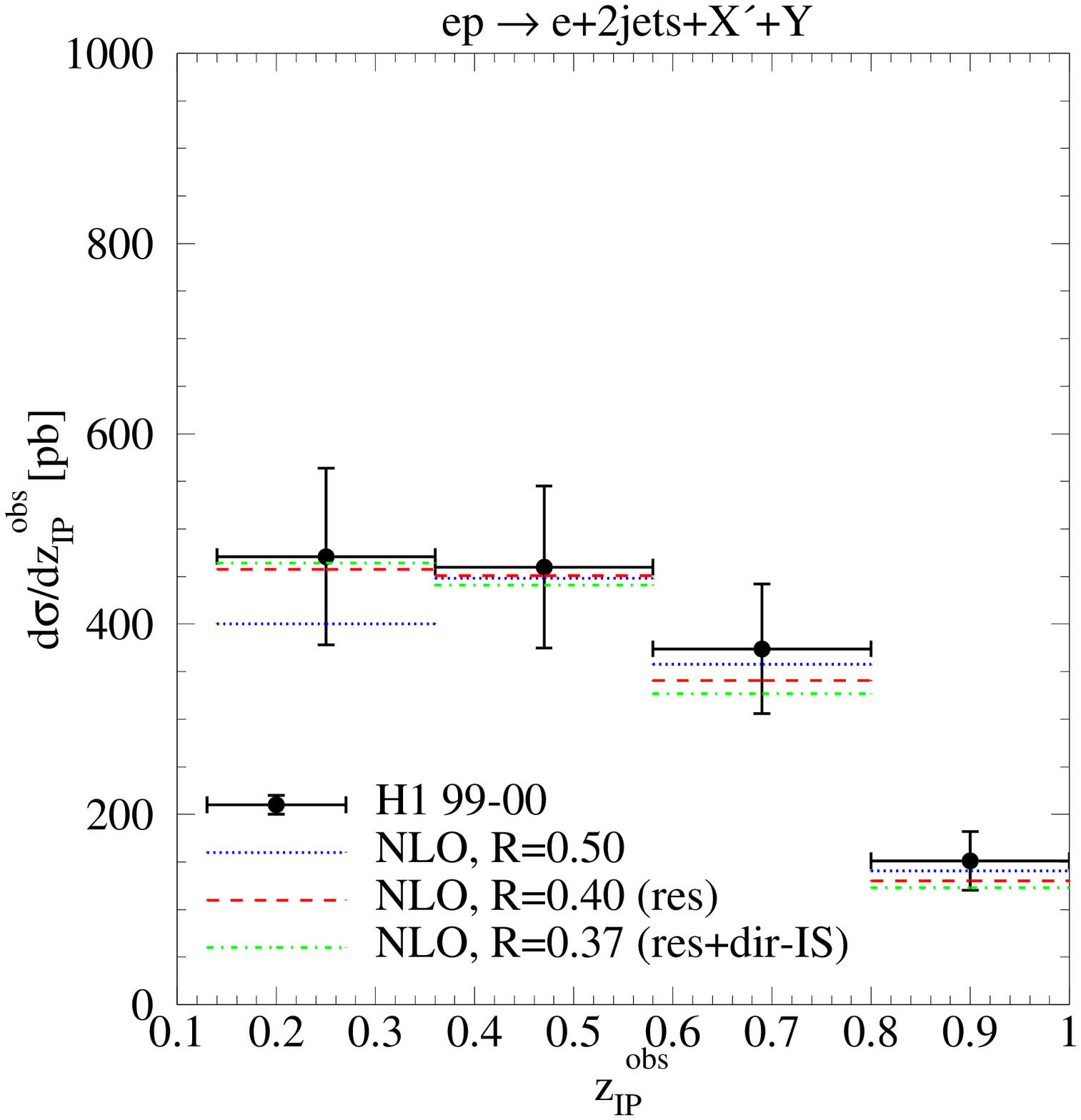}
 \includegraphics[width=0.325\columnwidth]{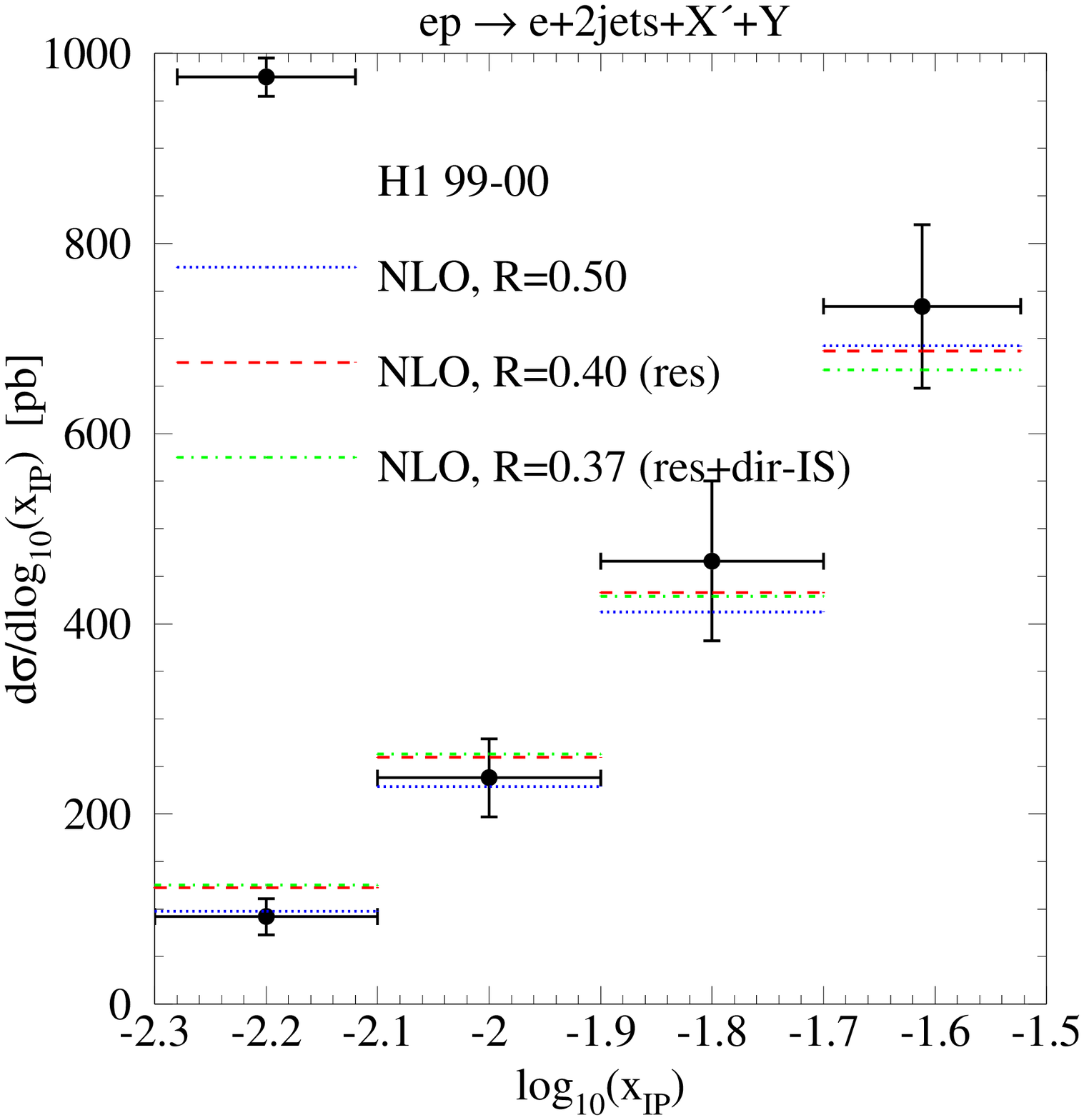}
 \includegraphics[width=0.325\columnwidth]{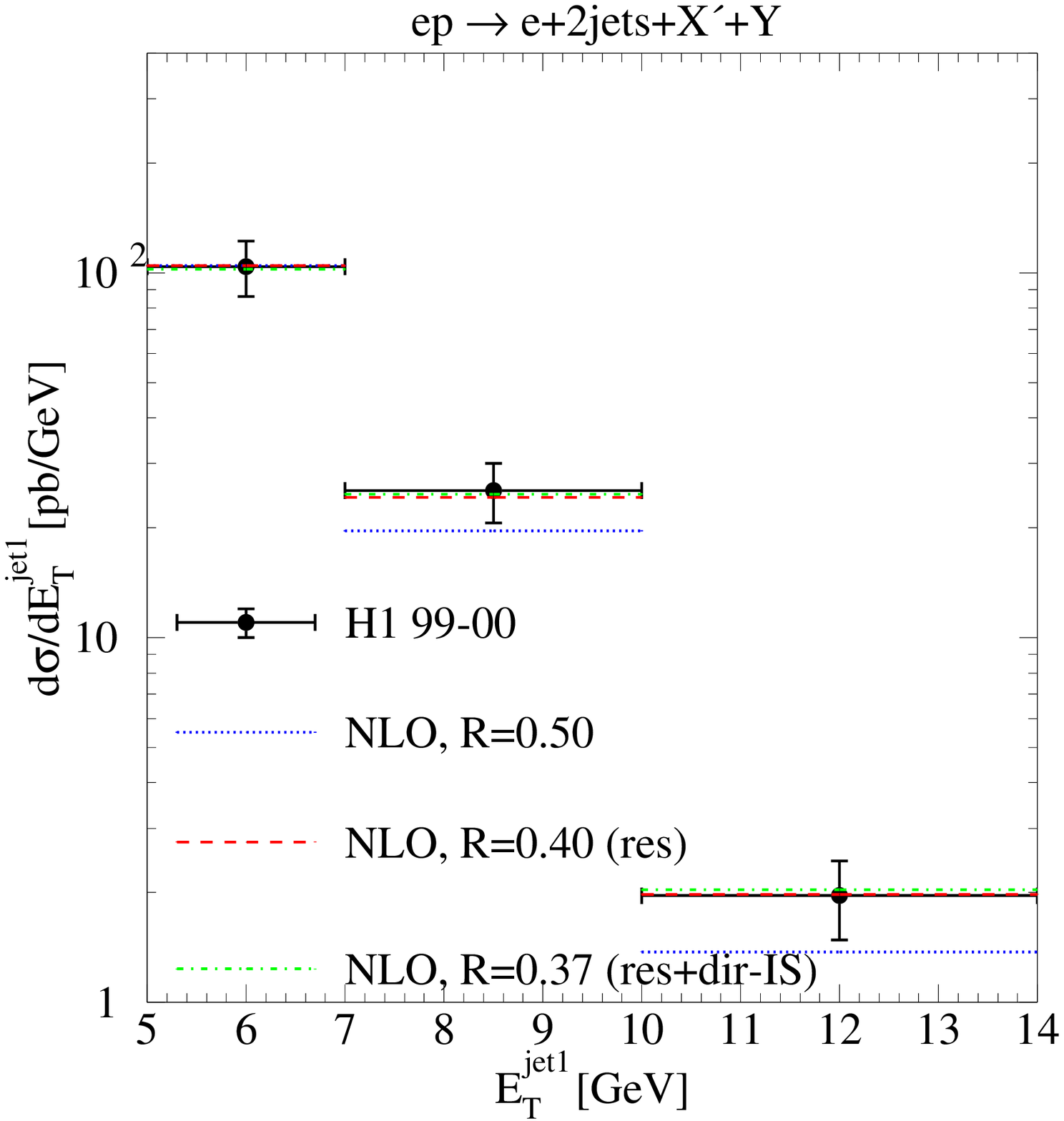}
 \includegraphics[width=0.325\columnwidth]{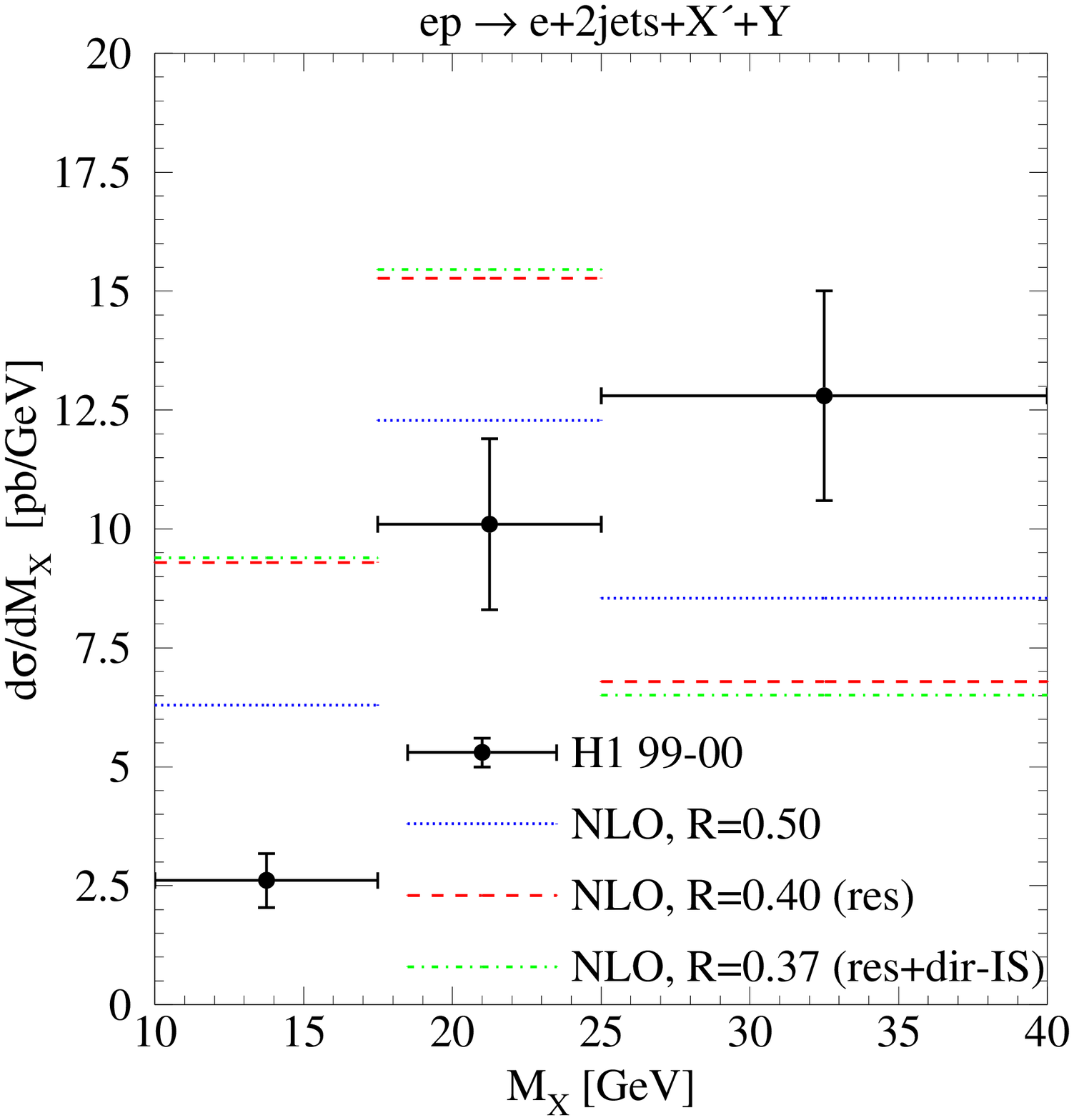}
 \includegraphics[width=0.325\columnwidth]{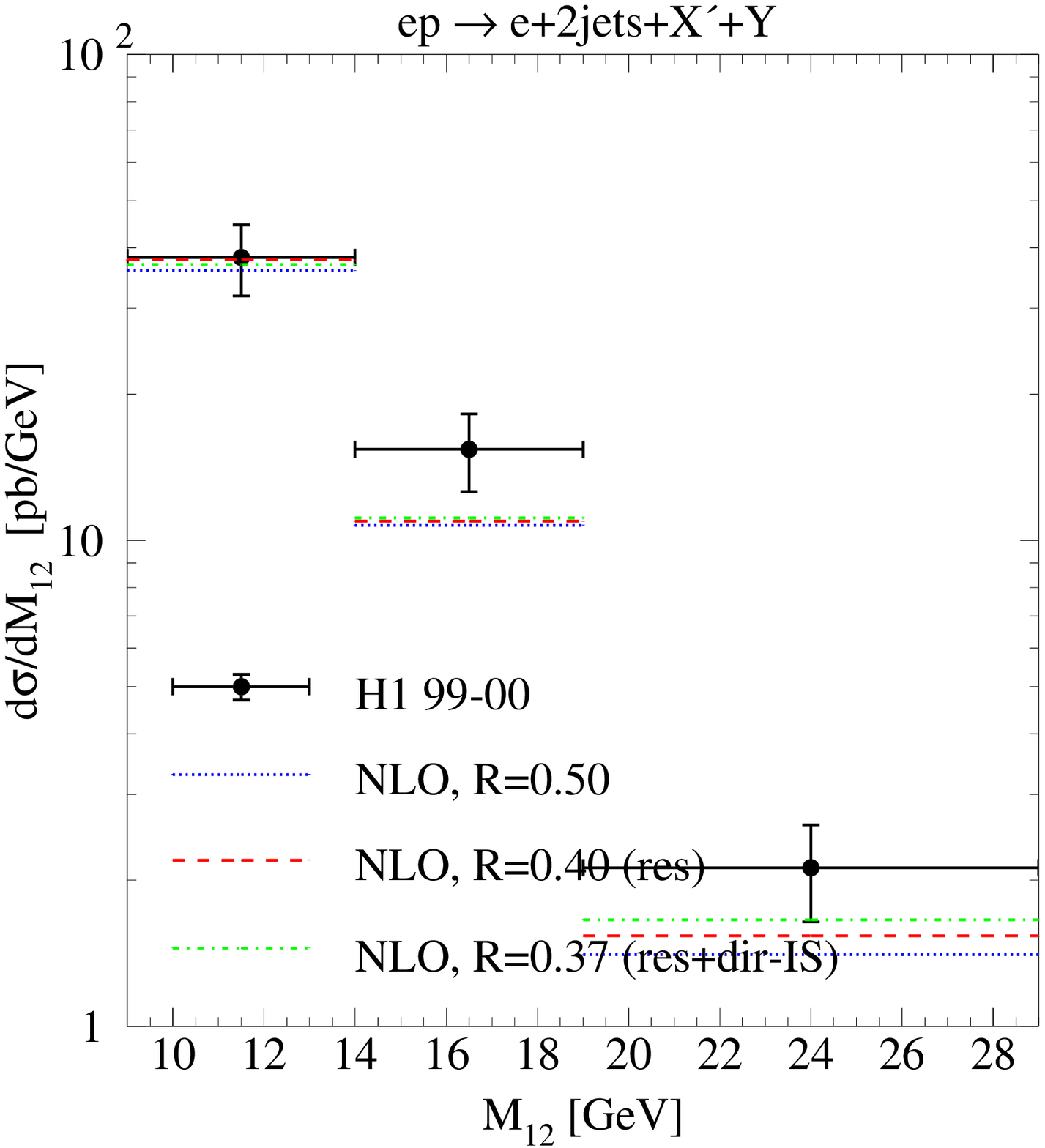}
 \includegraphics[width=0.325\columnwidth]{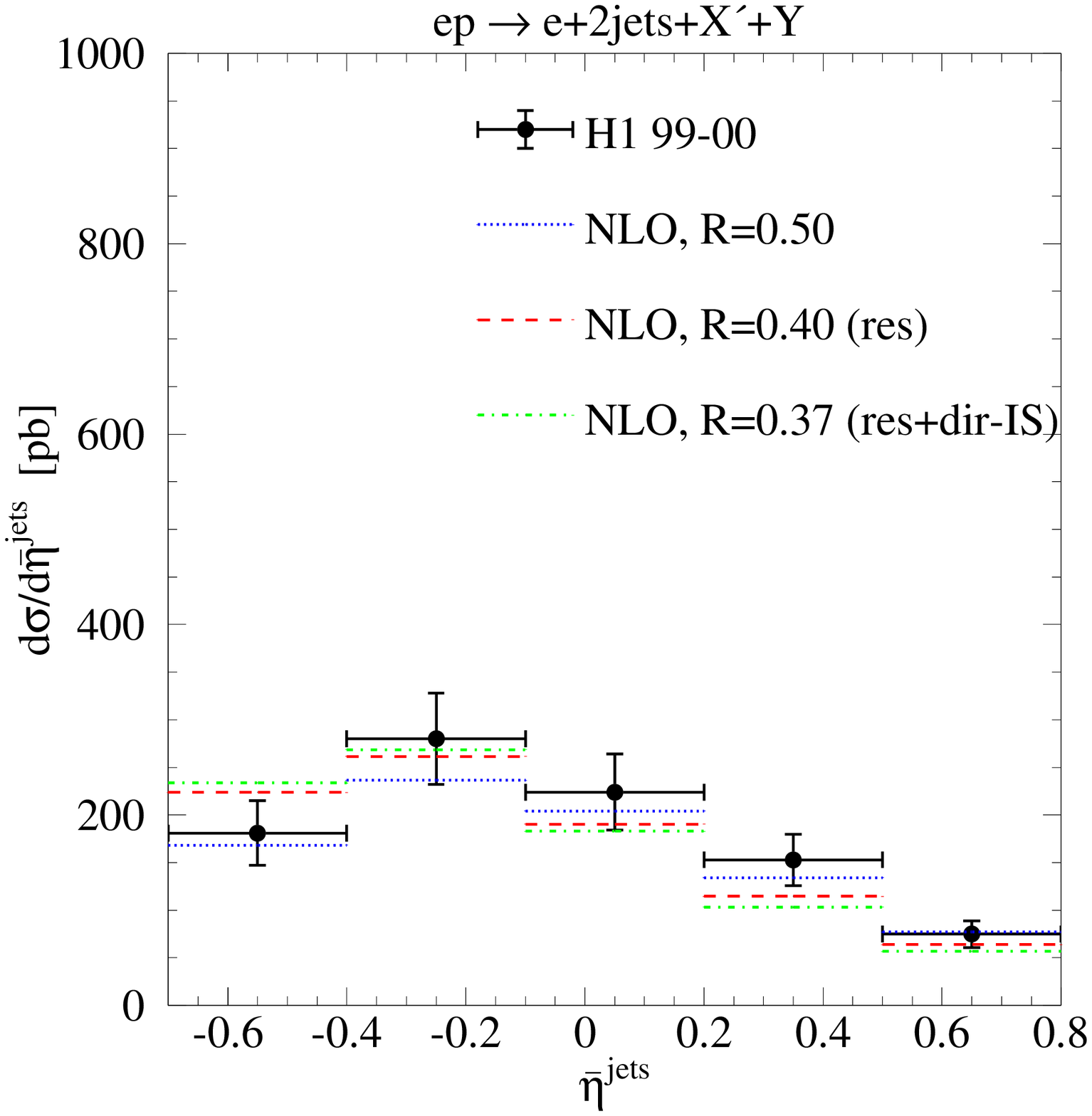}
 \includegraphics[width=0.325\columnwidth]{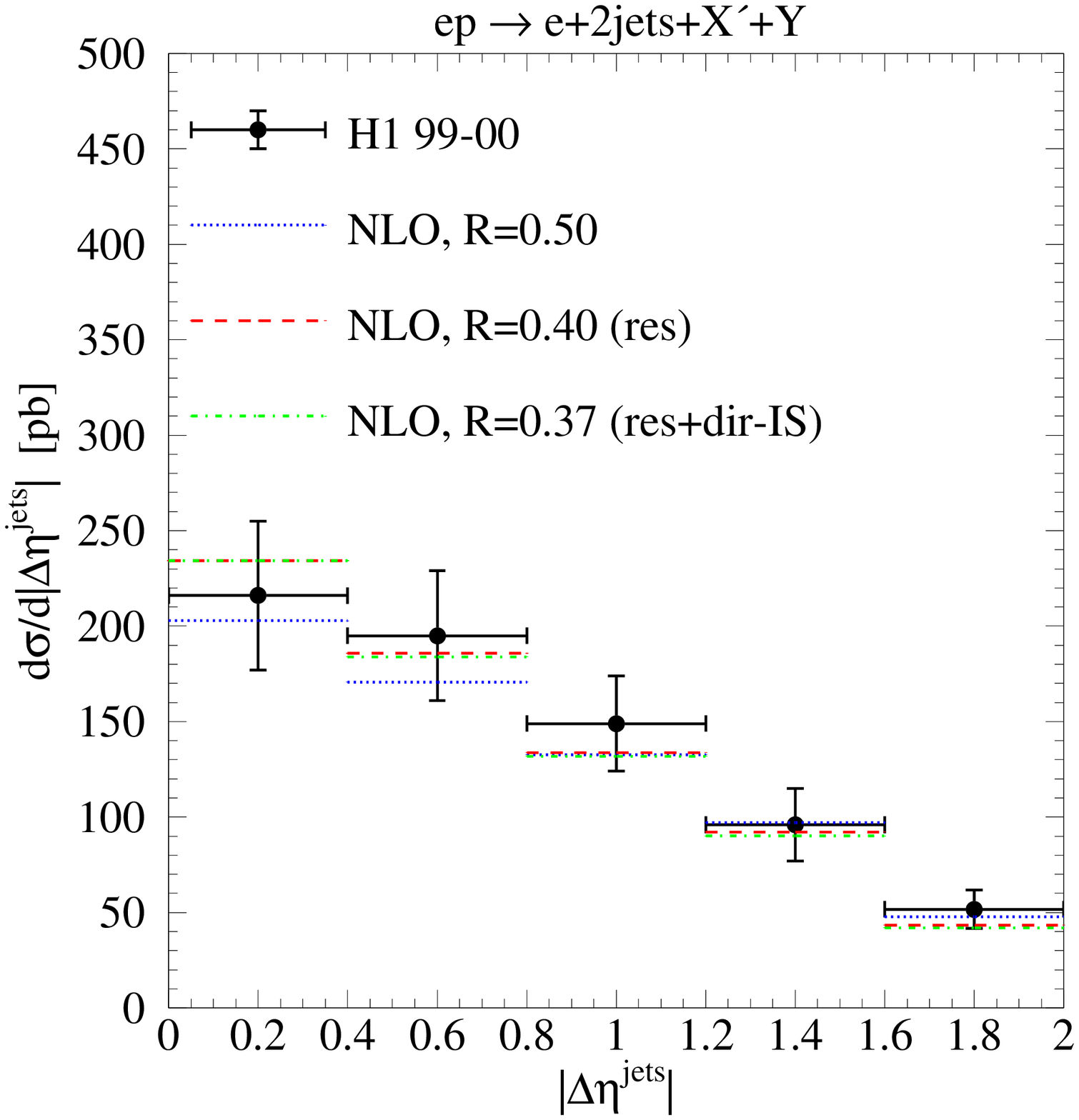}
 \includegraphics[width=0.325\columnwidth]{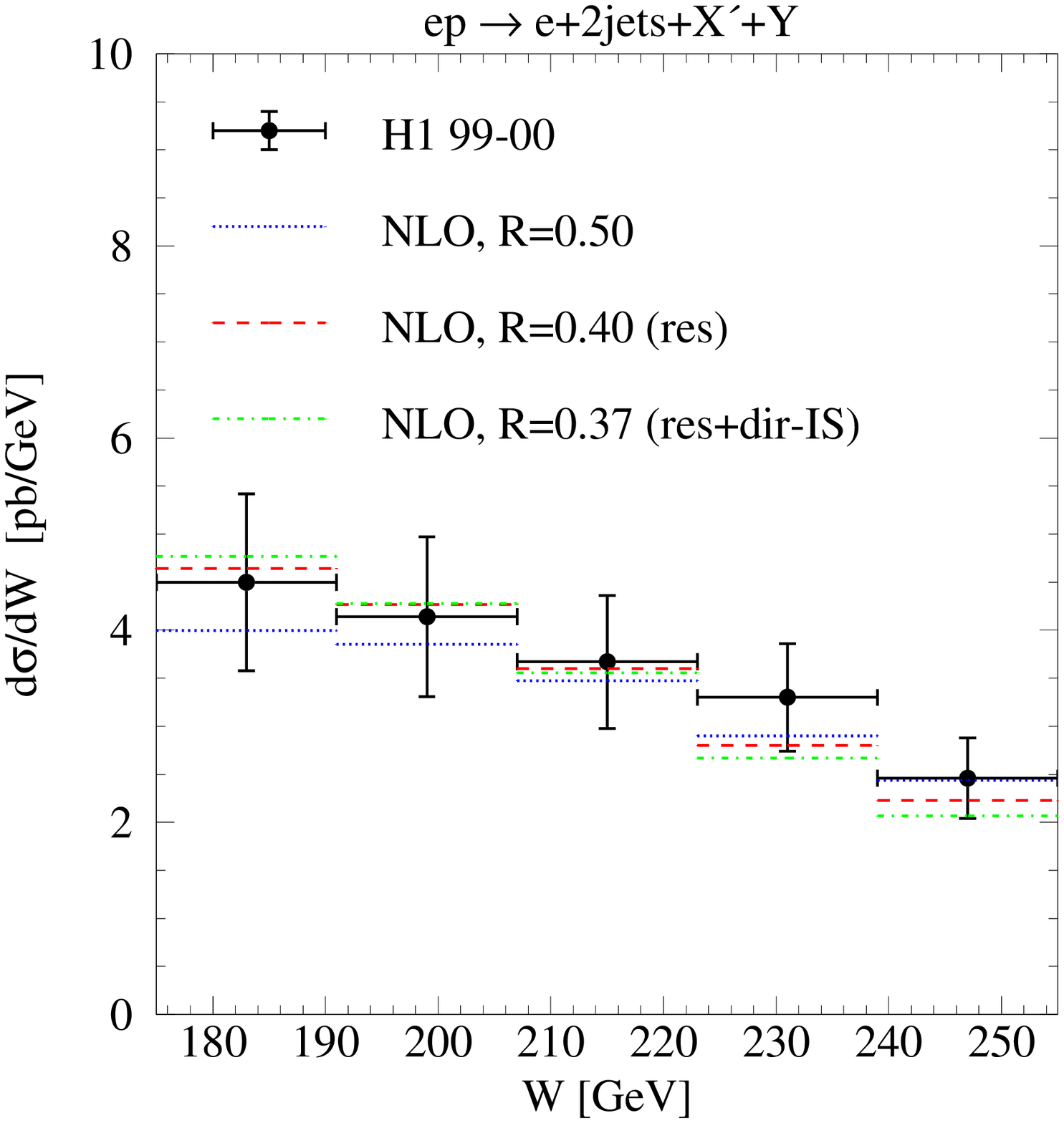}
 \caption{\label{fig:4}Differential cross sections for diffractive dijet
 photoproduction as measured by H1 with low-$E_T^{jet}$ cuts and compared to
 NLO QCD with global, resolved, and resolved/direct-IS suppression. Note that
 some of the theoretical predictions coincide with the experimental values.
}
\end{figure}
%
we have also plotted the prediction for the global suppression (direct and
resolved) with $R = 0.50$, already shown in Figs.\ 3a-i.
Looking at the Figs.\ 4a-i we can distinguish three
groups of results from the comparison with the data. In the first group, the 
cross sections as functions of $z_{\p}^{obs}$, $log_{10}(x_{\p})$,
$M_{12}$, $\bar{\eta}^{jets}$,  $|\Delta \eta| ^{jets}$ and $W$, the agreement 
with the global suppression ($R = 0.50$) and the resolved suppression
($R = 0.40$ or $R = 0.37$) is comparable. In the second group, namely for 
$d\sigma/dE_T^{jet1}$, the agreement is better for the resolved suppression 
only. In the third group, $d\sigma/dx_{\gamma}^{obs}$ and $d\sigma/dM_X$, the
agreement with the resolved suppression is worse than with the global 
suppression. In particular, for $d\sigma/dx_{\gamma}^{obs}$, which is usually
considered as the characteristic distribution for distinguishing global versus 
resolved suppression, the agreement with resolved suppression does not improve.
Unfortunately, this cross section has the largest hadronic corrections of the 
order of $(20-30)\%$ \cite{Aaron:2010su}. Second, also for the usual photoproduction of
dijets the comparison between data and theoretical results has similar problems
in the large $x_{\gamma}^{obs}$-bin \cite{36}, although the $E_T^{jet}$ cut is
much larger there. On the other hand, for the cross sections
$d\sigma/dE_T^{jet1}$ (and $d\sigma/dM_{12}$)
the agreement improves considerably (and somewhat) with the suppression 
of the resolved part only (note the logarithmic scale in Fig.\ 4d). Here, of 
course, we must admit that the suppression factor could be $E_T$-dependent,
although we see no theoretical reason for such a dependence.
 
We also checked for two distributions whether the predictions for resolved 
suppression depend on the chosen diffractive PDFs. For this purpose we have 
calculated for the two cases $d\sigma/dz_{\p}^{obs}$ and $d\sigma/dE_T^{jet1}$ 
the cross sections with the `H1 2006 fit A' parton distributions
\cite{4}. The results are compared in Figs.\ 5a and b to the results with
%
\begin{figure}
 \centering
 \includegraphics[width=0.495\columnwidth]{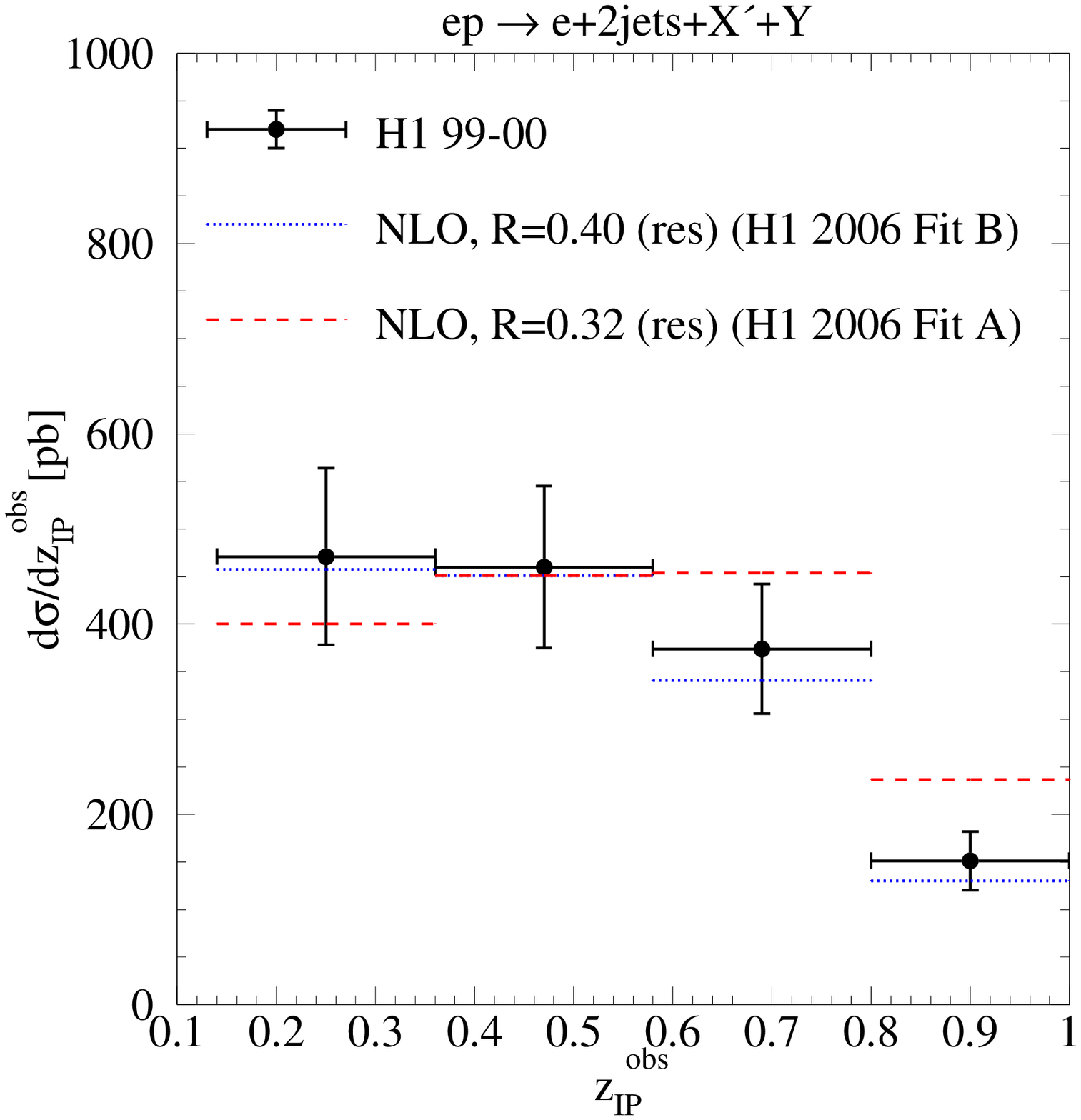}
 \includegraphics[width=0.495\columnwidth]{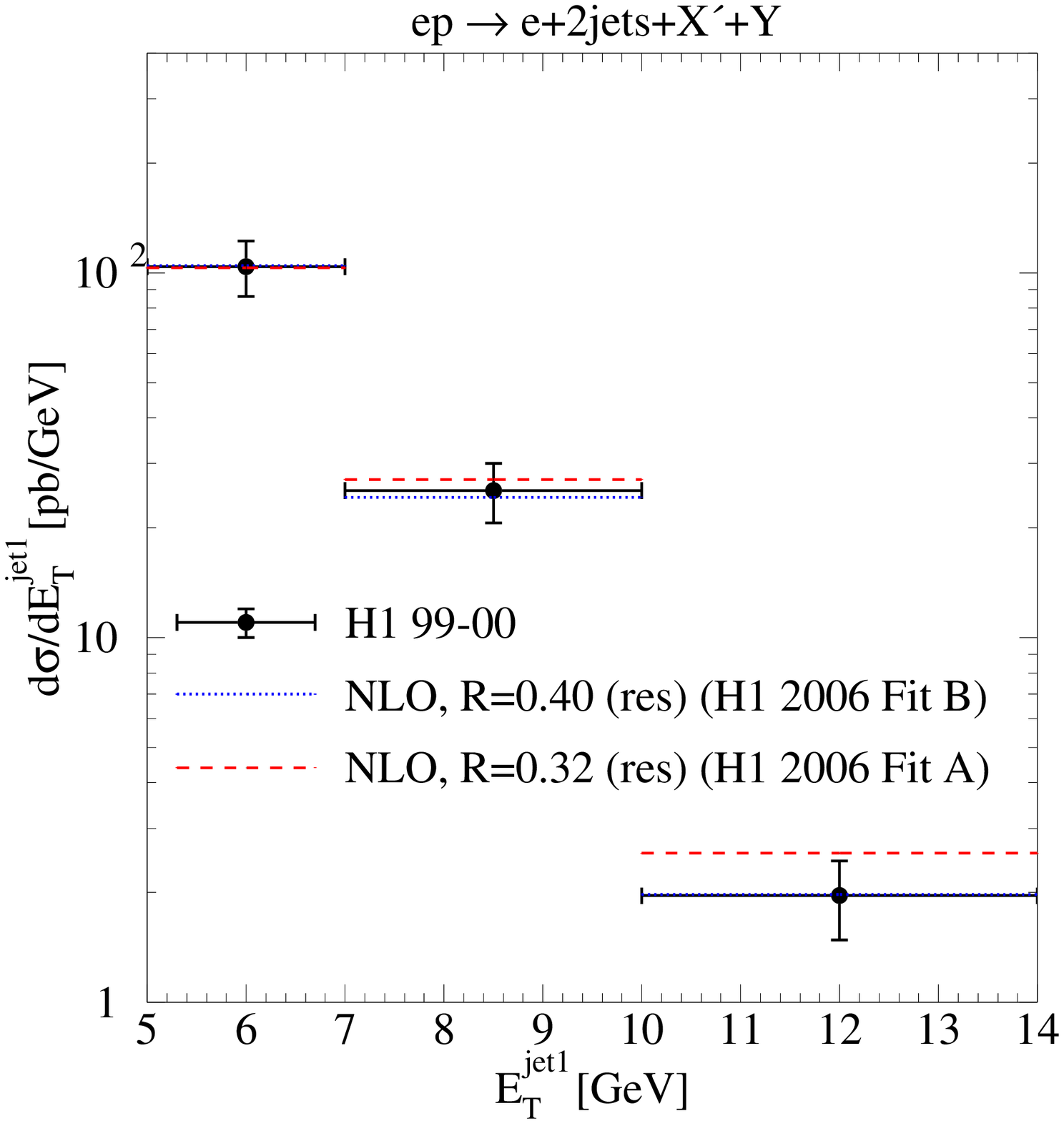}
 \caption{\label{fig:5}Differential cross sections for diffractive dijet
 photoproduction as measured by H1 with low-$E_T^{jet}$ cuts and compared to
 NLO QCD with resolved suppression and two different DPDFs.}
\end{figure}
%
the `H1 2006 fit B' and the experimental data. Of course, since the `H1 2006
fit A' PDFs have a larger gluon component at large $z$, the cross sections are
larger and therefore need a larger suppression of $R = 0.32$. Note that in the
published low-$E_T^{jet}$ H1 analysis as well as in the comparison presented here
the contribution from the largest $z_{\p}^{obs}$-bin has been removed from all
other distributions. From Figs.\
5a and b we conclude that the dependence on the chosen DPDFs is then weaker, but
that `H1 2006 fit B' is still favored over `H1 2006 fit A'. In total, we are
tempted to conclude from the comparisons in  Figs.\ 4a-i that the predictions
with a resolved-only (or resolved+direct-IS) suppression are consistent with the
new low-$E_T^{jet}$ H1 data \cite{Aaron:2010su}.

The same comparison of the high-$E_T^{jet}$ data of H1 \cite{29} with the 
various theoretical predictions is shown in the following figures. The global
suppression factor is obtained again from a fit to the smallest $E_T^{jet1}$-bin.
It is equal to $R = 0.62 \pm 0.16$, again in agreement with the H1 result
$R=0.62$ $\pm$ 0.03 (stat.) $\pm$ 0.14 (syst.) \cite{29}
obtained with our theoretical cross sections. The comparisons of the same
cross sections as in the low-$E_T^{jet}$ comparison are shown in Figs.\
6a-i for the two cases $R = 1$ (no suppression) and $R = 0.62$
%
\begin{figure}
 \centering
 \includegraphics[width=0.325\columnwidth]{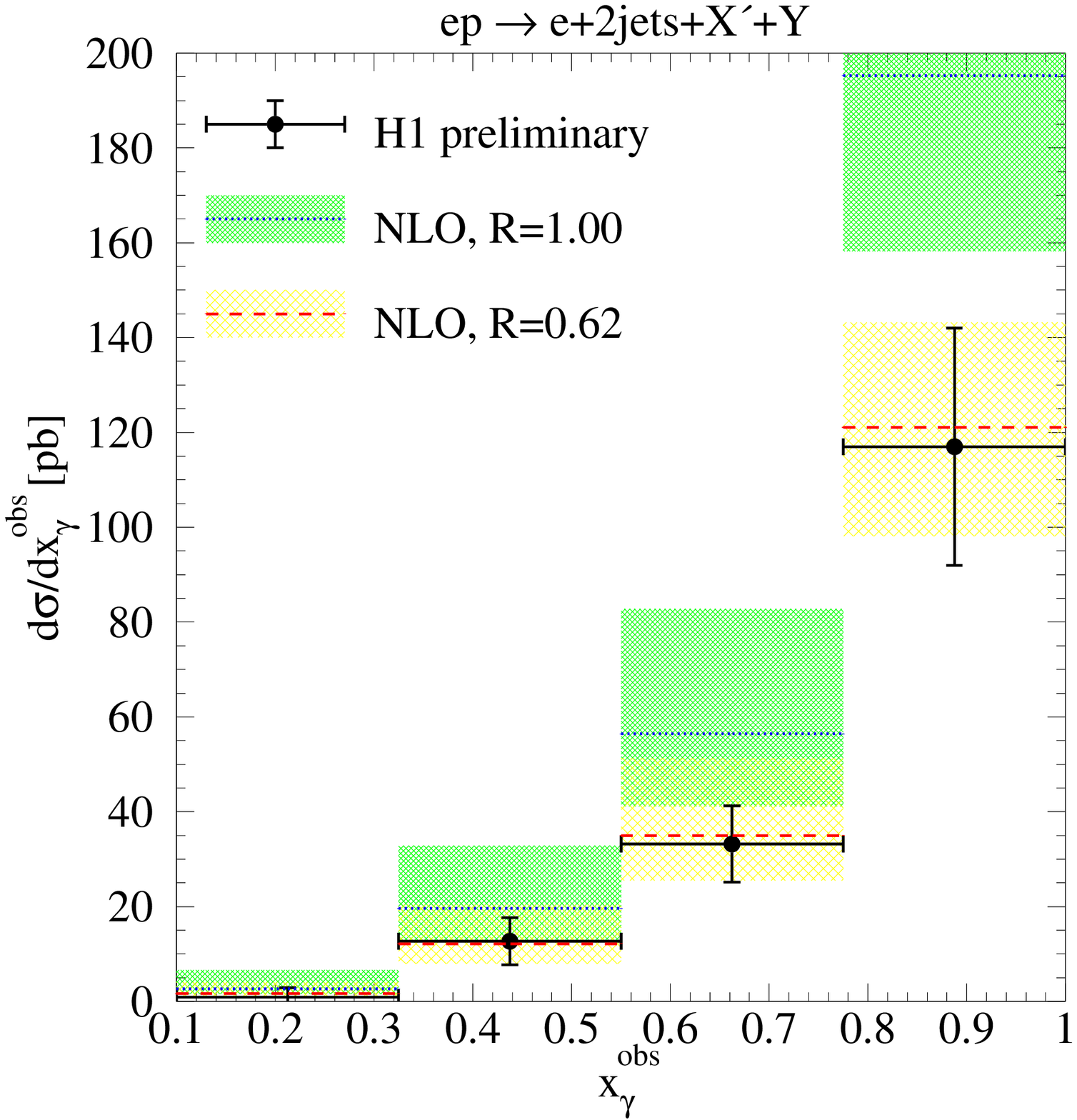}
 \includegraphics[width=0.325\columnwidth]{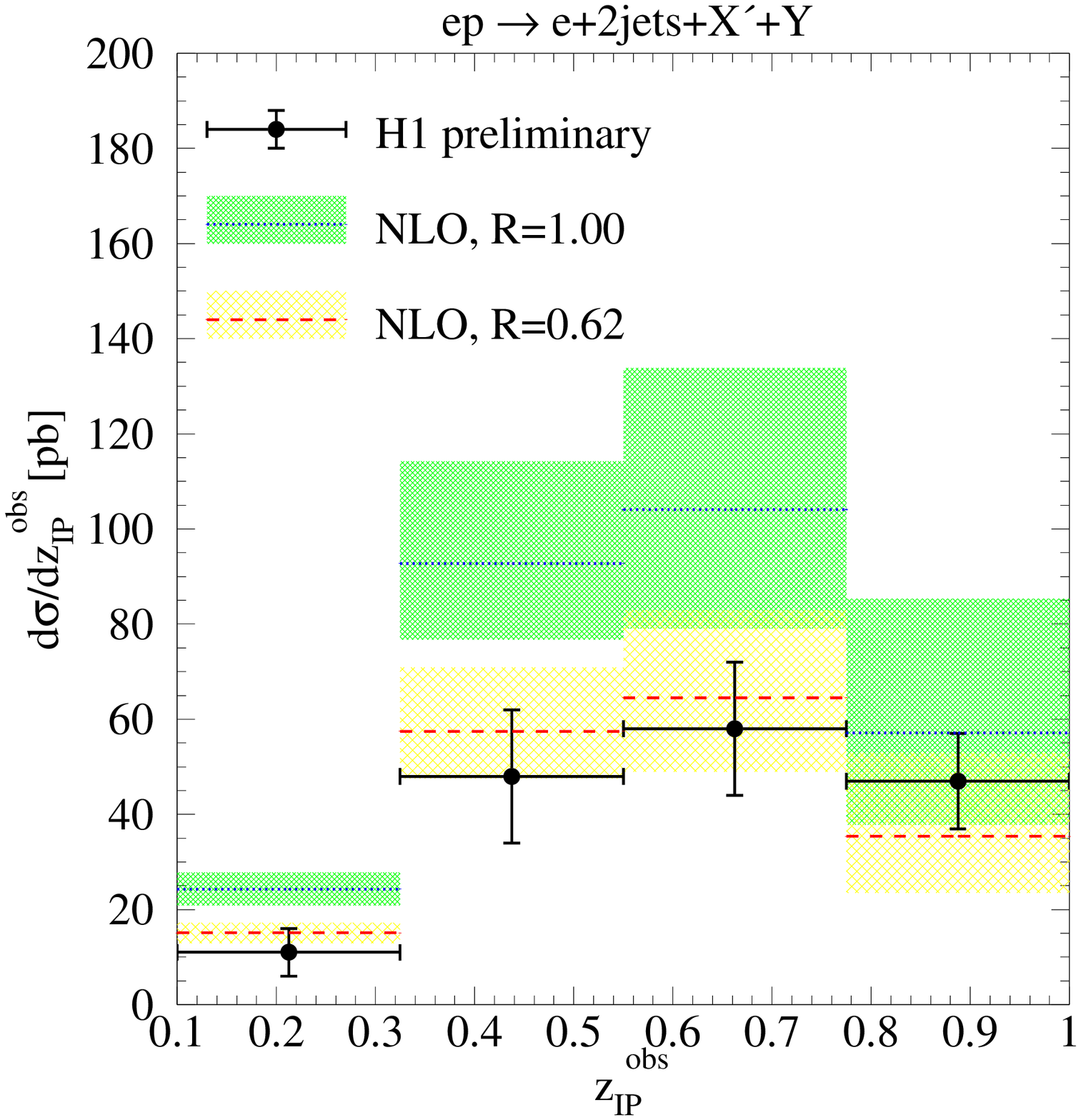}
 \includegraphics[width=0.325\columnwidth]{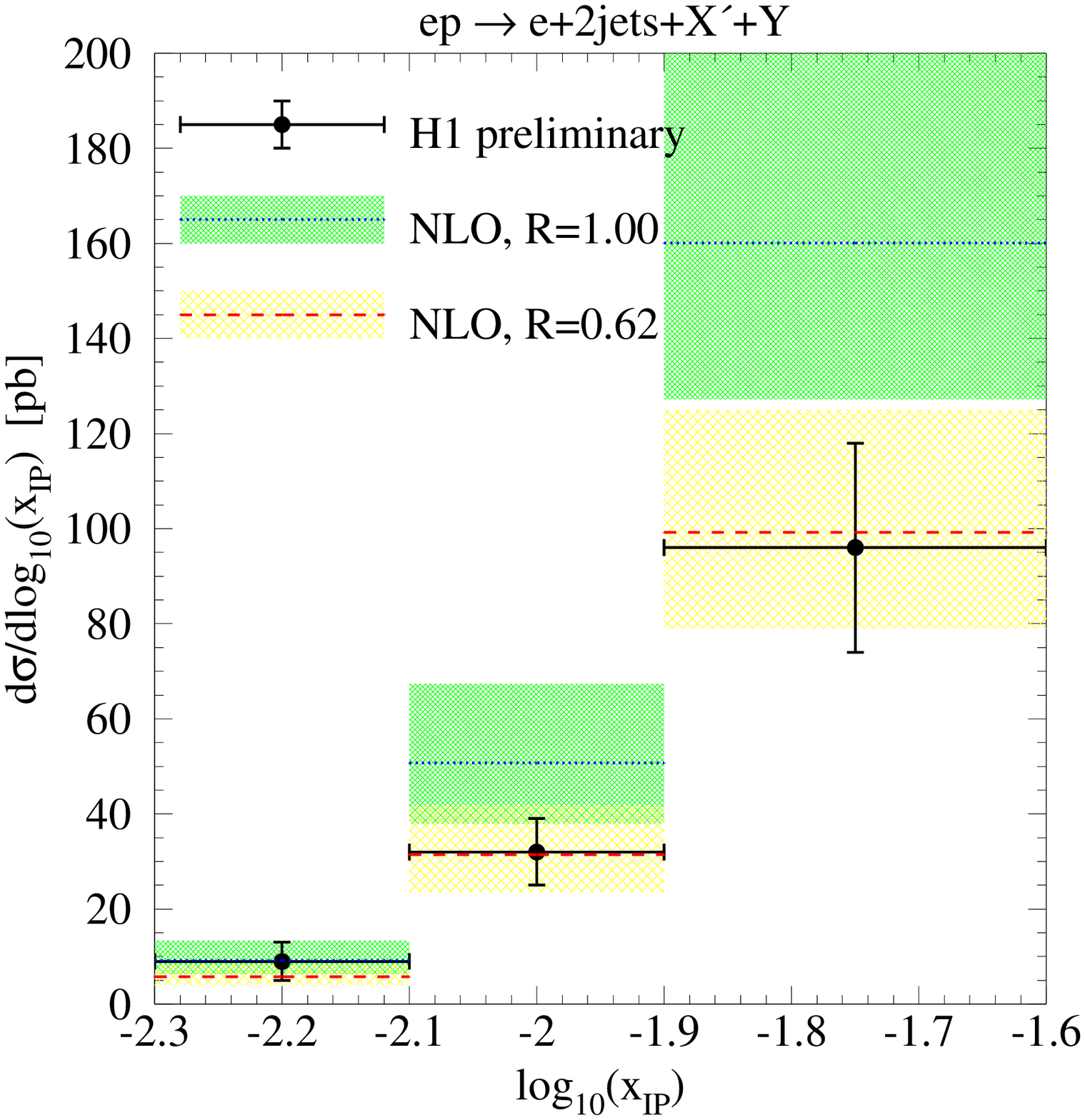}
 \includegraphics[width=0.325\columnwidth]{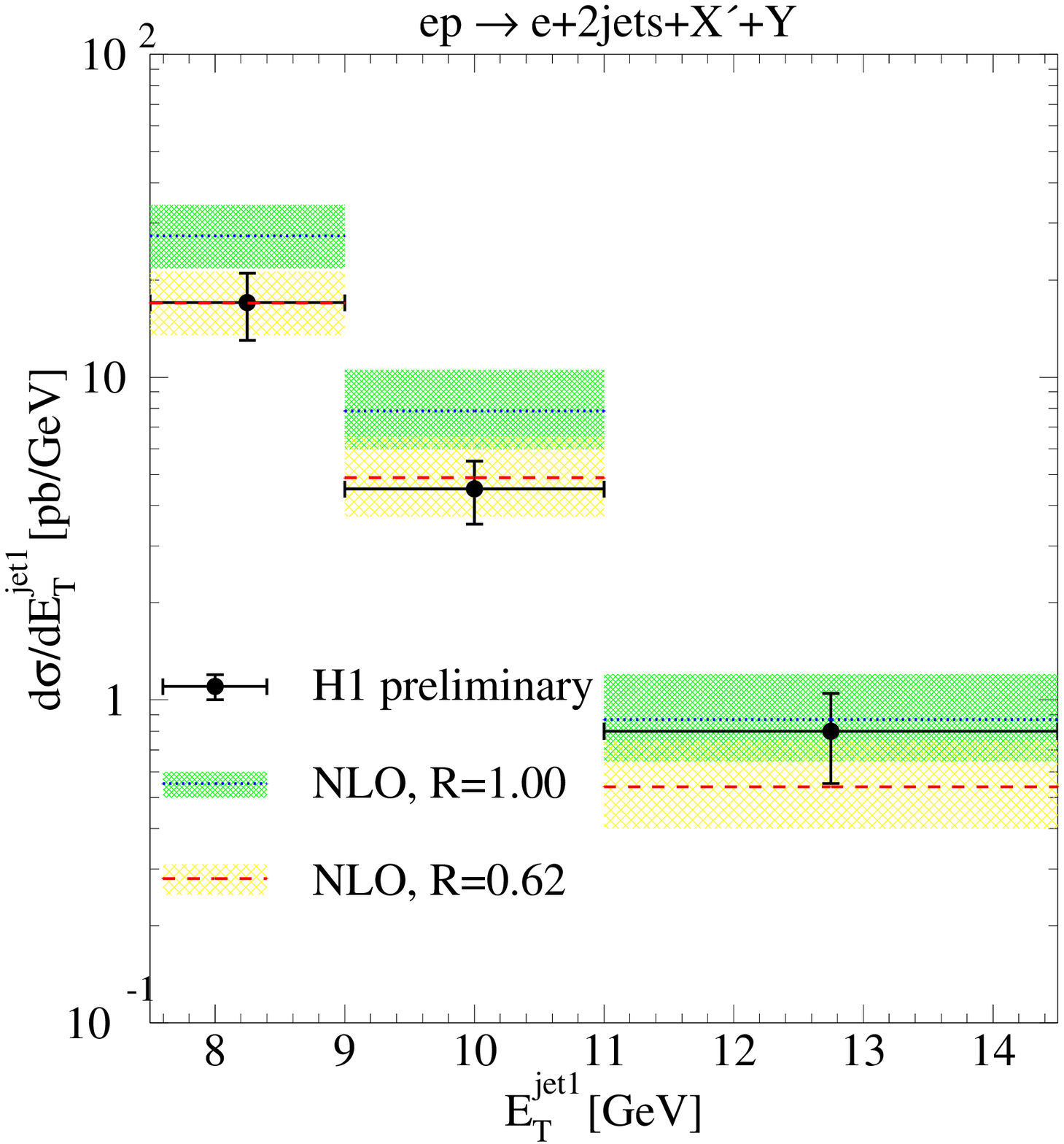}
 \includegraphics[width=0.325\columnwidth]{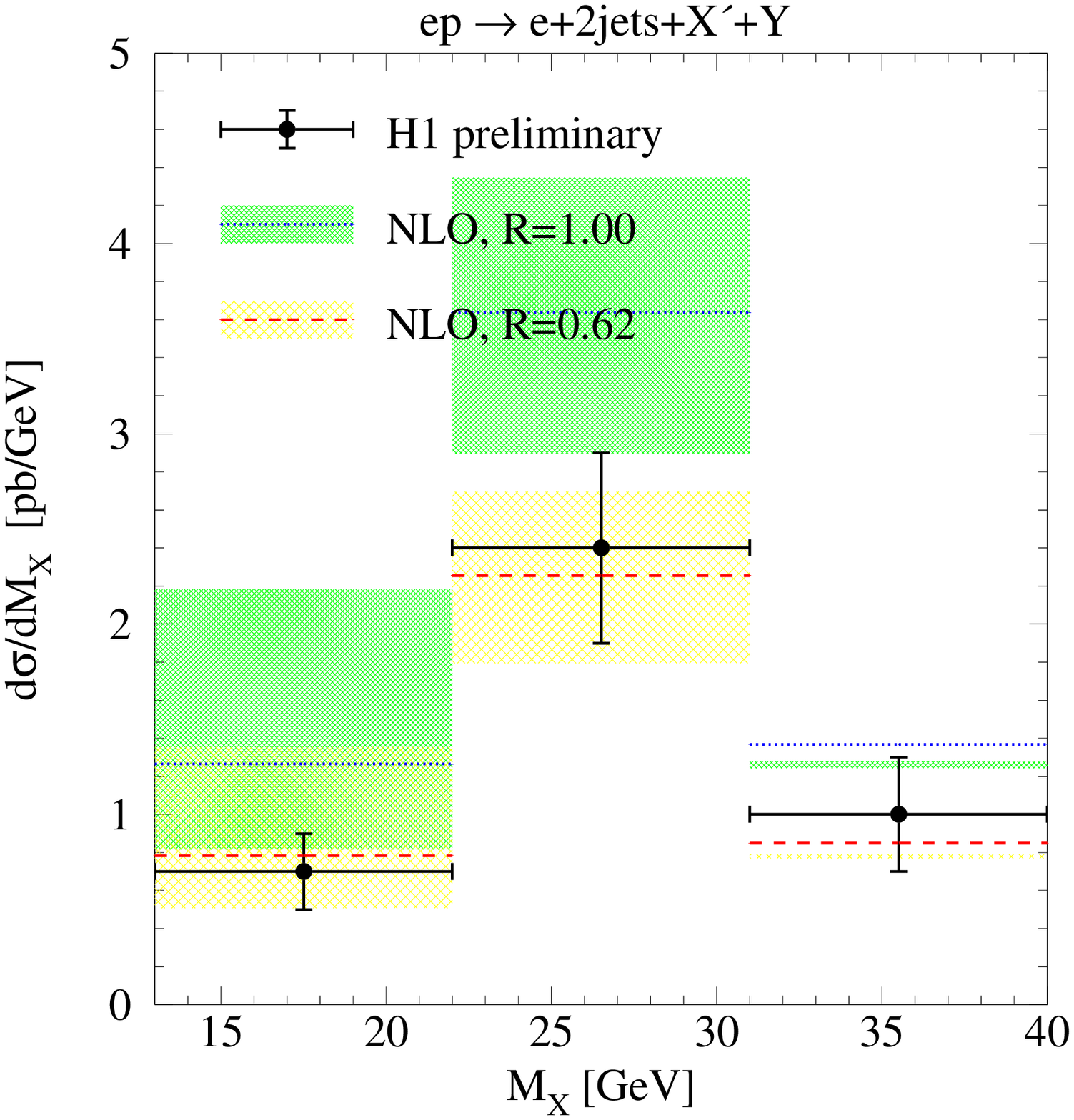}
 \includegraphics[width=0.325\columnwidth]{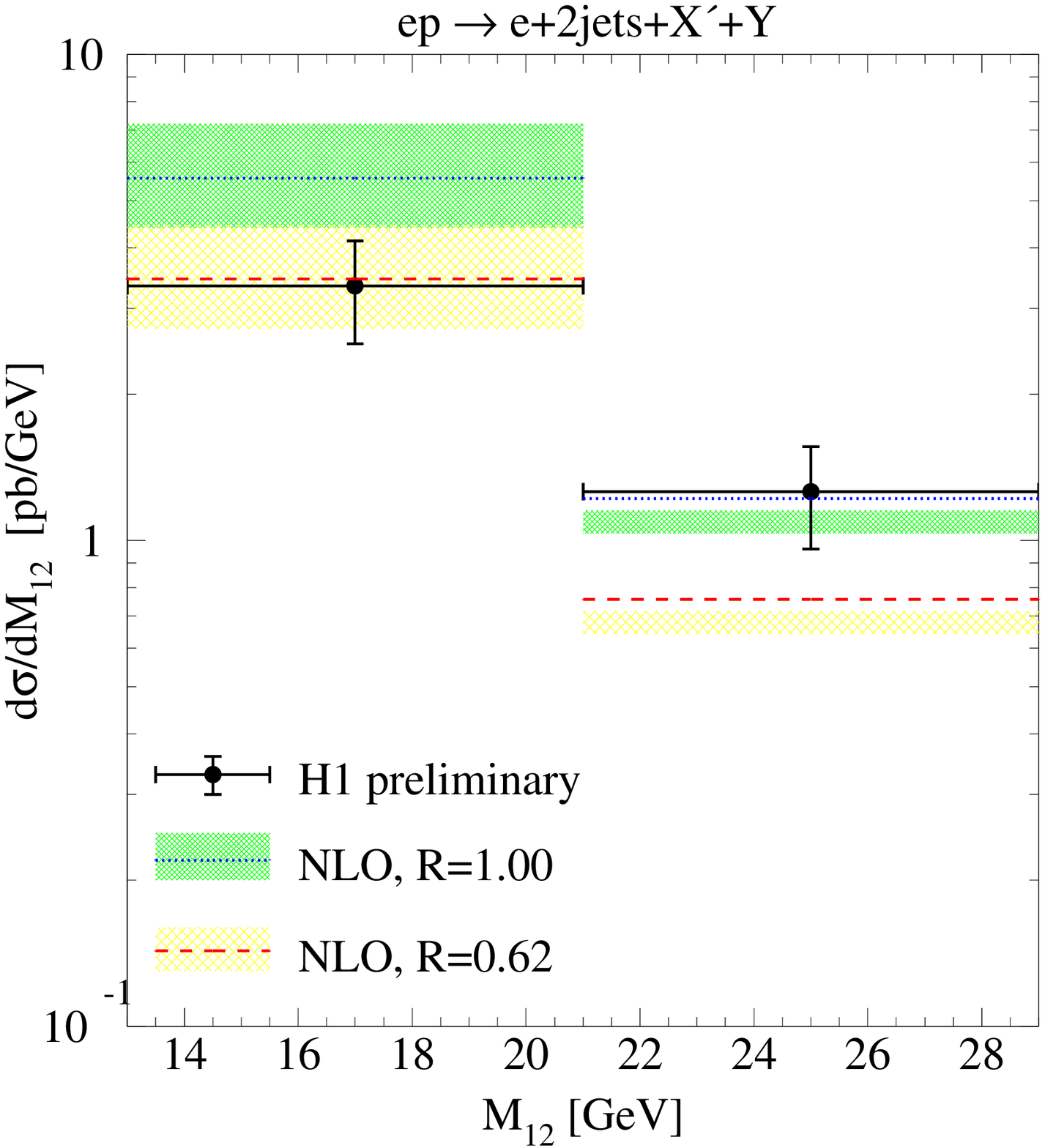}
 \includegraphics[width=0.325\columnwidth]{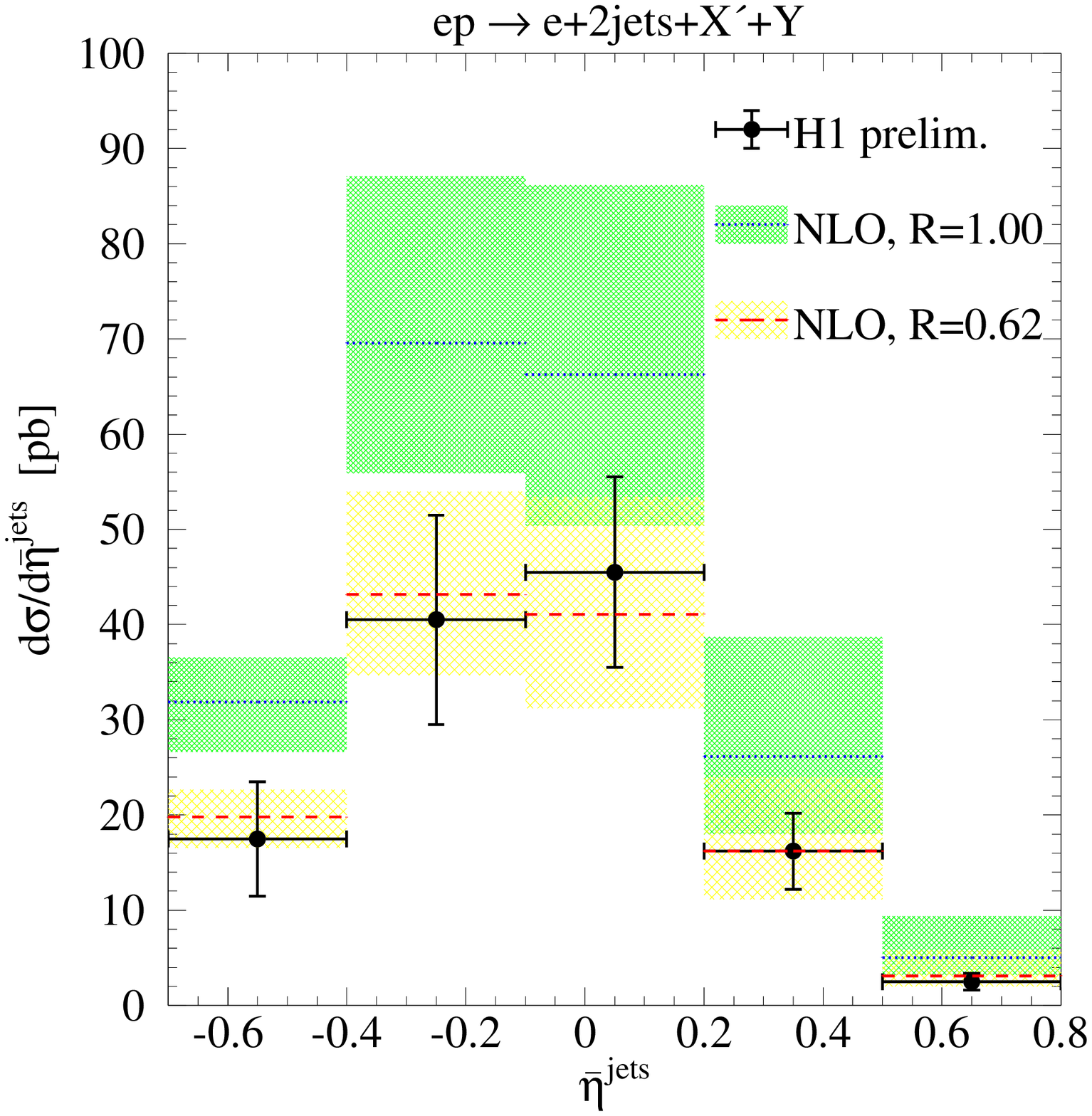}
 \includegraphics[width=0.325\columnwidth]{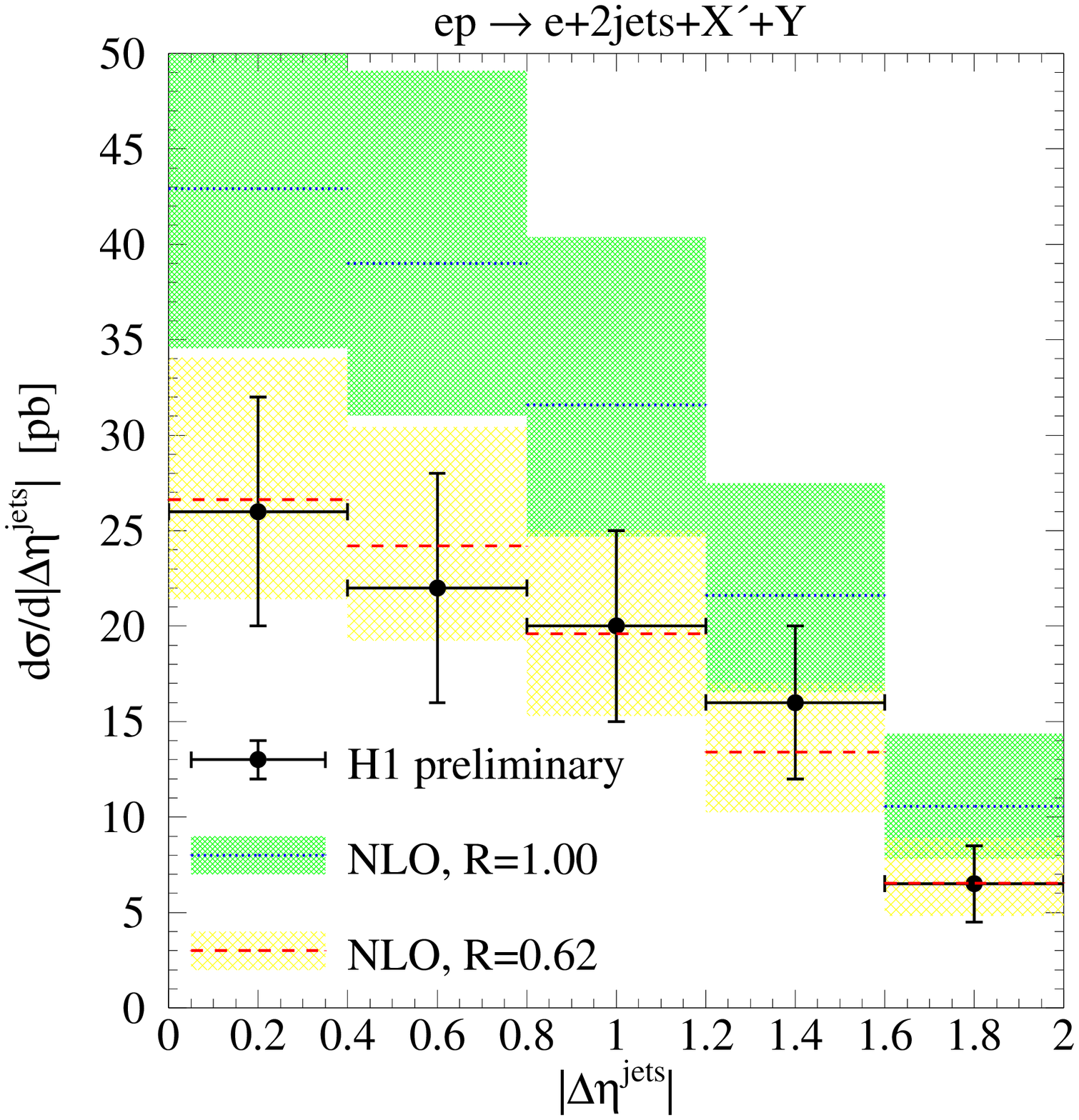}
 \includegraphics[width=0.325\columnwidth]{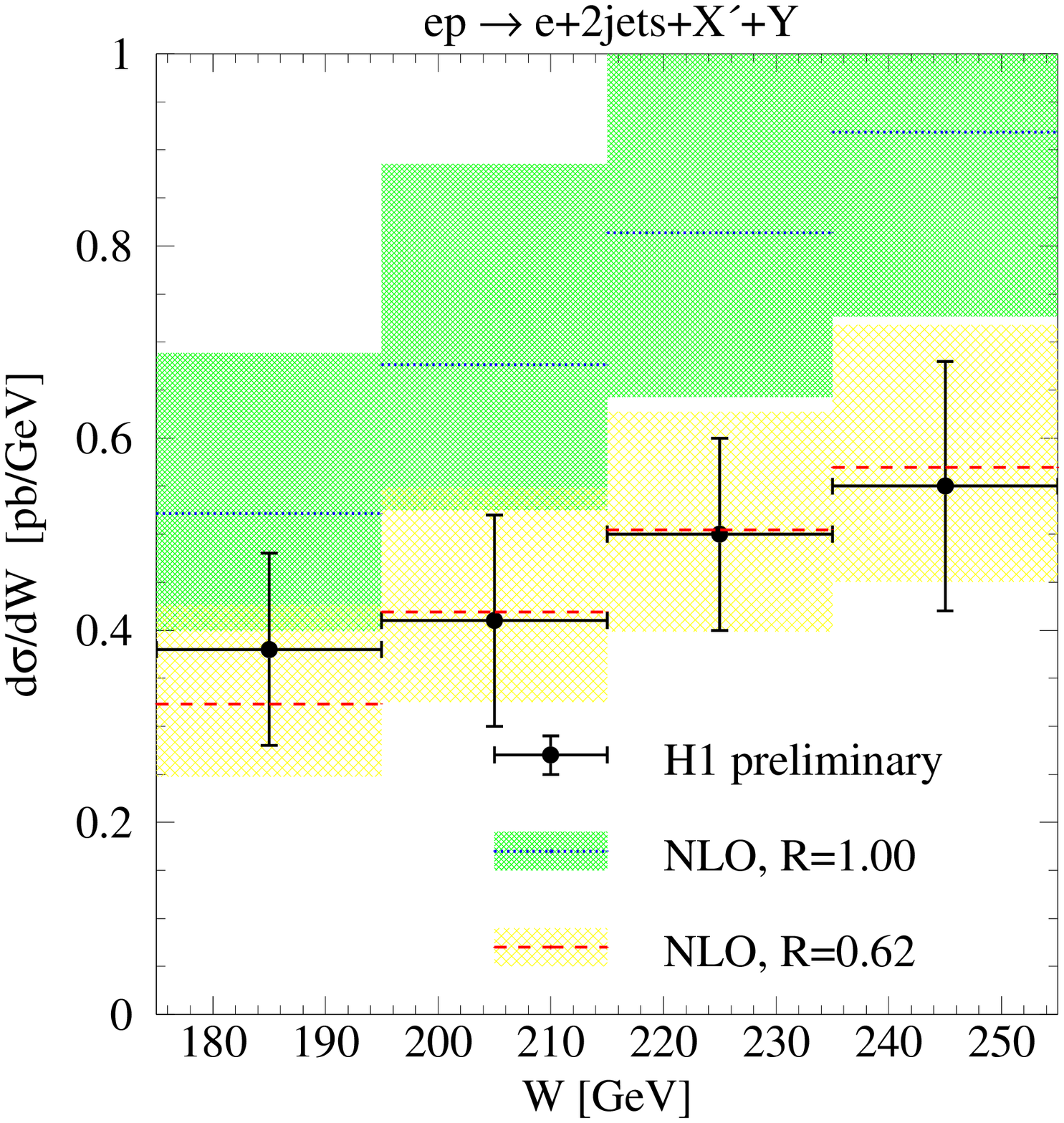}
 \caption{\label{fig:6}Differential cross sections for diffractive dijet
 photoproduction as measured by H1 with high-$E_T^{jet}$ cuts and compared to 
 NLO QCD without ($R=1$) and with ($R=0.62$) global suppression 
 (color online).}
\end{figure}
%
(global suppression). As before with the exception of $d\sigma/dE_T^{jet1}$ 
and $d\sigma/dM_{12}$ in Fig.\ 6d and Fig.\ 6f, most of the data points lie 
outside the $R = 1$ results with their
error bands and agree with the suppressed prediction with $R = 0.62$ inside
the respective errors. However, compared to the results in Figs.\ 3a-i
the distinction between the $R = 1$ band and the $R = 0.62$ band
and the data is somewhat less pronounced. We also tested the prediction for the
resolved (resolved+direct-IS) suppression, which is shown in Figs.\
7a-i. The suppression factor fitted to the smallest $E_T^{jet1}$-bin
%
\begin{figure}
 \centering
 \includegraphics[width=0.325\columnwidth]{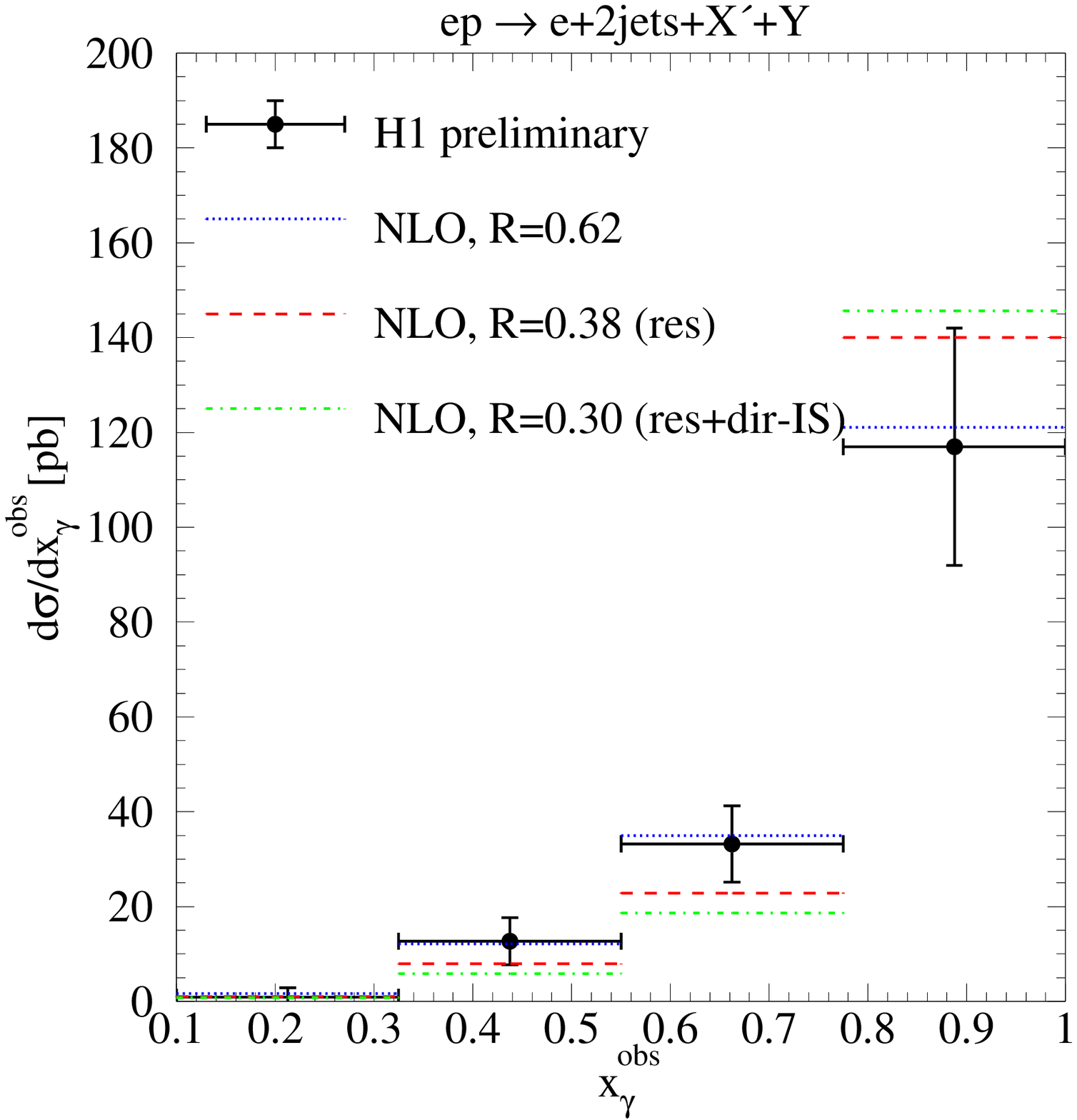}
 \includegraphics[width=0.325\columnwidth]{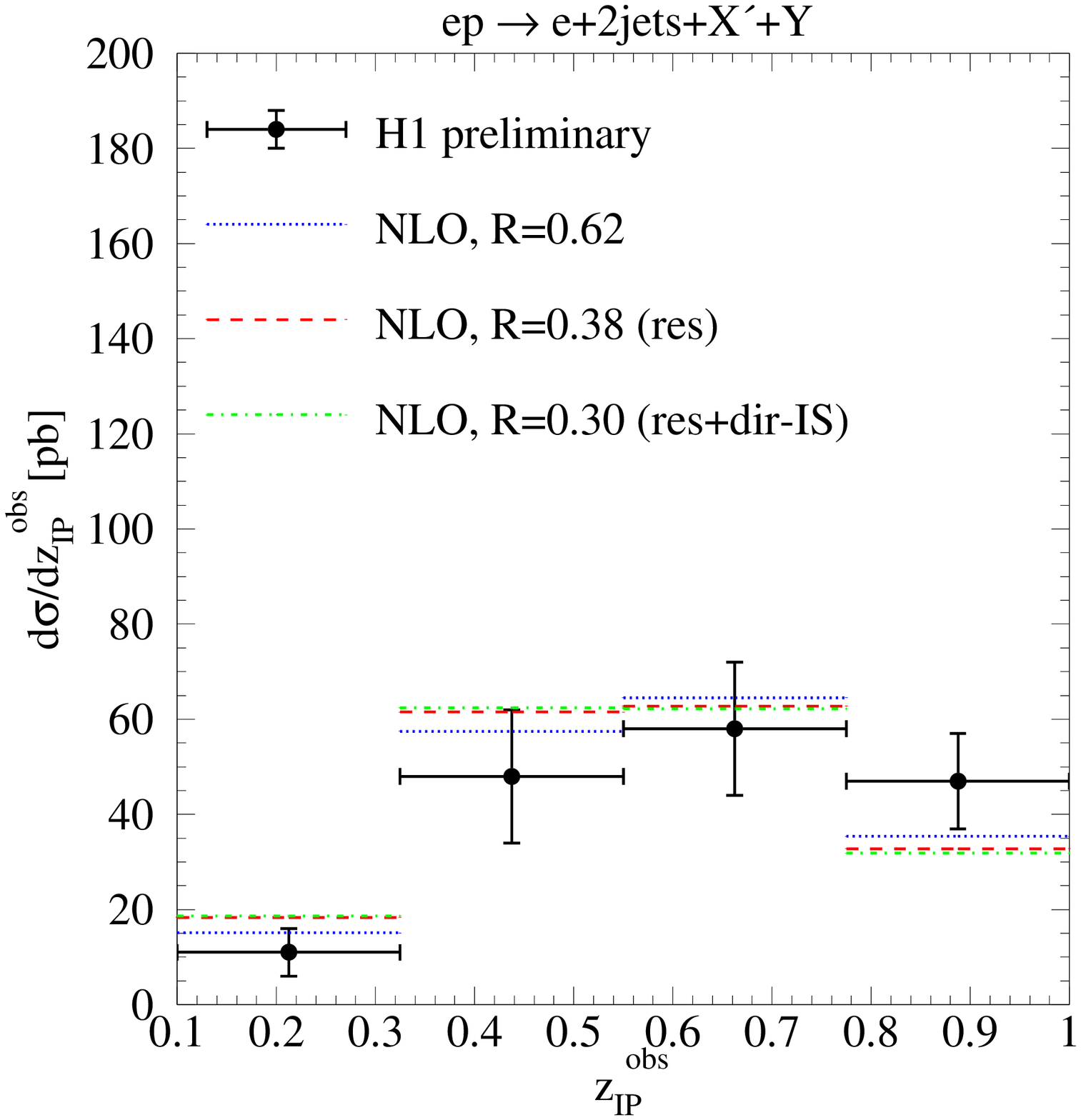}
 \includegraphics[width=0.325\columnwidth]{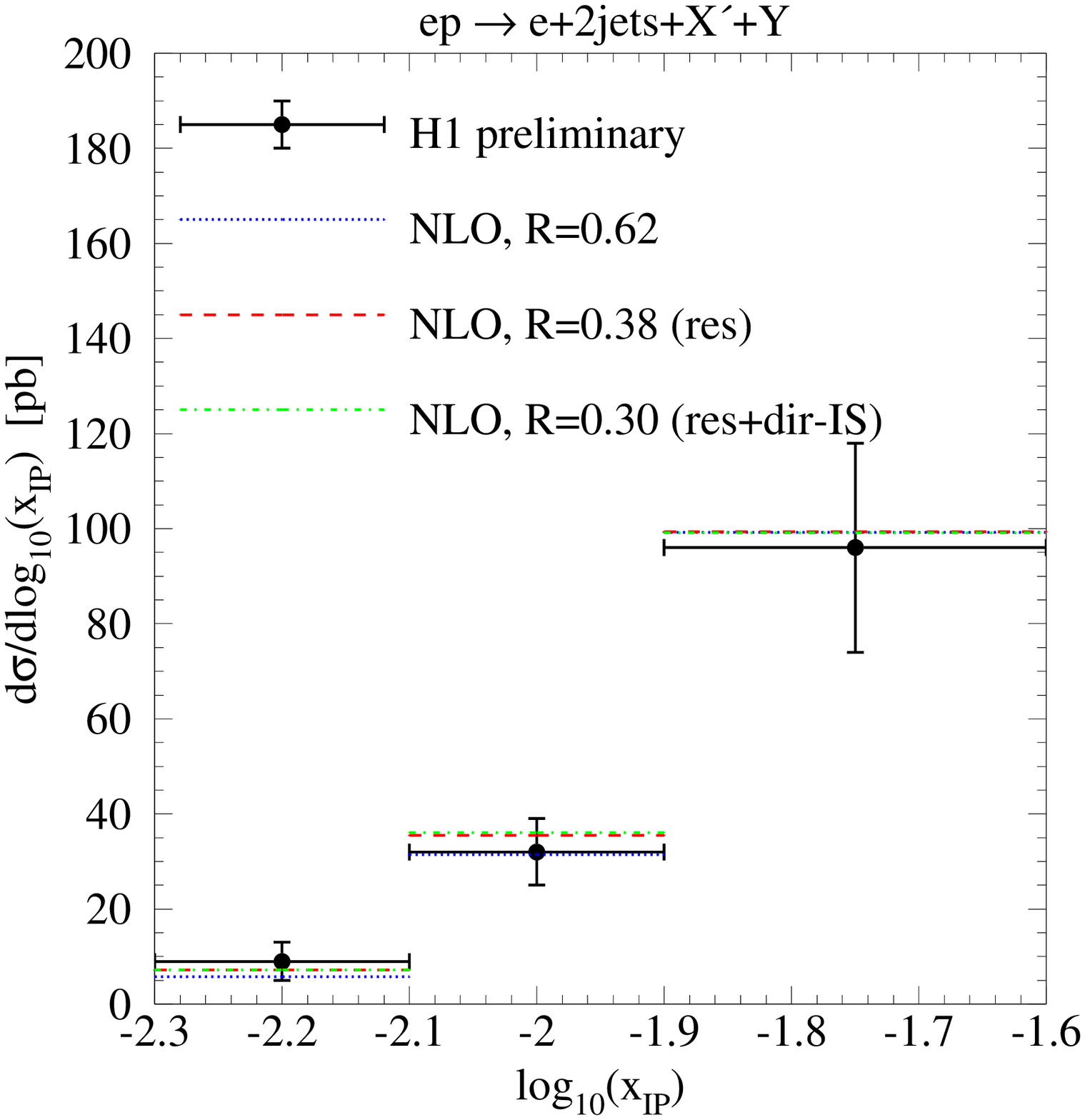}
 \includegraphics[width=0.325\columnwidth]{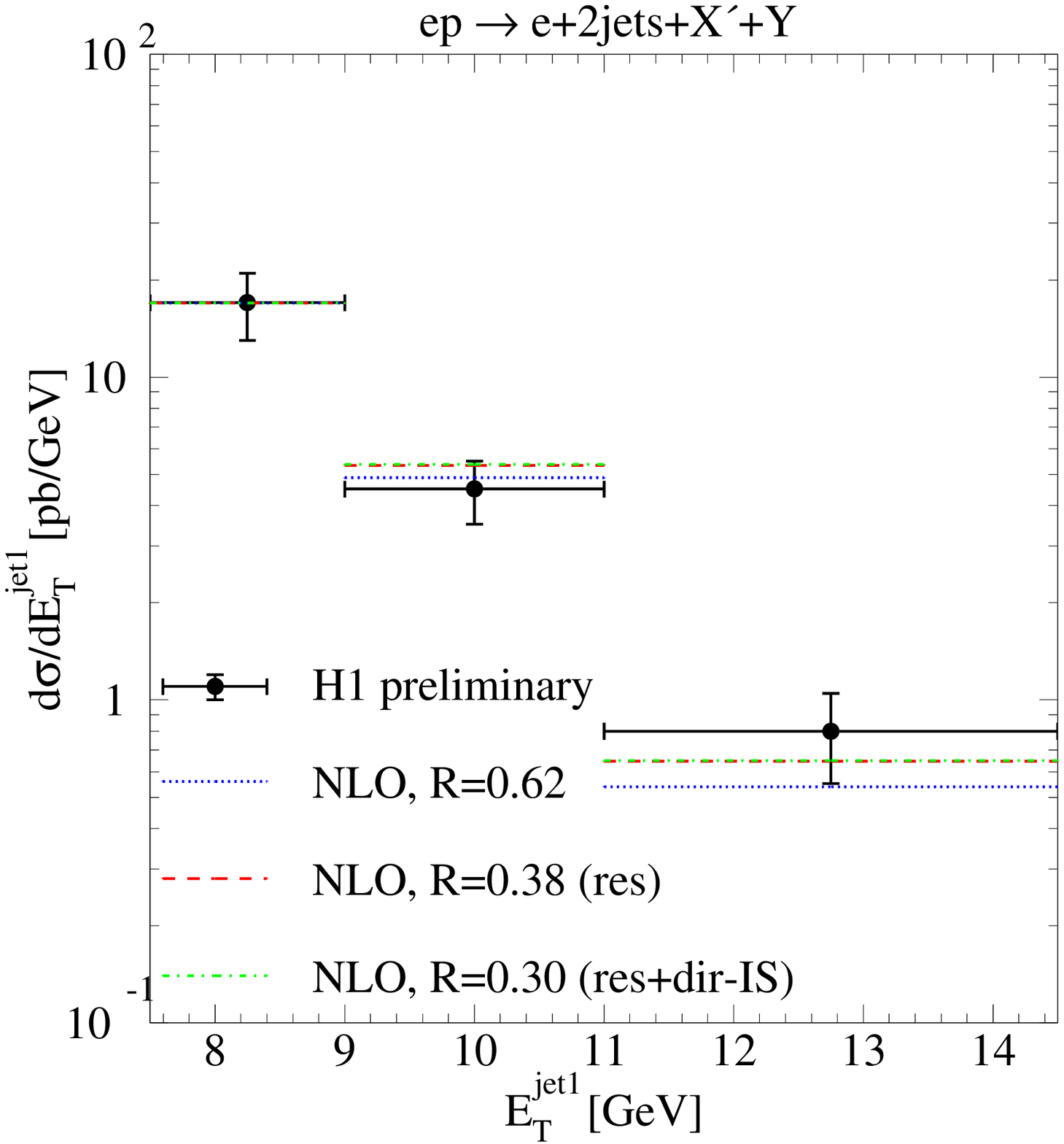}
 \includegraphics[width=0.325\columnwidth]{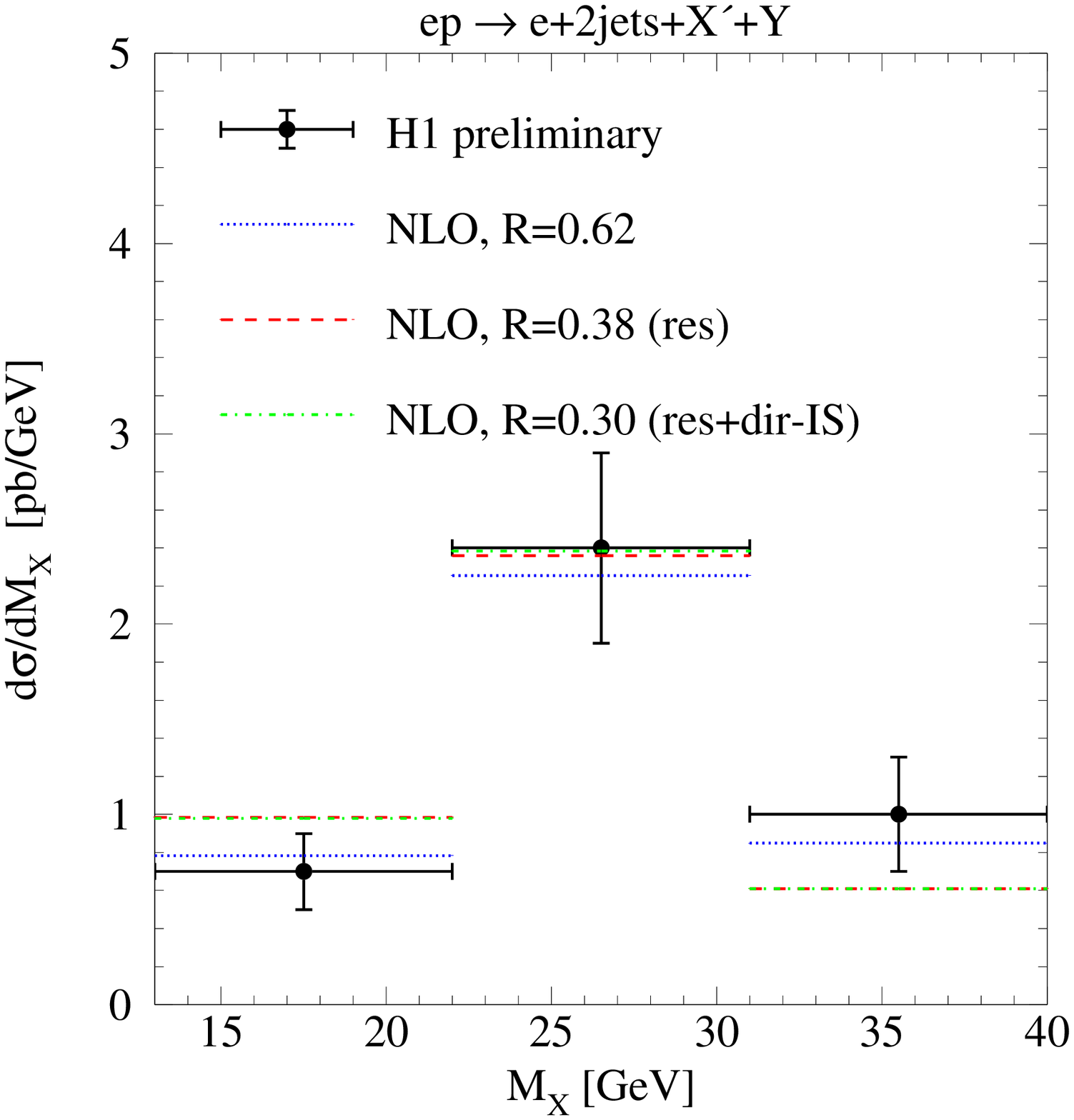}
 \includegraphics[width=0.325\columnwidth]{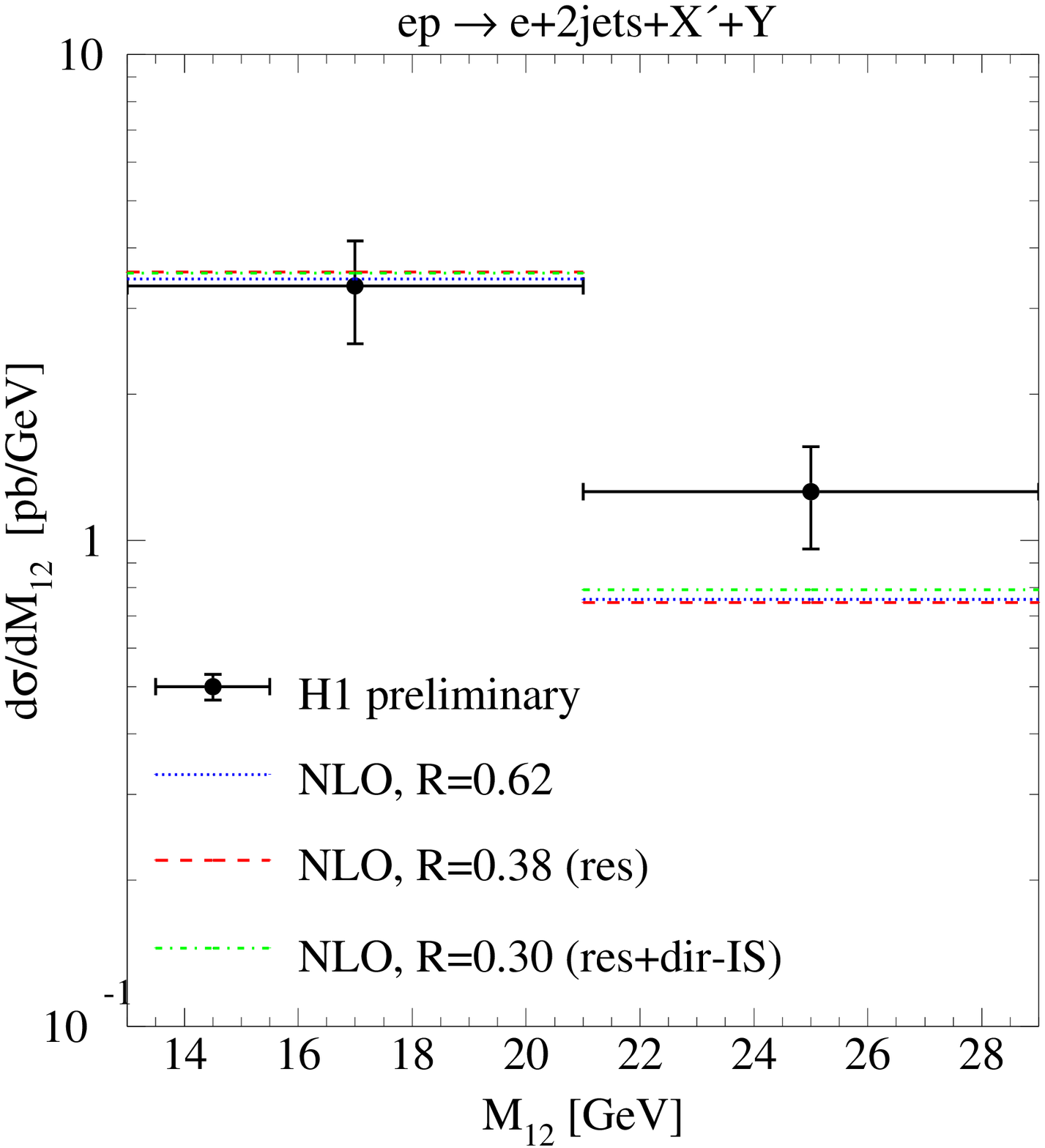}
 \includegraphics[width=0.325\columnwidth]{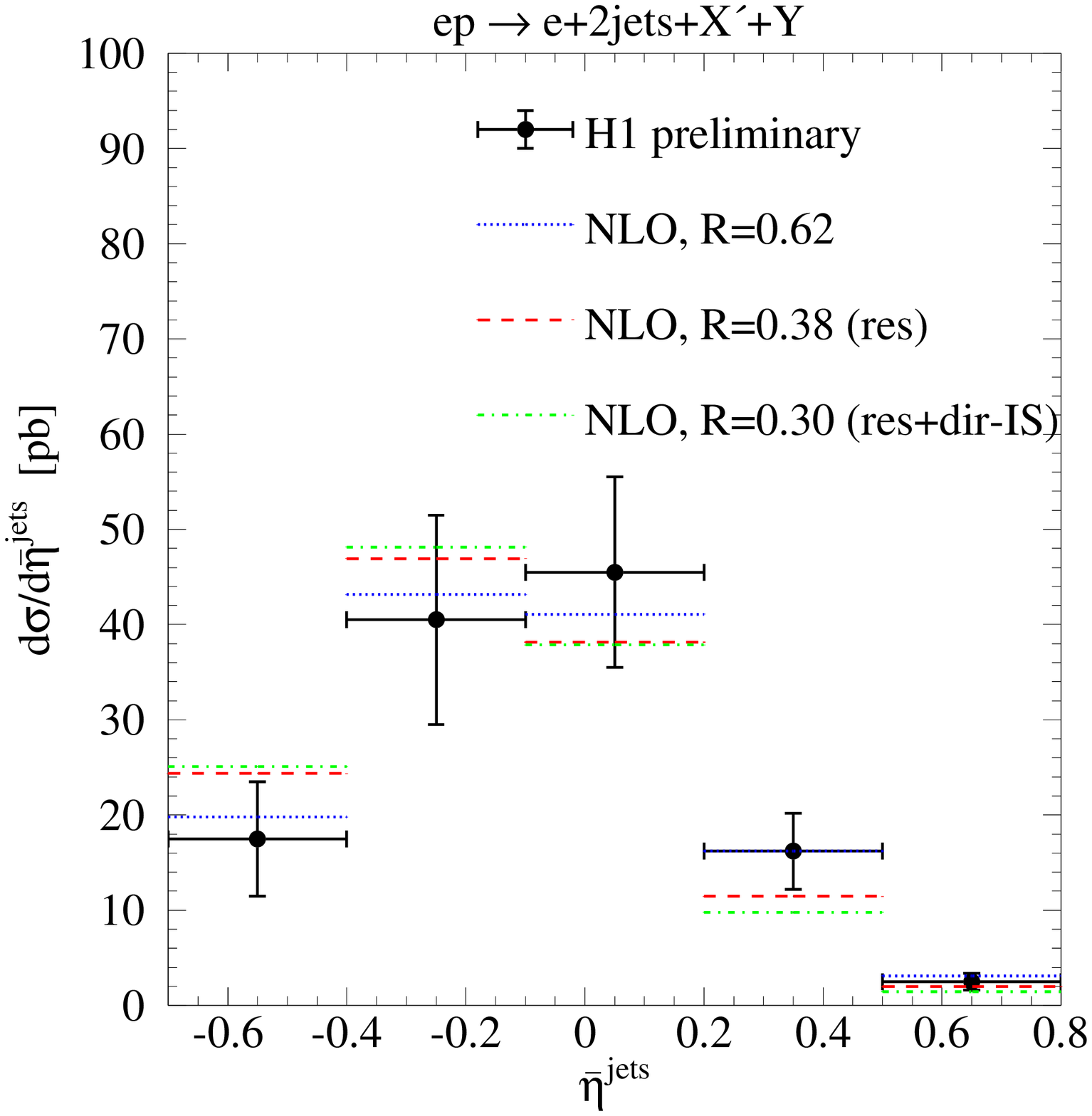}
 \includegraphics[width=0.325\columnwidth]{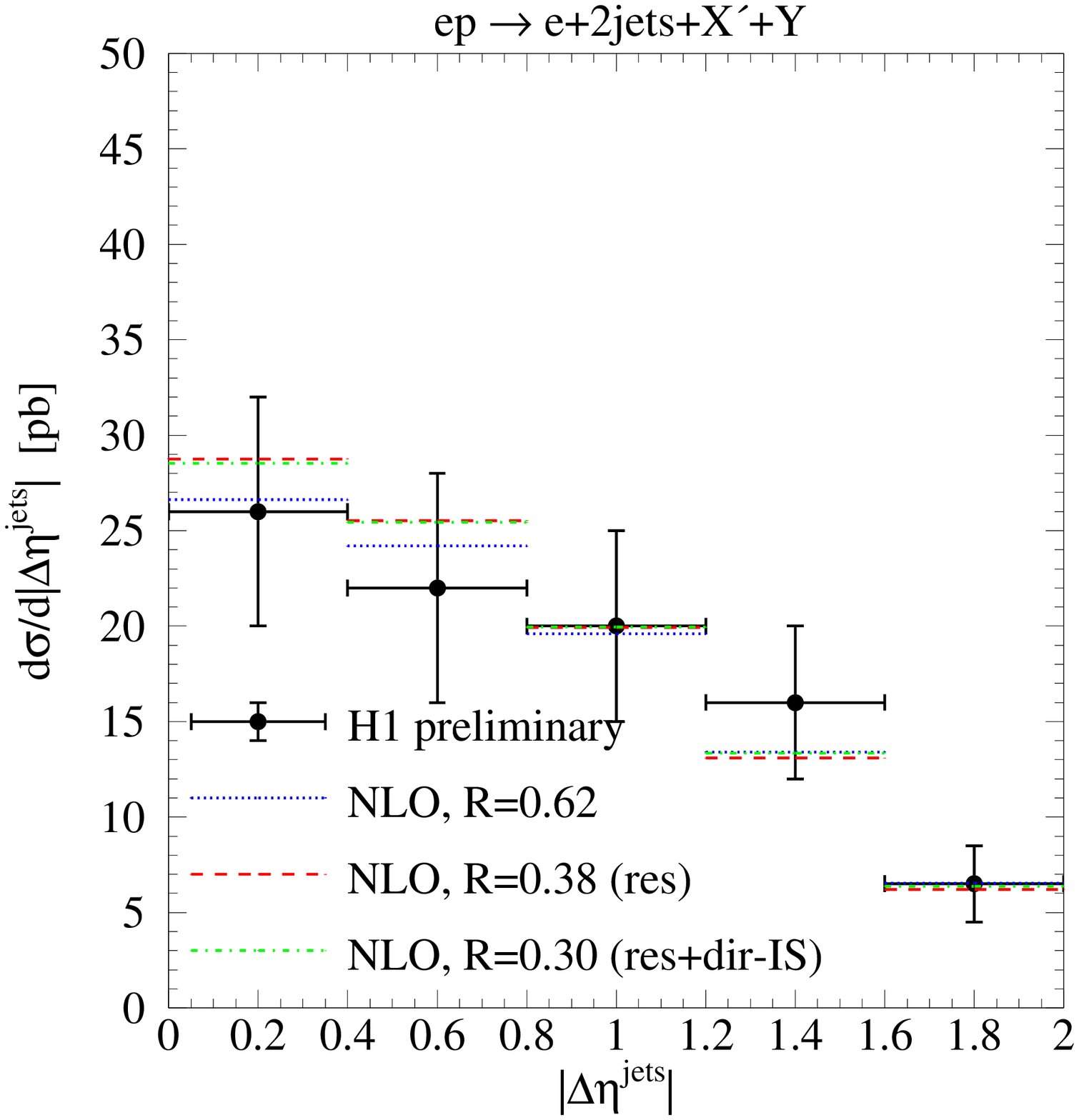}
 \includegraphics[width=0.325\columnwidth]{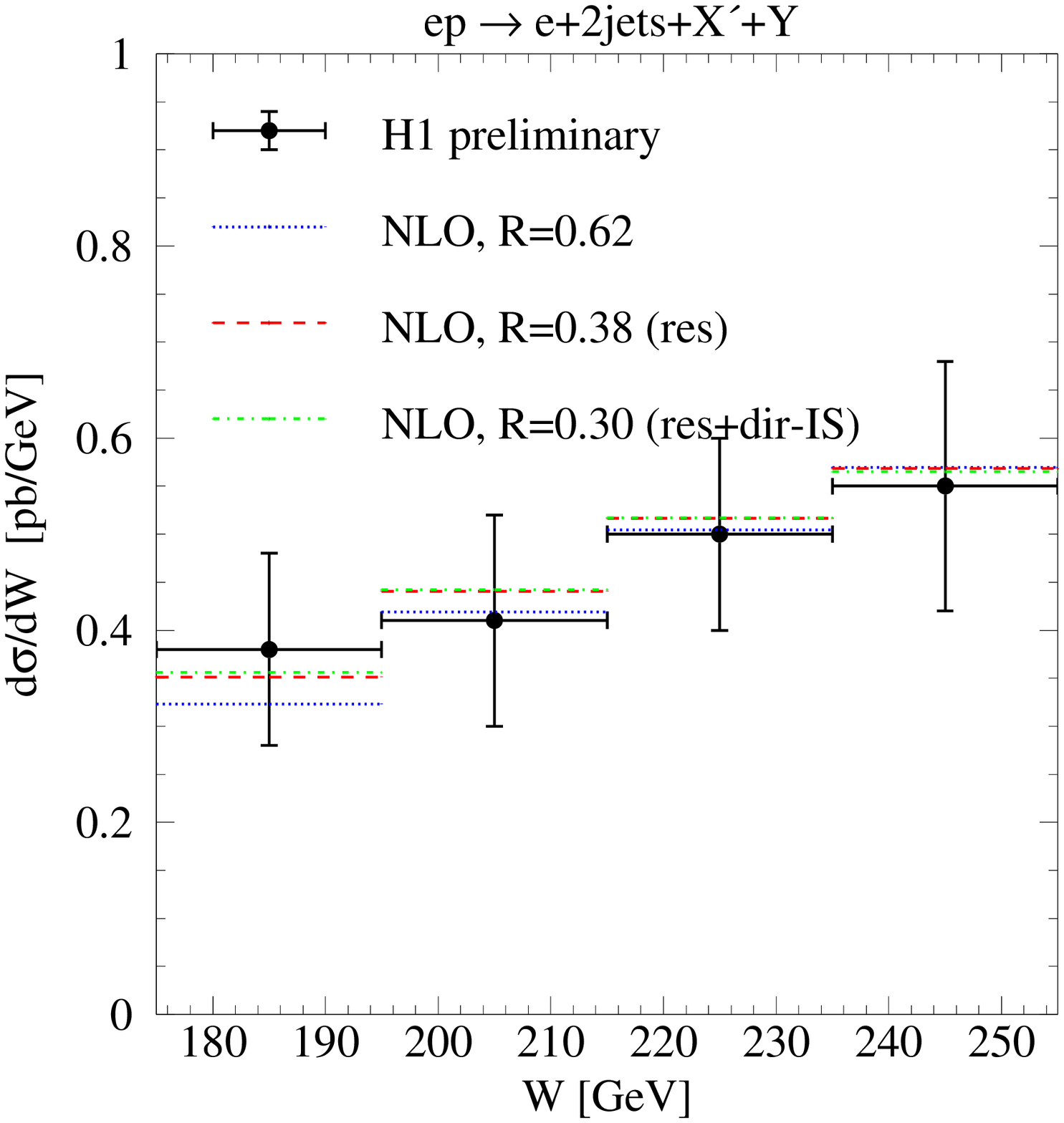}
 \caption{\label{fig:7}Differential cross sections for diffractive dijet
 photoproduction as measured by H1 with high-$E_T^{jet}$ cuts and compared to
 NLO QCD with global, resolved, and resolved/direct-IS suppression.}
\end{figure}
%
came out as $R = 0.38$ (res) and $R = 0.30$ (res+dir-IS). In most of the
comparisons it is hard to observe any preference for the direct plus
resolved (global) suppression against the resolved suppression only. We
remark that the suppression factor for the global suppression is increased
by $24\%$, if we go from the low-$E_T^{jet}$ to the high-$E_T^{jet}$ data,
whereas for the resolved suppression the difference is only $5\%$. Under the
assumption that the suppression factor should not depend on
$E_T^{jet1}$, we would conclude that the version with the resolved suppression
would be preferred. In Figs.\ 8a and b we tested the resolved
%
\begin{figure}
 \centering
 \includegraphics[width=0.495\columnwidth]{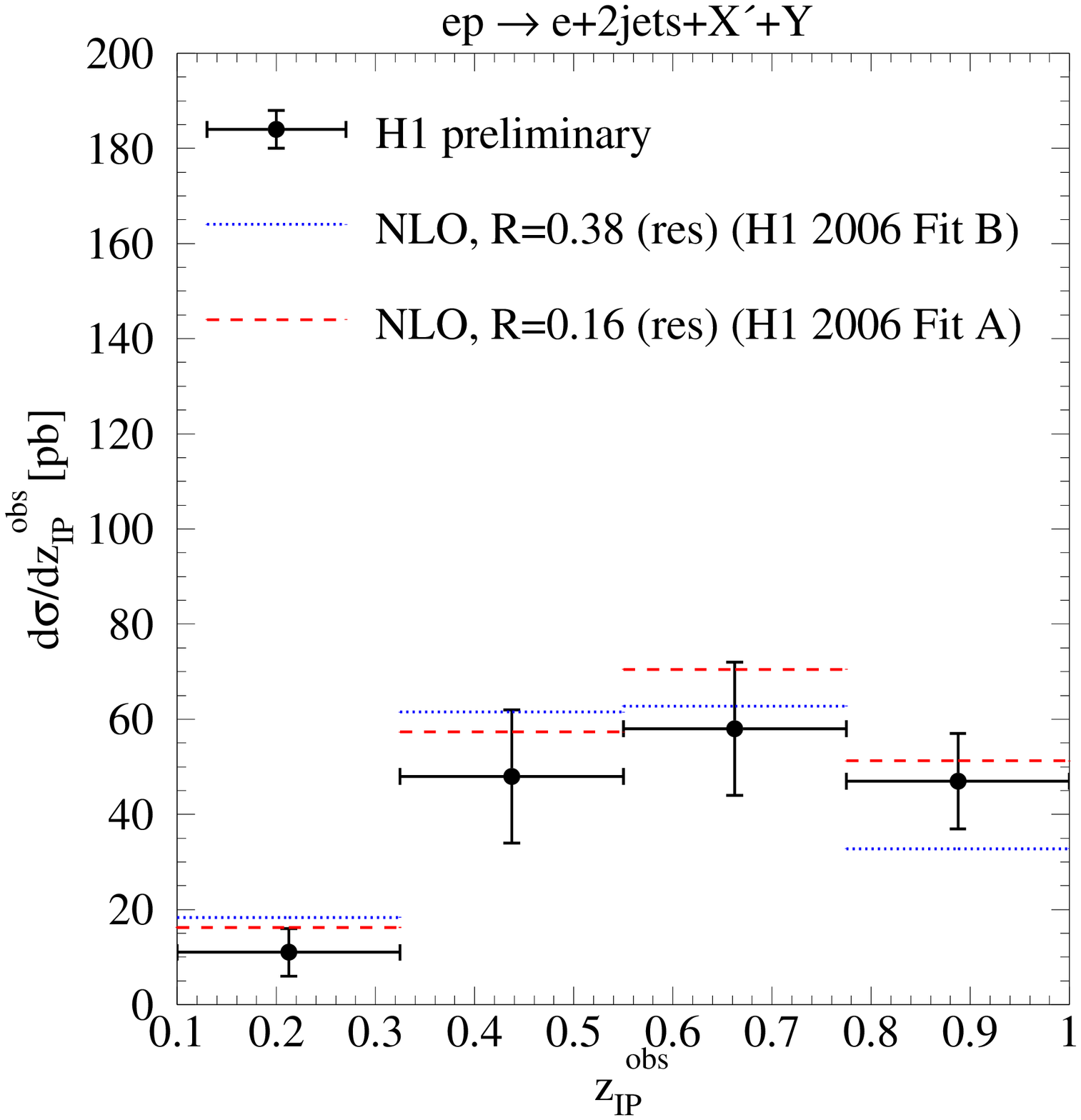}
 \includegraphics[width=0.495\columnwidth]{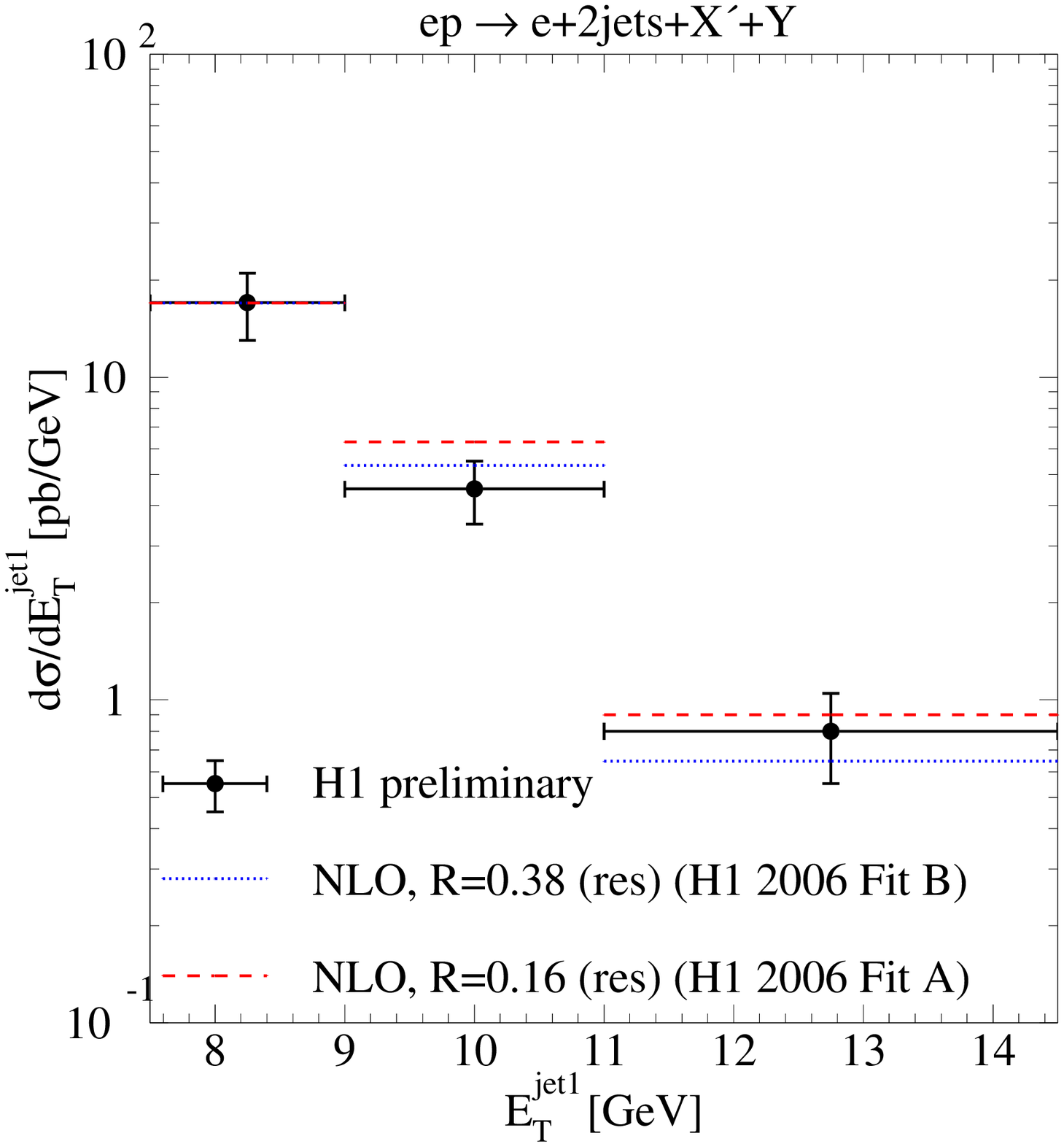}
 \caption{\label{fig:8}Differential cross sections for diffractive dijet
 photoproduction as measured by H1 with high-$E_T^{jet}$ cuts and compared to
 NLO QCD with resolved suppression and two different DPDFs.}
\end{figure}
%
suppression model against the choice of the two DPDFs, fit A versus fit B,
with the result that this dependence is weak if we adjust the suppression
factor, which is $R = 0.16$ for the `H1 2006 fit A'. The general conclusions
from the high-$E_T^{jet}$ comparison are very much the same as from the
low-$E_T^{jet}$
comparison. A global suppression is definitely observed and the version with
resolved suppression explains the data almost as well as with the global
suppression.

\section{Comparison with ZEUS data}

In this section we shall compare our predictions with the final analysis
of the ZEUS data, which was published just recently \cite{28}, in order to see 
whether they are consistent with the large-$E_T^{jet}$ data of H1. The
kinematic cuts are almost the same as in the high-$E_T^{jet}$  H1 measurements.
They are given in Tab.\ 2. The only major difference to the H1 cuts in
%
\begin{table}
 \caption{Kinematic cuts applied in the most recent ZEUS analysis of
 diffractive dijet photoproduction \cite{28}.}
 \begin{tabular}{c}
 ZEUS cuts \\
 \hline
 $Q^2<$ 1 GeV$^2$  \\
 0.2 $< y <$ 0.85     \\
 $E_T^{jet1} >$ 7.5 GeV \\
 $E_T^{jet2} >$ 6.5 GeV \\
 $-1.5 < \eta^{jet1(2)} < 1.5$ \\
 $x_{\p} <$ 0.025 \\
 $|t| <$ 5 GeV$^2$ \\
 \end{tabular}
\end{table}
%
Tab.\ 1 is the larger range in the variable $y$.
Therefore the ZEUS cross sections
will be larger than the corresponding H1 cross sections. The different cuts
on $Q^2$ and $|t|$  have little influence. For example, the larger $|t|$-cut 
in Tab.\ 2 as compared to Tab.\ 1 increases the cross section only by $0.2\%$.
The constraint on $M_Y$ is not explicitly given in the ZEUS publication 
\cite{28}. They give the cross section for the case that the diffractive final 
state consists only of the proton. For this they correct their measured cross
section by subtracting in all bins the estimated contribution of a
proton-dissociative background of $16\%$. When comparing to the theoretical
predictions they do the reverse and multiply the cross section with the
factor $0.87$, in order to correct for the proton-dissociative contributions,
which are contained in the DPDFs `H1 2006 fit A' and `H1 2006 fit B' by
requiring $M_Y < 1.6$ GeV. We do not follow this procedure. Instead we leave 
the theoretical cross sections unchanged, i.e.\ they contain a
proton-dissociative contribution with $M_Y < 1.6$ GeV, and multiply the ZEUS 
cross sections by $1.15$ to include the proton-dissociative contribution. 
Since the ZEUS collaboration did measurements only for the high-$E_T^{jet}$
cuts, $E_T^{jet1(2)} > 7.5$ (6.5) GeV,
we can only compare to those. In this comparison 
we shall follow the same strategy as before. We first compare to the 
predictions with no suppression ($R = 1$) and then determine a suppression
factor by fitting $d\sigma/dE_T^{jet1}$ to the smallest $E_T^{jet1}$-bin.
Then we compare to the cross sections as a function of the seven observables 
$x_{\gamma}^{obs}$, $z_{\p}^{obs}$, $x_{\p}$, $E_T^{jet1}$, $y$, $M_X$ and
$\eta^{jet1}$
instead of the nine variables in the H1 analysis. The distribution in $y$ is 
equivalent to the $W$-distribution in \cite{29}. The theoretical predictions
for these differential cross sections with no suppression factor ($R = 1$) are
shown in Figs.\ 9a-g, together with their scale errors and compared to
%
\begin{figure}
 \centering
 \includegraphics[width=0.325\columnwidth]{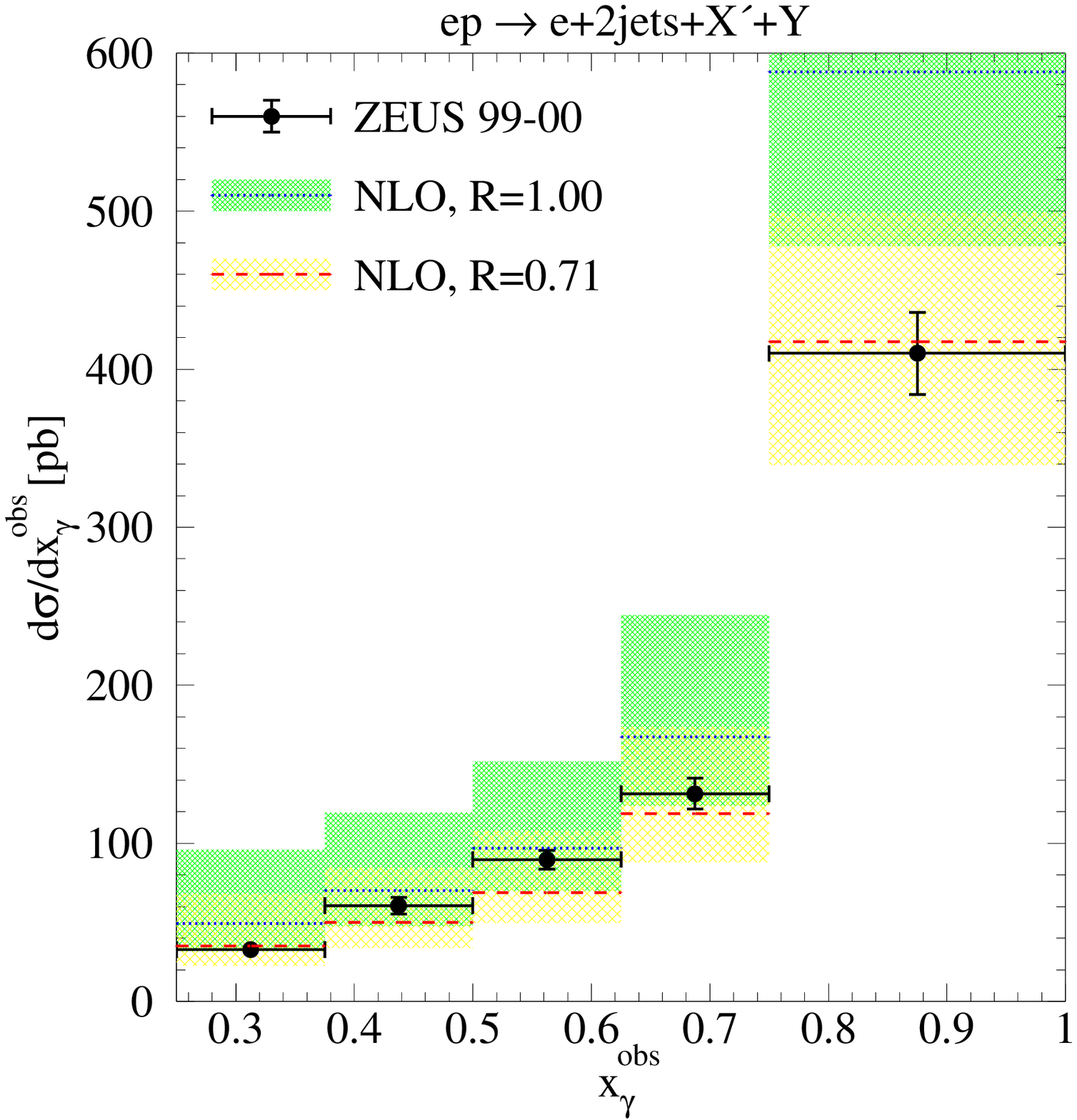}
 \includegraphics[width=0.325\columnwidth]{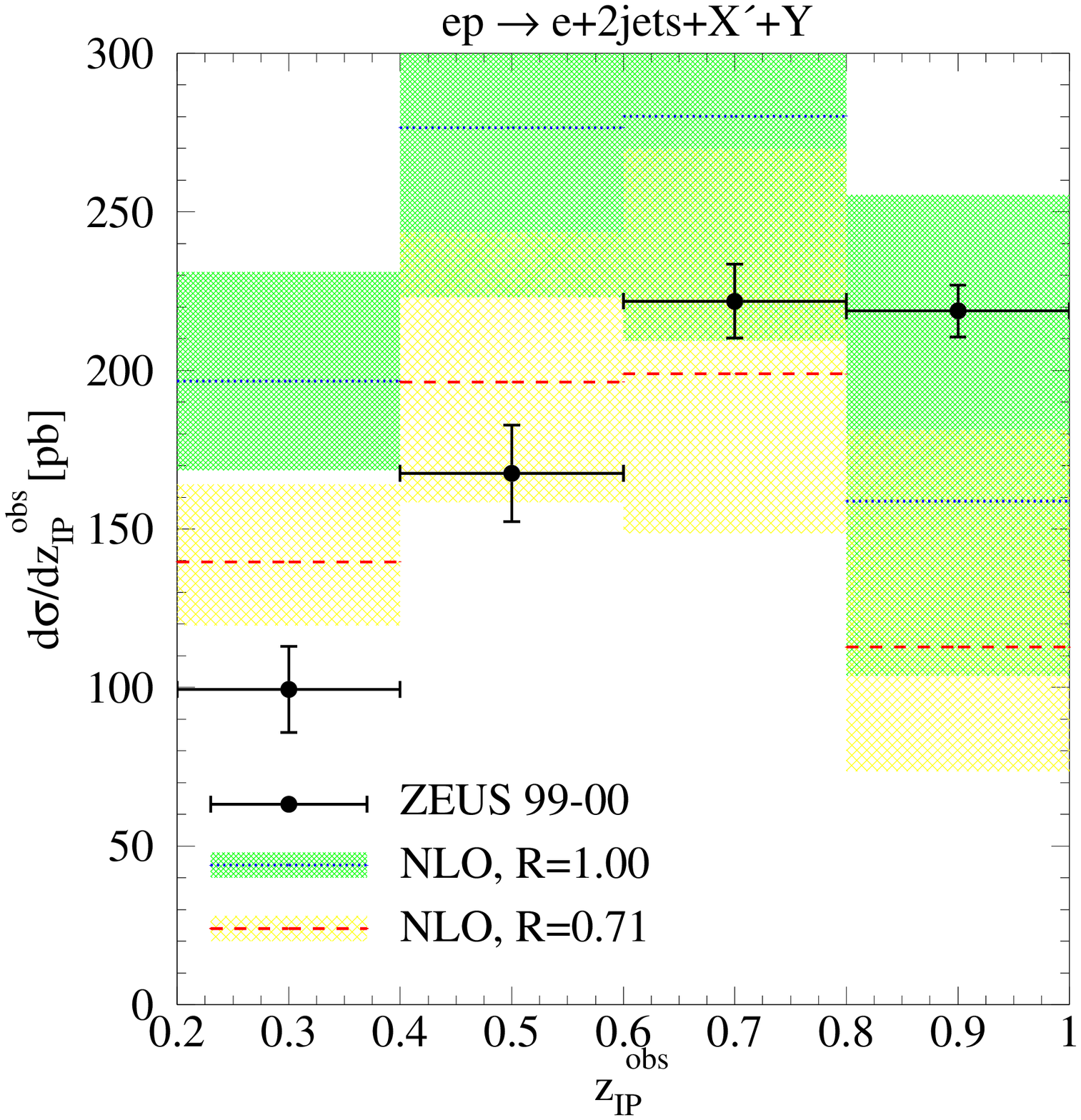}
 \includegraphics[width=0.325\columnwidth]{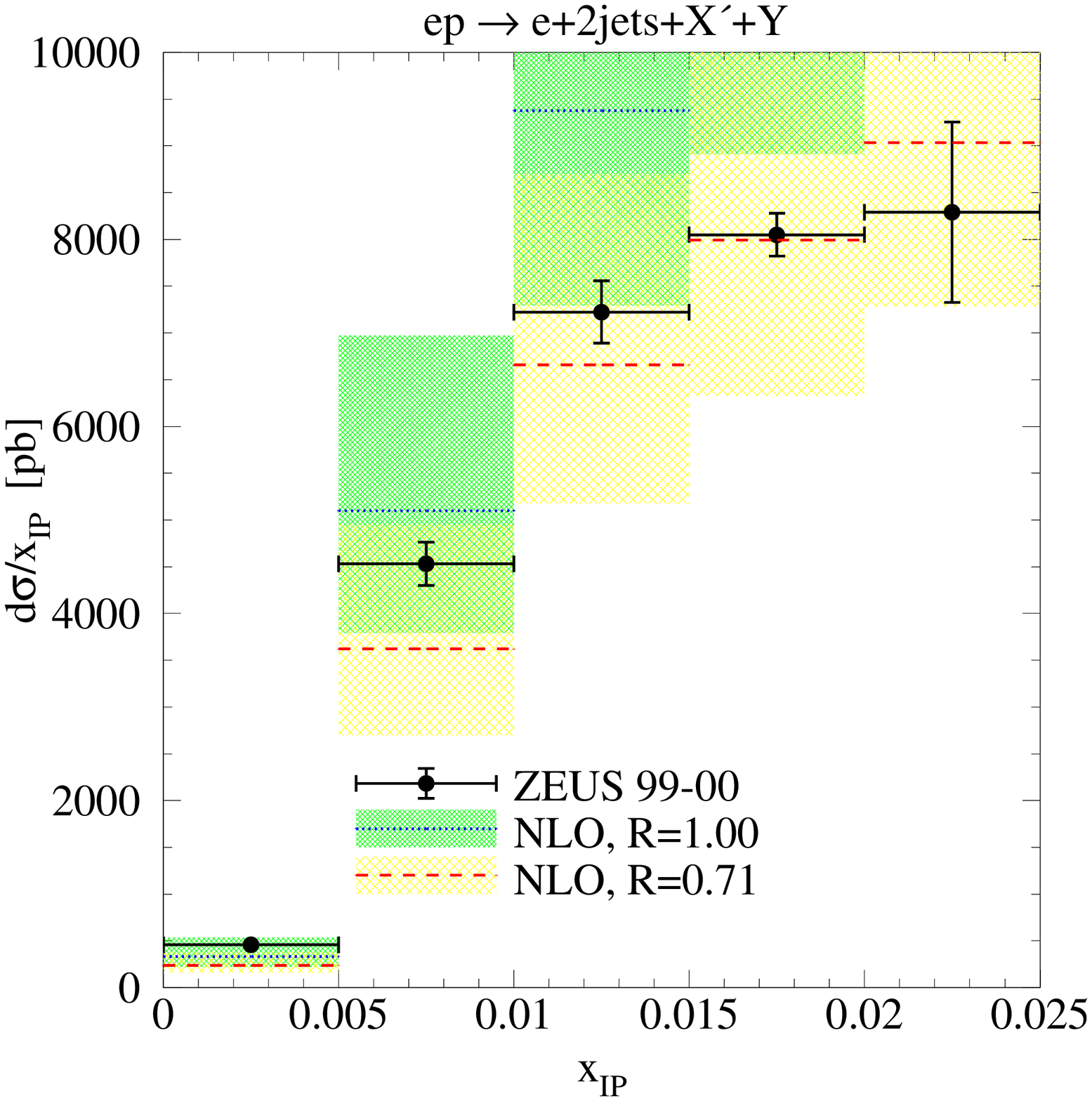}
 \includegraphics[width=0.325\columnwidth]{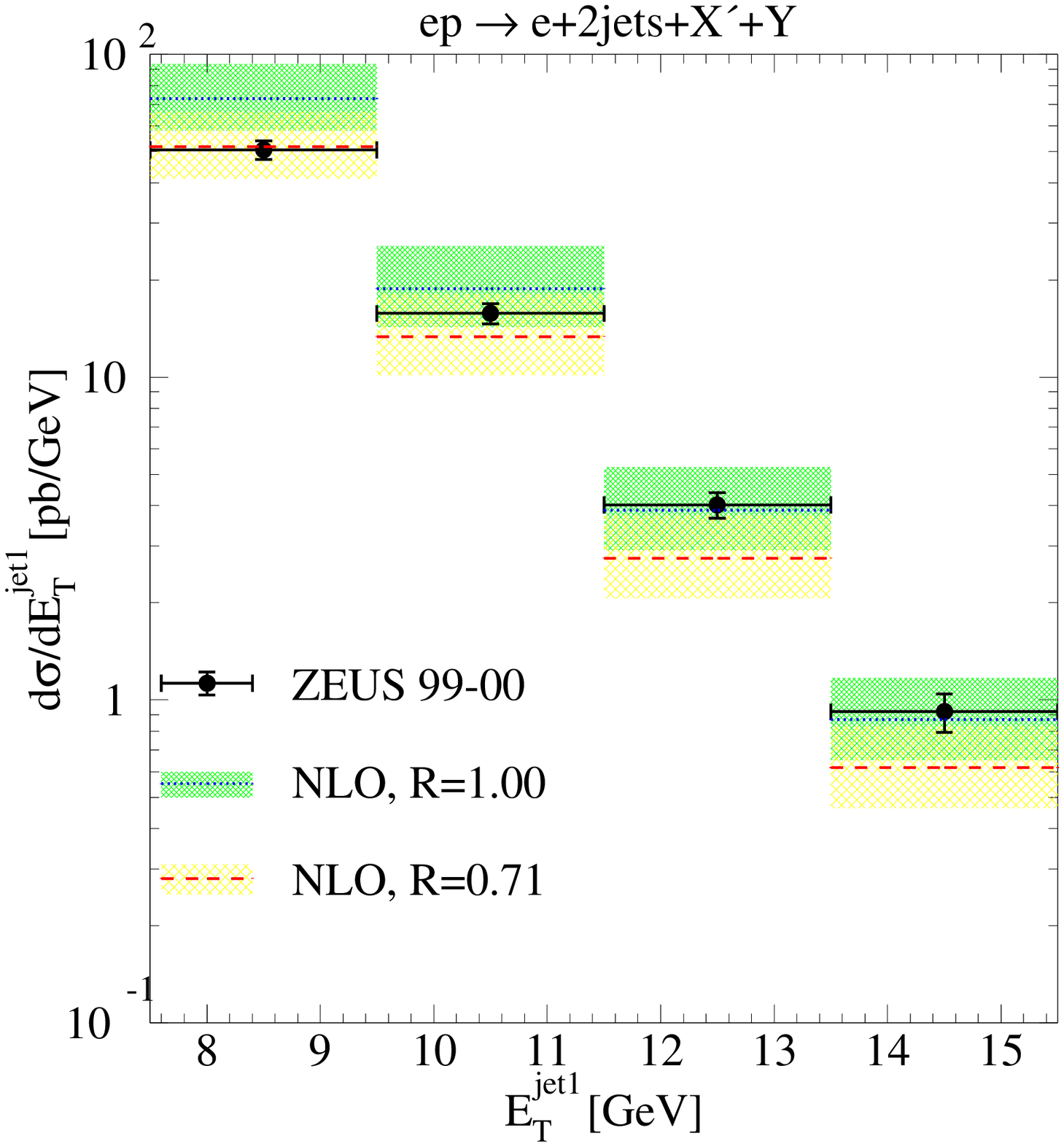}
 \includegraphics[width=0.325\columnwidth]{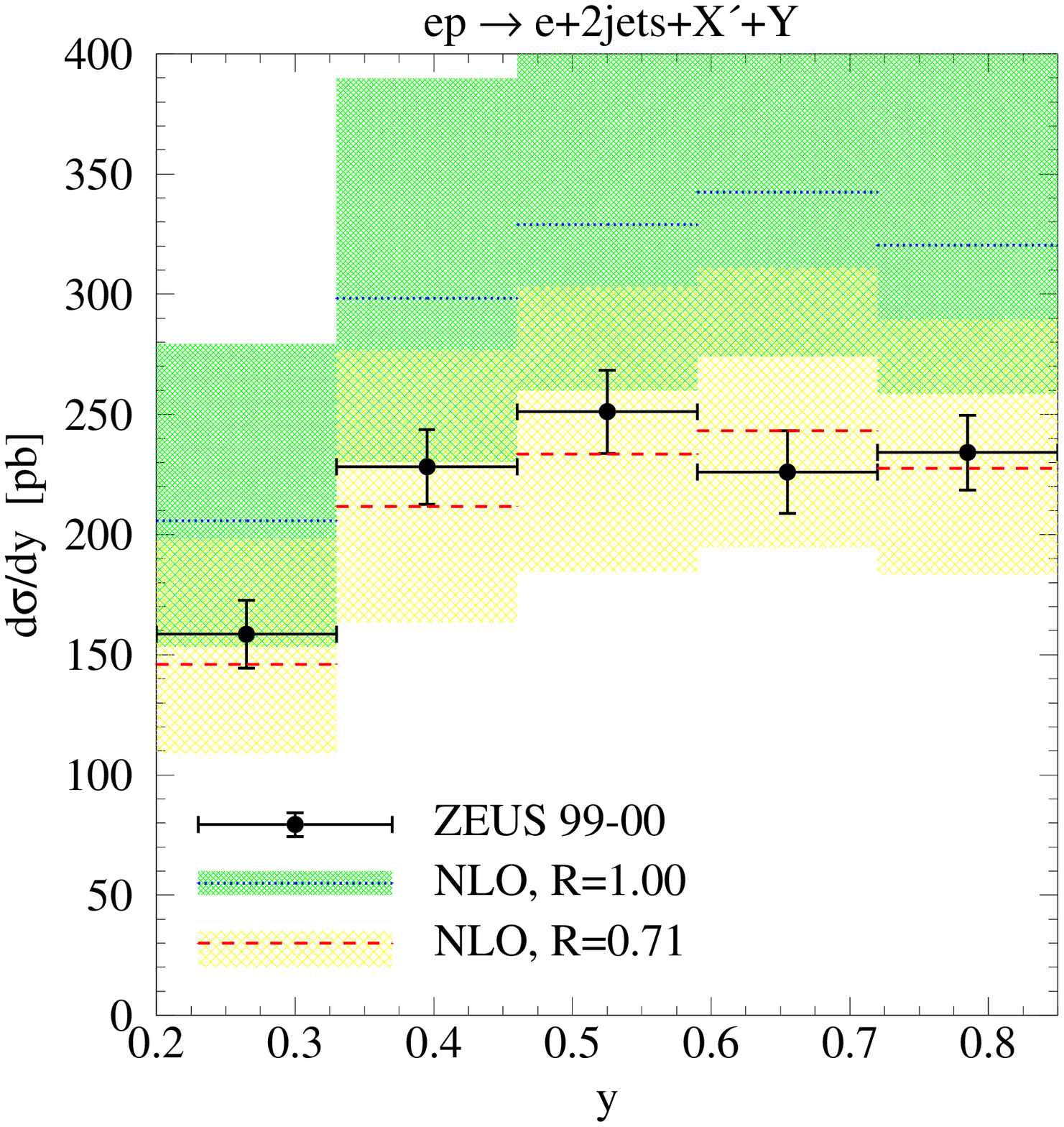}
 \includegraphics[width=0.325\columnwidth]{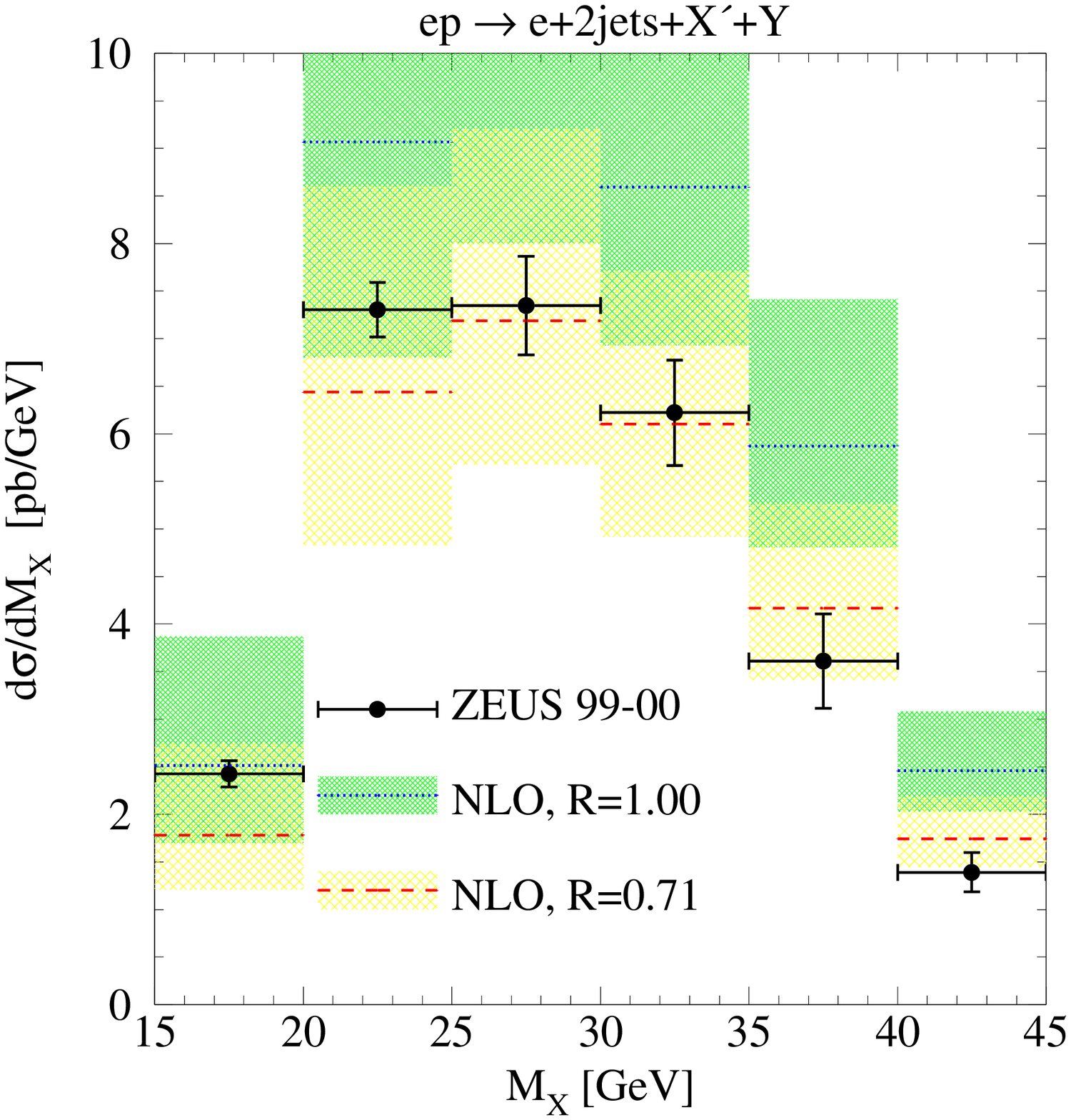}
 \includegraphics[width=0.325\columnwidth]{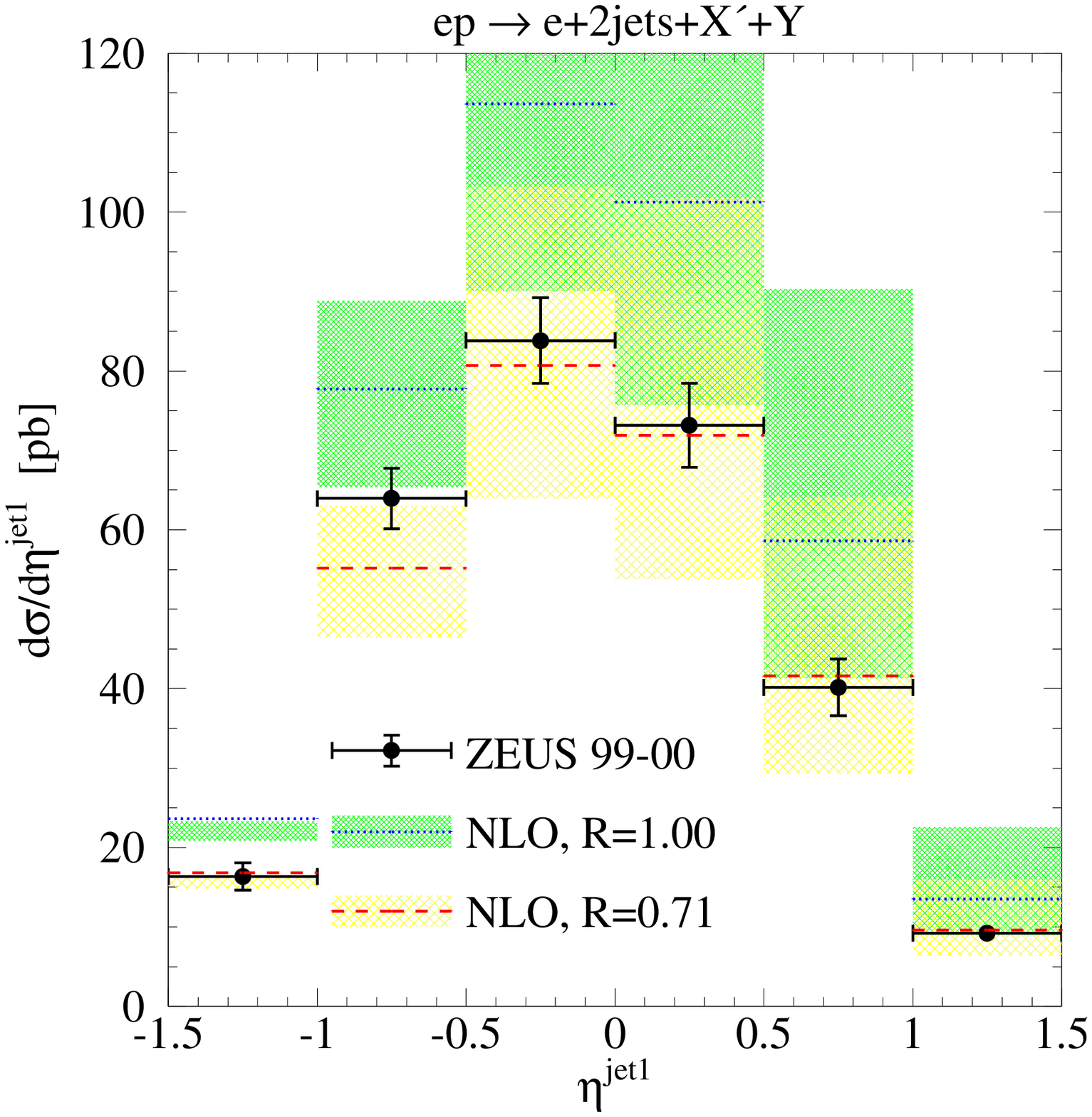}
 \caption{\label{fig:9}Differential cross sections for diffractive dijet
 photoproduction as measured by ZEUS and compared to
 NLO QCD without ($R=1$) and with ($R=0.71$) global suppression 
 (color online).}
\end{figure}
%
the ZEUS data points. Except for the $x_{\gamma}^{obs}$- and
$E_T^{jet1}$-distributions, most of the data points lie outside the
theoretical error bands for $R = 1$. In particular, in Figs.\ 9b, c, e, f and
g, 2, 3, 4, 4 and 5 points lie outside. This means that most of the data
points disagree with the unsuppressed prediction. Next, we determine the
suppression factor from the measured $d\sigma/dE_T^{jet1}$ at the lowest 
$E_T^{jet1}$-bin, 7.5 GeV $<E_T^{jet1}<9.5$ GeV, and obtain 
$R = 0.71$. As a curiosity, we remark that this factor is larger by a factor 
of $1.15$ than the suppression factor from the analysis of the 
high-$E_T^{jet}$ data from H1. This factor is exactly equal to the correction 
factor we had to apply to restore the dissociative proton 
contribution. Without this correction factor the suppression factor following 
from the ZEUS analysis would be in perfect agreement with the factor in the H1
analysis. Taking the total experimental error of $\pm7\%$ from the experimental
cross section $d\sigma/dE_T^{jet1}$ in the first bin
into account, the ZEUS suppression factor is $0.71\pm0.05$ to be 
compared to $0.62\pm0.14$ in the H1 analysis \cite{29}, so that both 
suppression factors agree inside the experimental errors.

If we now check how the predictions for $R = 0.71$ compare to the data
points inside the theoretical errors, we observe from Figs.\ 9a-g
that with the exception of $d\sigma/dz_{\p}^{obs}$ and 
$d\sigma/dE_T^{jet1}$ the majority of the data points
agree with the predictions. This is quite consistent with the H1 analysis,
discussed in the previous section, and leads to the conclusion that
also the ZEUS data agree much better with the suppressed predictions than
with the unsuppressed prediction. In particular, the global suppression
factor agrees with the global suppression factor obtained from the
analysis of the H1 data inside the experimental error.

Similarly as in the previous section we compared the ZEUS data also with the
assumption that the suppression results only from the resolved cross
section. Here we consider again the two versions: (i) only resolved
suppression (res) and (ii) resolved plus direct suppression of the
initial-state
singular part (res+dir-IS). For these two models we obtain the
suppression factors $R = 0.53$ and $R = 0.45$, respectively, where these
suppression
factors are again obtained by fitting the data point at the first bin of
$d\sigma/dE_T^{jet1}$. The comparison to the global suppression with $R=0.71$ 
and to the data is shown in Figs.\ 10a-g.
%
\begin{figure}
 \centering
 \includegraphics[width=0.325\columnwidth]{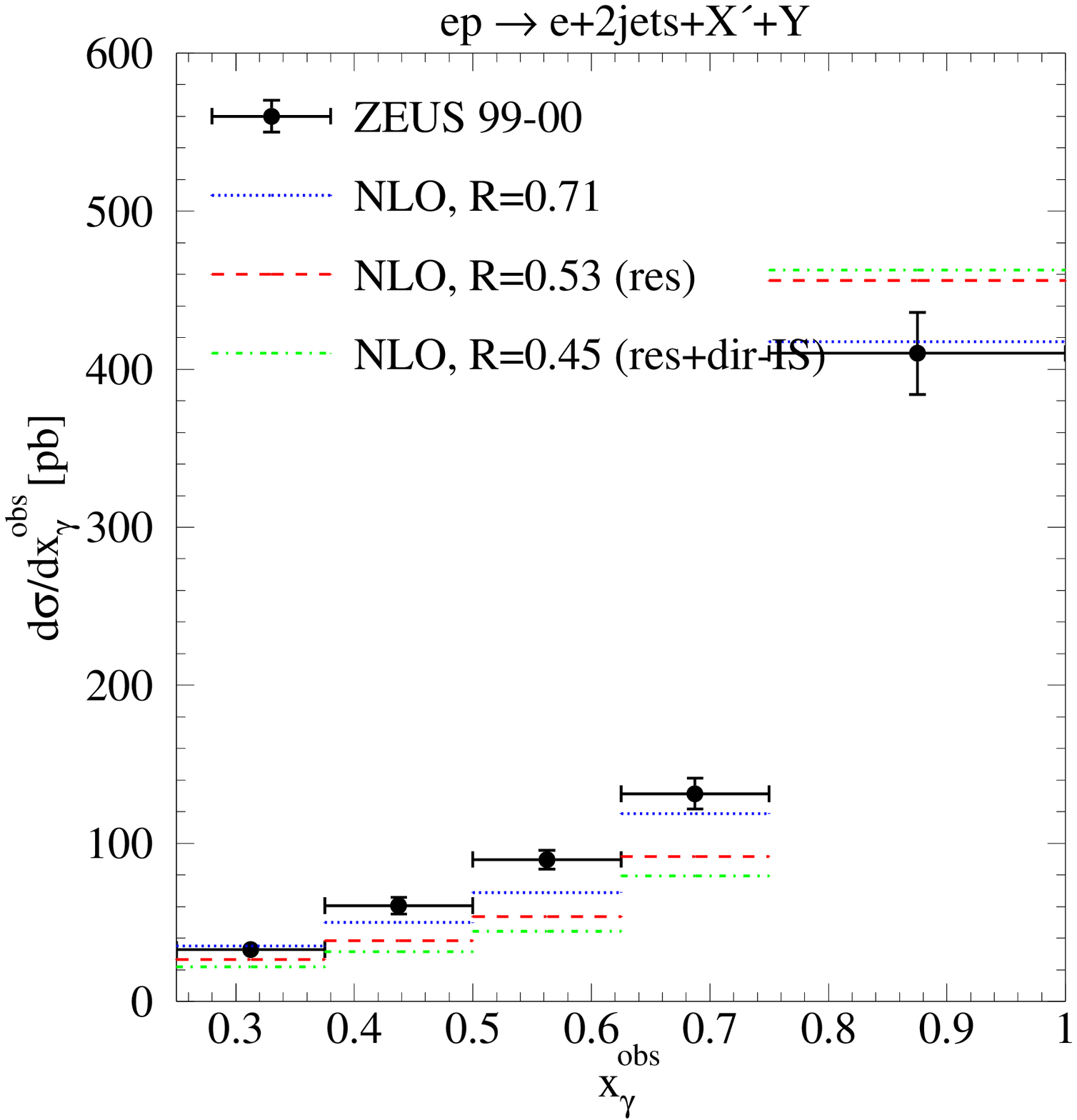}
 \includegraphics[width=0.325\columnwidth]{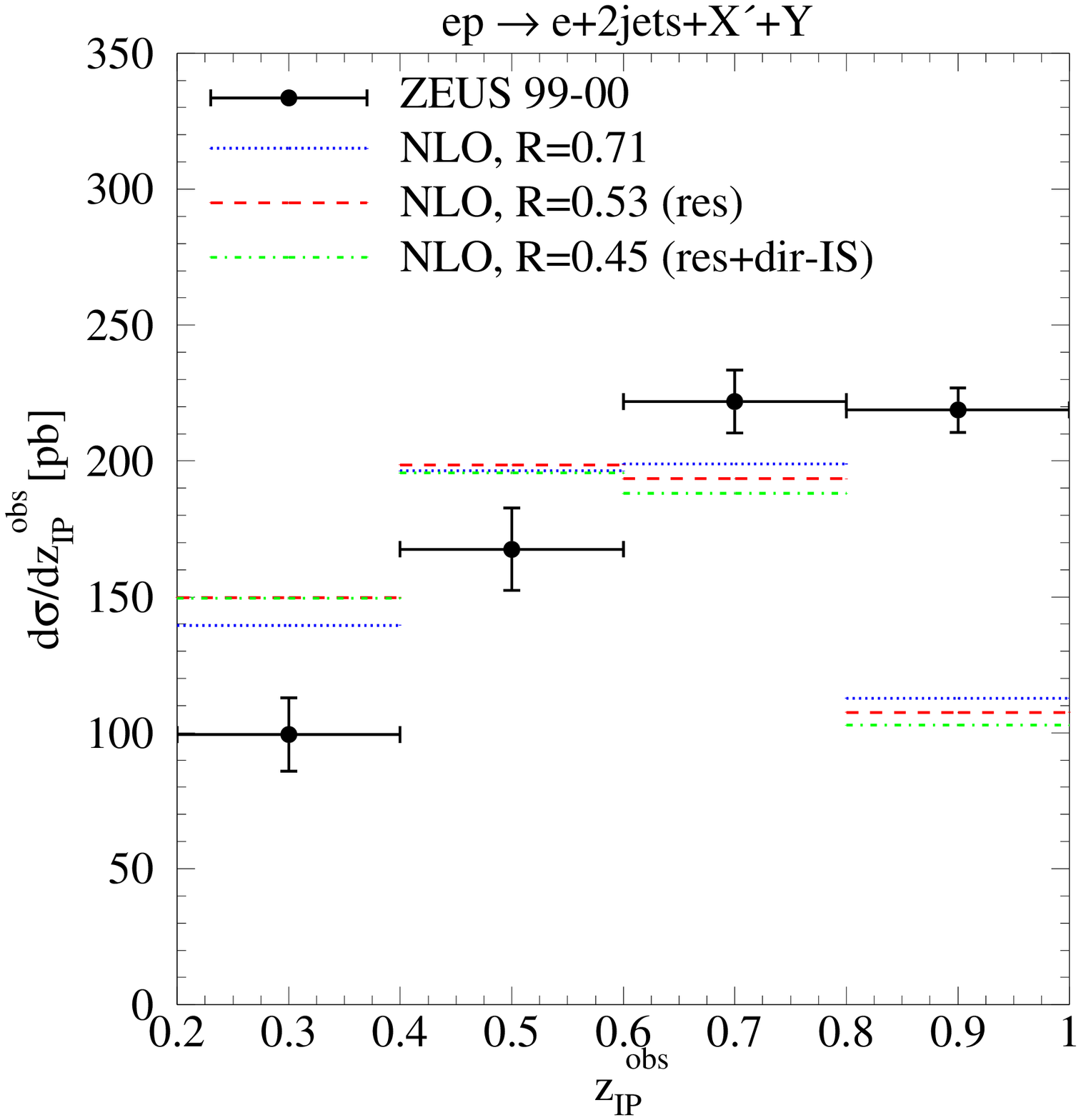}
 \includegraphics[width=0.325\columnwidth]{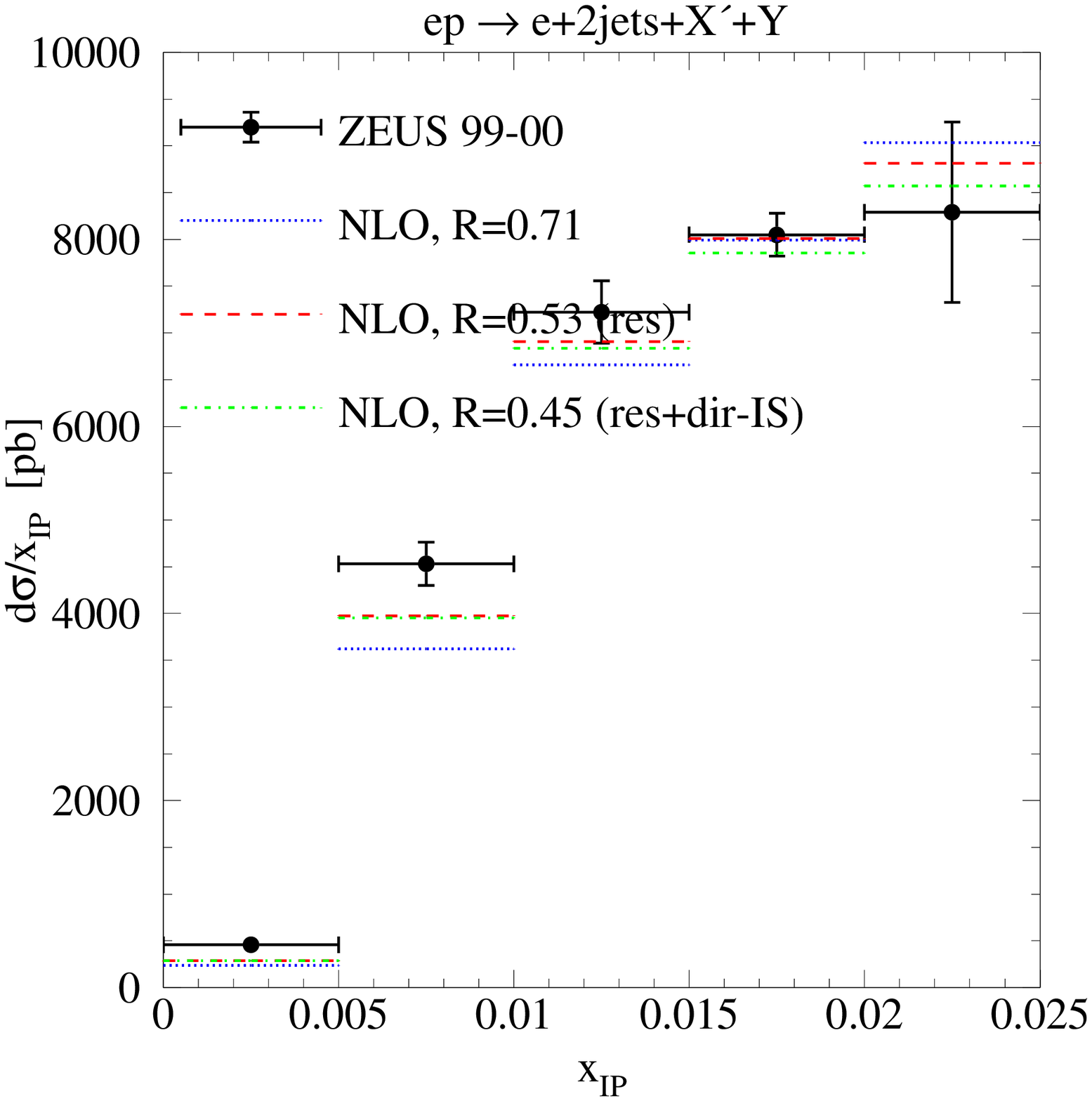}
 \includegraphics[width=0.325\columnwidth]{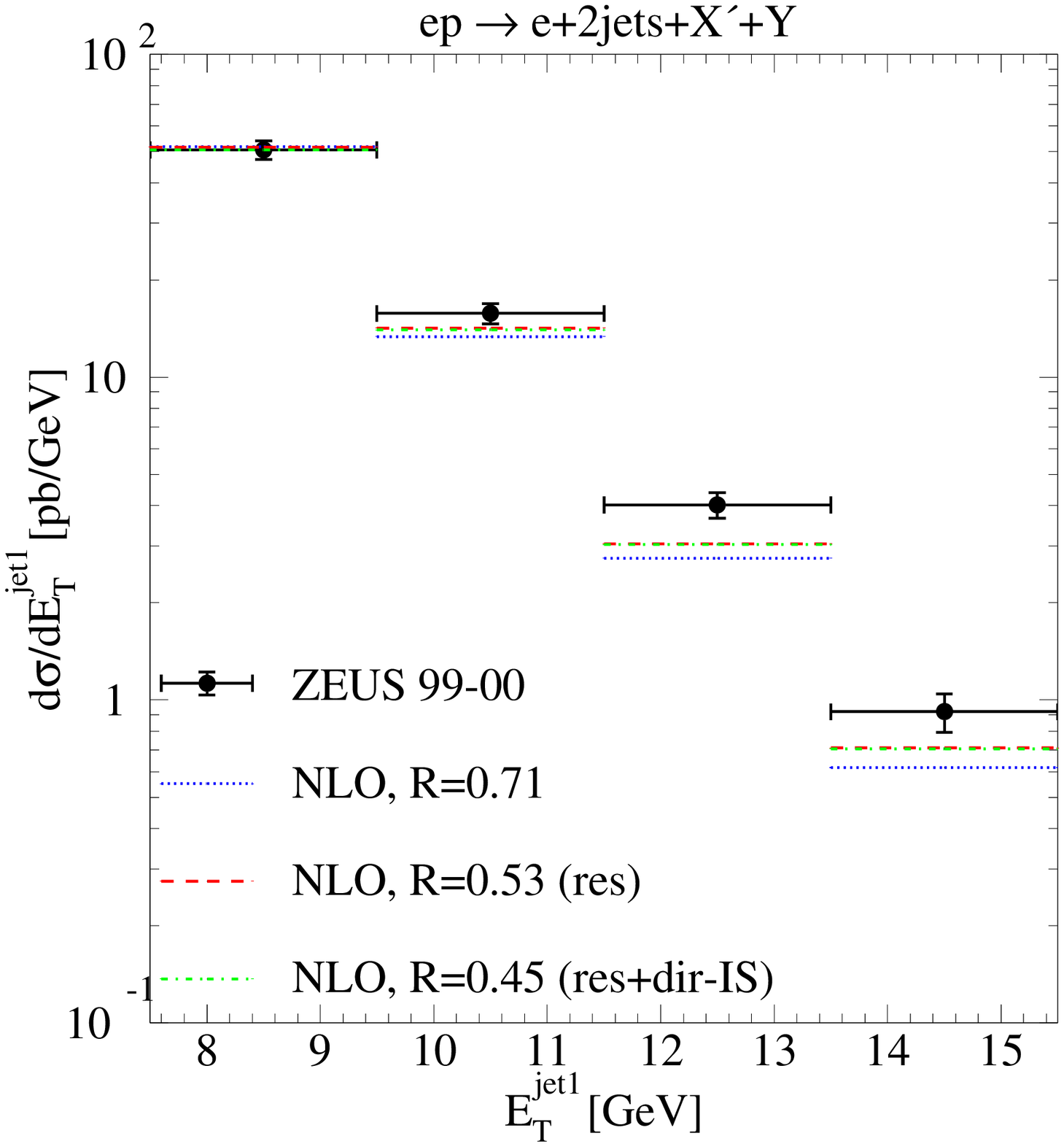}
 \includegraphics[width=0.325\columnwidth]{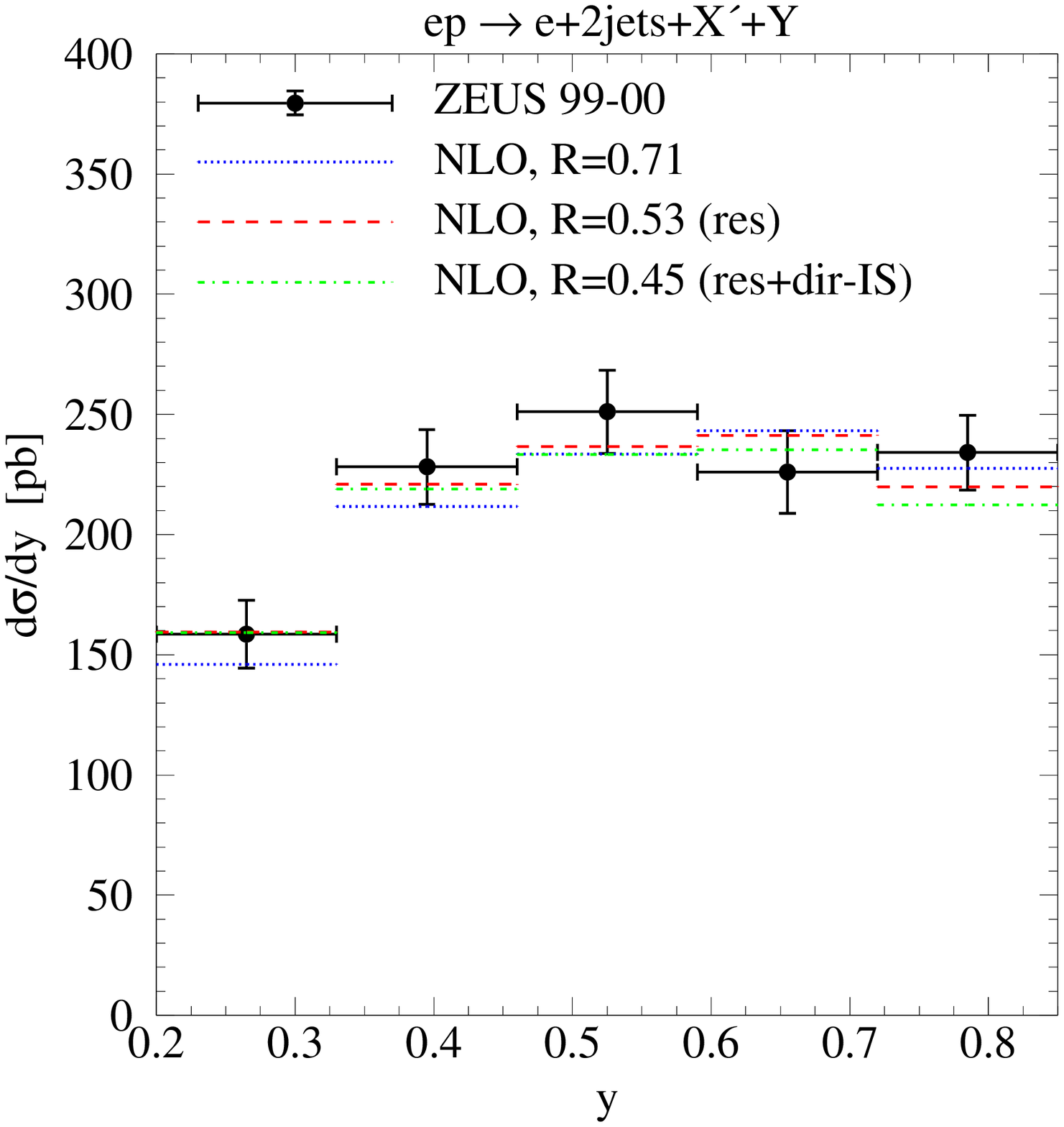}
 \includegraphics[width=0.325\columnwidth]{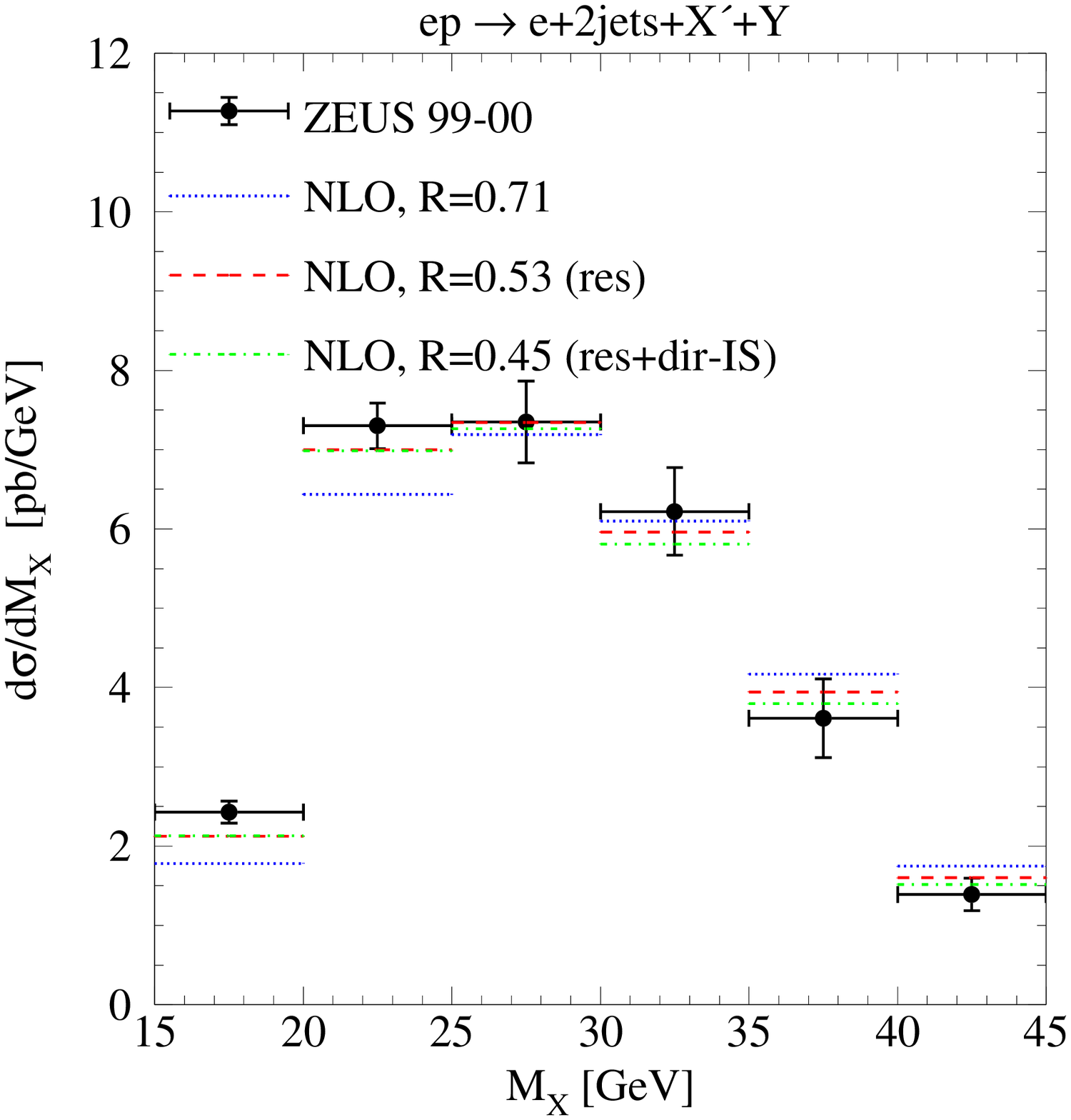}
 \includegraphics[width=0.325\columnwidth]{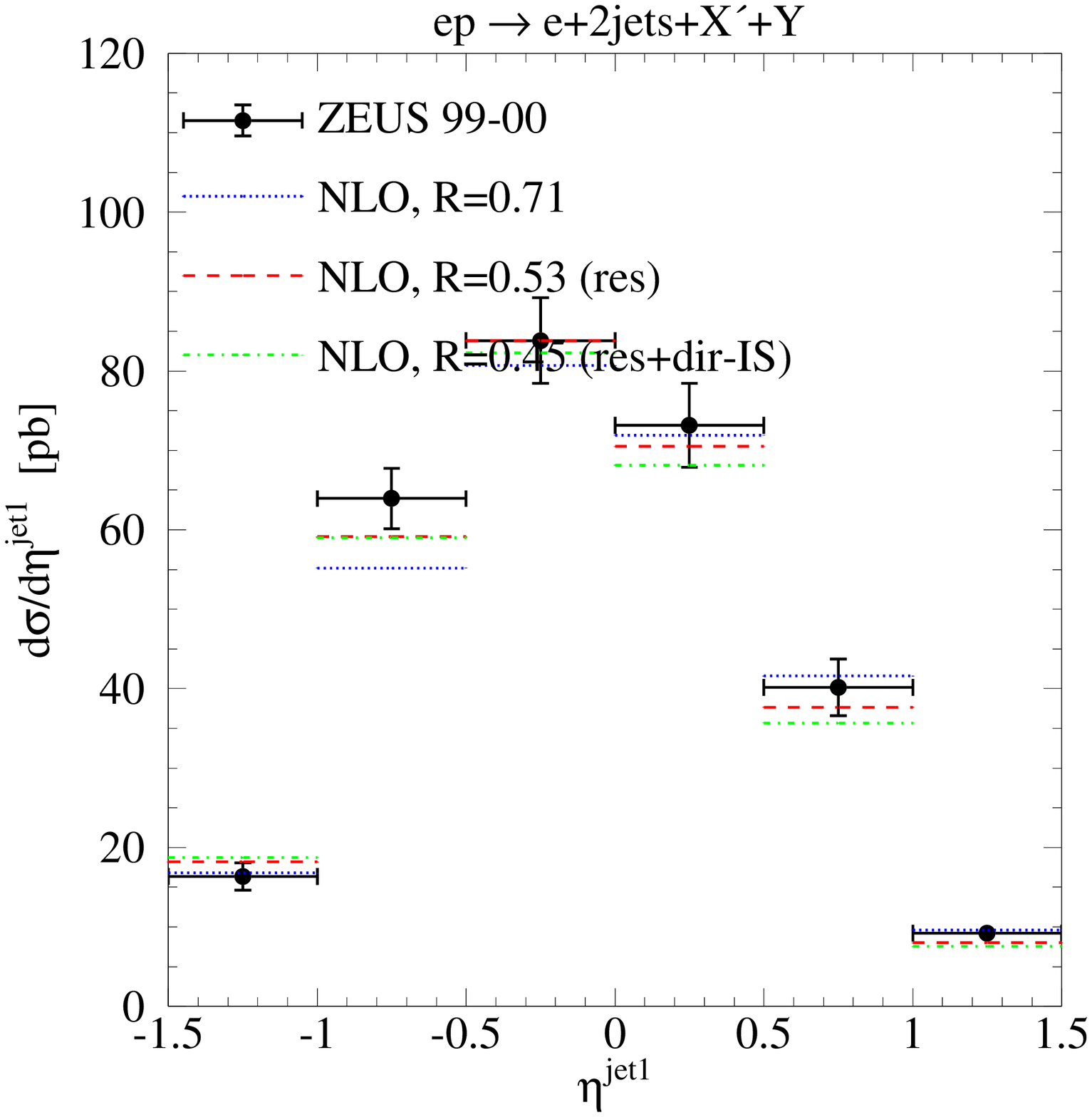}
 \caption{\label{fig:10}Differential cross sections for diffractive dijet
 photoproduction as measured by ZEUS and compared to
 NLO QCD with global, resolved, and resolved/direct-IS suppression.}
\end{figure}
%
In general, we observe that the
difference between global suppression and resolved suppression is small,
i.e.\ the data points agree with the resolved suppression as well as with the
global suppression. 

In Figs.\ 11a and b the difference between `H1 2006 fit B' and `H1 2006 fit A'
%
\begin{figure}
 \centering
 \includegraphics[width=0.495\columnwidth]{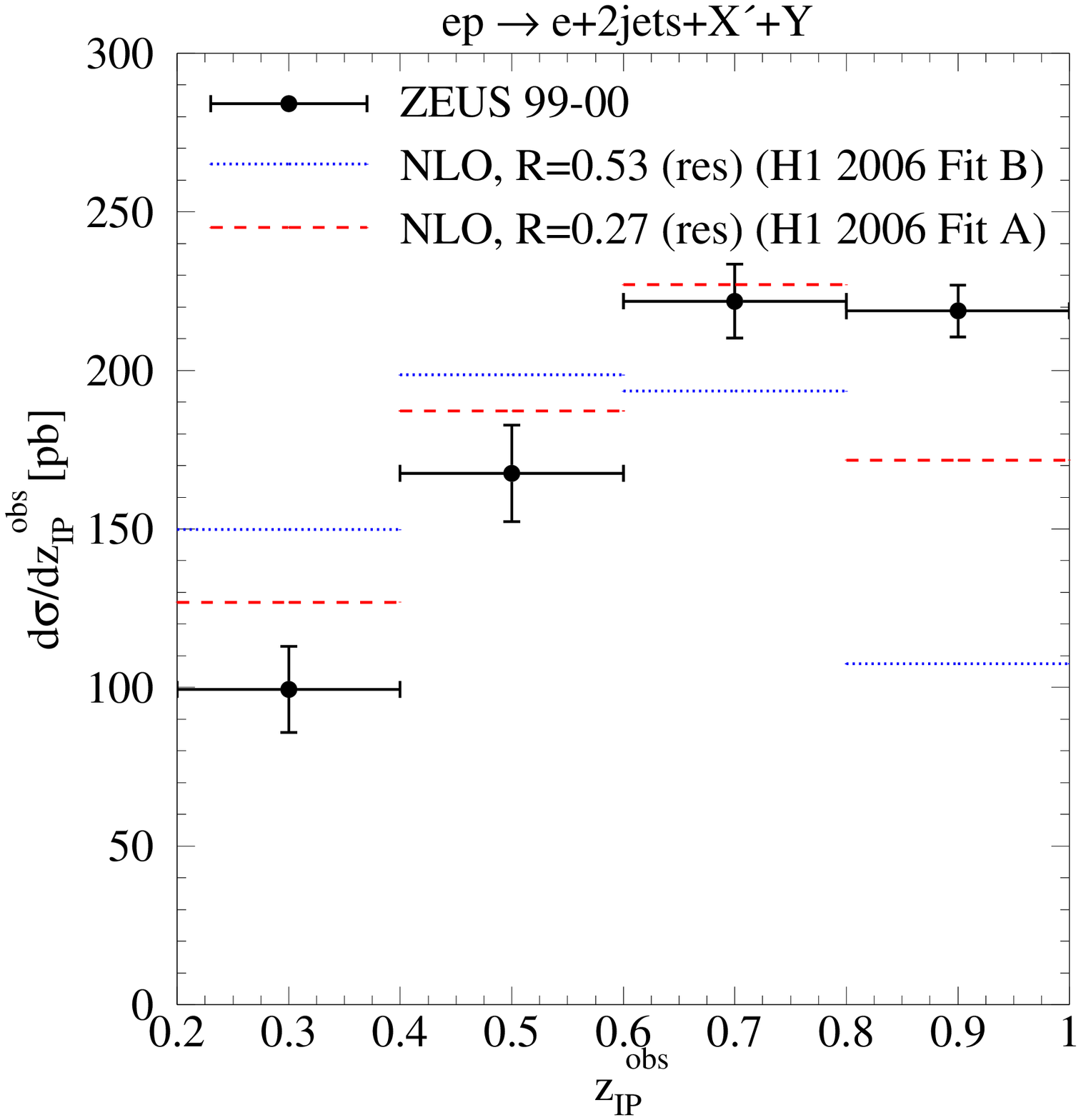}
 \includegraphics[width=0.495\columnwidth]{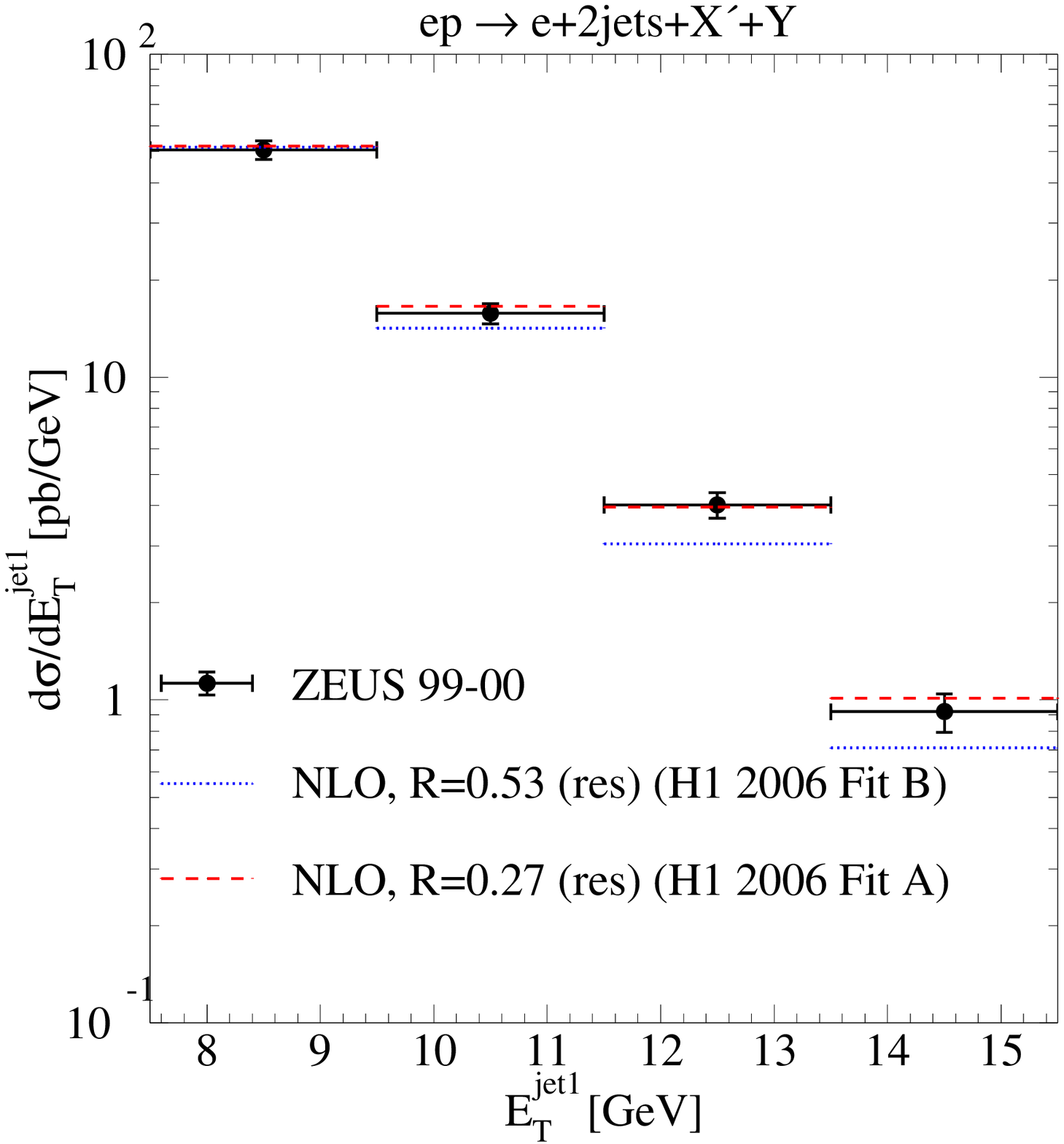}
 \caption{\label{fig:11}Differential cross sections for diffractive dijet
 photoproduction as measured by ZEUS and compared to
 NLO QCD with resolved suppression and two different DPDFs.}
\end{figure}
%
is shown again for the case of the resolved suppression. In both figures we
observe that the fit A suppression with the suppression factor $R = 0.27$
agrees better with the data than with the factor $R = 0.53$ for the
fit B suppression. In particular, for $d\sigma/dE_T^{jet1}$ the agreement with 
the three data points is perfect (note the logarithmic scale).

%
%
In our analysis of the ZEUS data so far we assumed that the measurements
with the large rapidity gap (LRG) method of ZEUS in \cite{28} are such that with
this method the same inclusive diffractive DIS cross section is measured as in
in the H1 measurement of this cross section with the LRG method [4], on
which the fits of the DPDFs 'H1 2006 fits A,B' are based. This is actually
not true, and this problem has been analysed by the ZEUS collaboration in
their publication, in which they present their data for the inclusive diffractive
DIS cross sections using different definitions for these cross sections
\cite{Zeus}.
They find that their LRG cross section has to be corrected by two factors in
order to make it agree with the H1 measurement of the diffractive DIS cross
section and with the prediction of this cross
section based on the `H1 2006 fit B'. First, a factor of $0.91\pm0.07$ was
estimated with the PYTHIA Monte Carlo, so that the ZEUS cross sections
correspond to a dissociative background with $M_Y < 1.6$ GeV. Second, even
then the so corrected ZEUS diffractive DIS cross section was still $13\%$ larger
than
the corresponding H1 diffractive DIS cross section. This amounts to a total
correction
factor of $0.79\pm0.06$. If we apply this correction factor to our ZEUS
suppression factors, we obtain a global (resolved-only) suppression of 0.56 $\pm$
0.05 rather than 0.71 (0.42 $\pm$ 0.04 rather than 0.53), i.e.\ suppression
factors which are closer to the results found for the similar high-$E_T^{jet}$ H1
analysis. Here, the errors refer only to the errors coming from the
renormalization of the ZEUS data and do not include the experimental stat./syst.\
and theoretical scale errors.

\section{Conclusion}

In this paper, we confronted the final HERA data on the diffractive
photoproduction of dijets, as published by the ZEUS \cite{28} and H1 \cite{Aaron:2010su,29}
collaborations, with our NLO QCD calculations in order to see if factorization
breaking effects in the resolved and eventually direct cross sections can be
established consistently from both data sets. The new comparison is even more
conclusive than the one published previously in our invited review
\cite{34}, as the proton beam energy of 920 GeV, the cuts on the jet
transverse energies of $E_T^{jet1(2)}> 7.5$ $(6.5)$ GeV, and the cut on the
momentum fraction carried by the diffractive exchange $x_{\p}<0.025$ are now
the same in both experiments, which was not the case before. At the same time,
some experimental cuts are still different. In particular, the momentum fraction
$y$ transferred by the electron to the hadronic system is larger for ZEUS than
for H1. We also re-computed in NLO QCD the cross sections for the lower
cuts on $E_T^{jet1(2)} > 5$ (4) GeV, which have been re-measured by the H1
collaboration \cite{Aaron:2010su,29} in order to establish consistency with the their
previous low-luminosity data set \cite{27}.
We found that the large majority of H1 and ZEUS data points lay below the
NLO QCD predictions, even when using the `H1 2006 fit B' diffractive PDFs with
small gluon density at large fractional momentum and taking into account the
experimental (statistical and systematic) and theoretical (scale variation)
errors. The data at larger $E_T^{jet1}$ (or $M_{12}$) tended to agree better with
the NLO QCD predictions than those at small $E_T^{jet1}$.

By fitting the lowest (and dominant) bins in the three $E_T^{jet1}$-distributions,
we established the amounts by which {\em both} the direct and resolved NLO QCD
cross
sections had to be reduced to find agreement with the data. These suppression
factors are shown in the second line (`global') of Tab.\ 3 for the
low-$E_T^{jet}$ and
%
\begin{table}
 \caption{Suppression factors for global and resolved-only suppression
 in the low-$E_T^{jet}$ and high-$E_T^{jet}$ analyses of H1 and the ZEUS
 analysis before and after rescaling their data by a factor of 1.15.}
 \begin{tabular}{c|c|c|c|c}
 Suppression factor & H1 low-$E_T^{jet}$ & H1 high-$E_T^{jet}$ & ZEUS
 & ZEUS renormalized \\
 \hline
 global        & 0.50 & 0.62 & 0.71 & 0.56 $\pm$ 0.05 \\
 resolved-only & 0.40 & 0.38 & 0.53 & 0.42 $\pm$ 0.04 \\
 res+dir-IS    & 0.37 & 0.30 & 0.45 & 0.36 $\pm$ 0.03 \\
 res, H1 2006 fit A & 0.32 & 0.16 & 0.27 & 0.21 $\pm$ 0.01 \\
 \end{tabular}
\end{table}
%
high-$E_T^{jet}$ analyses of H1, the ZEUS analysis (multiplied by a factor of 1.15
to allow for proton dissociation), and the ZEUS analysis renormalized by a factor
of $0.79\pm0.06$ for $M_Y<1.6$ GeV and correspondance with the H1 measurements
and DPDF fits \cite{Zeus}. The first, second and third factors agreed very
well with those found by the experimental collaborations when fitting multiple
distributions or total cross sections. The fourth factor (`ZEUS renormalized')
agrees better with the second factor, relevant for the similar high-$E_T^{jet}$
H1 analysis.
%
%
%
We also tested the hypotheses that factorization breaking is only present
in the resolved (third line) or the resolved and the related initial-state
singular part of the direct photoproduction cross sections (fourth line).
Both hypotheses gave very similar
results and described the data sets almost as well as the predictions with
global factorization breaking. The suppression factors applicable to just the
resolved cross section are shown in the third line of Tab.\ 3. As observed
previously \cite{18}, they agree very well with absorptive-model predictions
\cite{12}. We conclude that in this case the suppression factors do not show a
significant $E_T^{jet}$-dependence, in particular when renormalizing the ZEUS
data as described above. The fact that no $E_T^{jet}$-dependence
is visible here can, of course, be explained by the fact that the resolved cross
section falls more steeply with $E_T^{jet}$ than the direct one \cite{11}.
Finally, we investigated whether these conclusions depended on the diffractive
PDFs by comparing the results with resolved-only suppression of the `H1 2006
fit B' to those obtained with the `H1 2006 fit A' (last line in Tab.\ 3).
Since the latter has a larger
gluon density at large momentum fraction, the suppression had to be more
important. The fit A results then tended to describe the high-$E_T^{jet}$ H1 data
and the ZEUS data slightly better, in particular in the
$z_{\p}^{obs}$-distribution, which should be directly sensitive to the DPDFs, but
the low-$E_T^{jet}$ H1 data slightly worse. Unfortunately the experimental and
theoretical errors are still too large to draw any strong conclusions.

While the epoch of HERA experiments has now ended and an International
Linear Collider may not be built in the near future, it will be very interesting
to investigate diffractive physics at the LHC. Suprisingly, proton-proton and
heavy-ion collisions at the LHC can also be a source of high-energy photon
collisions, and this may open up a whole new field of investigation for
diffractive dijet photoproduction \cite{Klasen:2008ja}.

\end{document}